\documentclass[prl,aps,amssymb,twocolumn,superscriptaddress,notitlepage]{revtex4-2}
\usepackage{amsmath}
\usepackage{amssymb}
\usepackage{amsthm}
\usepackage{amsfonts}
\usepackage{listings}
\usepackage{physics}
\usepackage{enumerate}
\usepackage{latexsym}
\usepackage{psfrag}
\usepackage{bm}
\usepackage{graphicx}
\usepackage[caption=false]{subfig}
\usepackage{blkarray}
\usepackage{array}
\usepackage{color}
\usepackage[normalem]{ulem}
\usepackage{hyperref}
\hypersetup{
     colorlinks = true,
     linkcolor = magenta,
     citecolor = magenta
}
\usepackage{mathtools}

\begin{document}
\title{Quantum Transport Protected by Acceleration From Nonadiabaticity and Dissipation} 

\author{Arnab Chakrabarti}
\thanks{These two authors contributed equally.}
\affiliation{AMOS and Department of Chemical and Biological Physics, Weizmann Institute of Science, Rehovot -- 7610001, Israel}
\affiliation{Department of Physics, Rajiv Gandhi University, Rono Hills, Doimukh -- 791112, Arunachal Pradesh, India}
\email{arnab.chakrabarti@rgu.ac.in}

\author{Biswarup Ash}
\thanks{These two authors contributed equally.}
\affiliation{Department of Physics of Complex Systems, Weizmann Institute of Science, Rehovot -- 7610001, Israel}
\affiliation{Department of Physics, University of Michigan, Ann Arbor, MI 48109, USA}

\author{Igor Mazets}
\affiliation{Vienna Center for Quantum Science and Technology (VCQ),\\ Atominstitut, TU Wien, 1020 Vienna, Austria}
\affiliation{\textcolor{black}{Wolfgang Pauli Institut c/o Fakult\"at f\"ur Mathematik, Universit\"at Wien, 1090 Vienna, Austria}}

\author{Xi Chen}
\affiliation{Instituto de Ciencia de Materiales de Madrid (CSIC), Cantoblanco, E-28049 Madrid, Spain}

\author{Gershon Kurizki}
\affiliation{AMOS and Department of Chemical and Biological Physics, Weizmann Institute of Science, Rehovot -- 7610001, Israel}

\begin{abstract}

We put forth a hitherto unexplored  control strategy that enables high-fidelity fast transport of an unstable quantum wavepacket even in the presence of bath-induced dissipation. The wavepacket, which is confined within any shallow (anharmonic) potential trap is steered in acceleration, so as to maximize the transfer fidelity. This strategy can generally optimize any non-Markovian bath-dressed continuous-variable system dynamics. It can simultaneously cope with wavepacket leakage via non-adiabatic transitions and bath-induced dissipation in an optimal fashion. It can outperform methods based on counterdiabatic \textcolor{black}{fields} (shortcuts to adiabaticity) particularly for fast non-adiabatic transport. Transport fidelity is maximized even for trajectories exceeding the speed of bath-excitation propagation, e.g. for supersonic transfer through phonon baths. This general approach is illustrated for optimized transfer of impurities in Bose-Einstein condensates. It is applicable to both dissipative and non-dissipative transfer of trapped atoms and ions and molecular reaction products.

\end{abstract}

\maketitle

%-----------------------------------------------------------------------------------
\section*{Introduction}

The ability to minimize the relaxation and decoherence of open quantum systems, so as to protect the fidelity of their coherent evolution, is a major challenge of quantum science and technology, at both fundamental and applied levels \cite{nielsenchuang, degen17, kkbook21}. The main thrust has been on decoherence control of discrete variables in qubit systems because of their central role in quantum information processing \cite{nielsenchuang}. Such control calls for intervention in the system evolution on non-Markovian time scales \cite{breuer02, kkbook21, clausen10, viola98}. 

Here we address the much less explored task of suppressing the leakage of unstable quantum wavepackets from finite-depth (hence anharmonic) trapping potentials to the continuum, a process akin to tunneling in nuclear alpha decay  \cite{gamow1928} and atomic traps \cite{kofman01, wilkinson1997experimental} or in superconducting devices \cite{barone04}. \textcolor{black}{The spread of stable quantum wavepackets can be suppressed by a resonant drive \cite{buchleitner2002non}. However, this method does not apply to the scenarios discussed here.} 

Because of the unrestricted number of degrees of freedom involved, the task of maintaining high fidelity of the initial wavepacket becomes much more challenging when the wavepacket is moving through a dissipative environment. Pertinent scenarios involve a trapped multiatom impurity moving through a condensate as a Bose polaron \cite{Coalson19, Lampo17, Mazets05}, or as a part of a quantum refrigeration cycle \cite{niedenzu19}; transport of ions in a trap \cite{jain24, sterk22, walther12, Rowe02}, trapped-atoms transported in vacuum (e.g. using tweezers) \cite{ibloch22, ibloch01} where dissipation vanishes but non-adiabatic leakage from the trap hampers fidelity. Analogous scenarios arise in molecular dissociation or collisions \cite{opatrny01, deb83} where the wavepacket is moving along a potential surface while being dissipated by other (rovibronic or electronic) degrees of freedom \cite{brumer12, tannor07}.

The fundamental dilemma in such scenarios is that the faster is the wavepacket transferred the less it is affected by the environment (bath), but, on the other hand, its nonadiabatic evolution increases its leakage out of the finite-depth trap. \textcolor{black}{To prevent such leakage, the standard recipe that comes to mind is the use of shortcuts to adiabaticity (STA), either by applying counterdiabatic fields (CDF) or by inverse engineering of the system Hamiltonian based on dynamical invariants  \cite{polkovnikov17, berry09, Torrontegui11, chen2011optimal, Chen11, Chen151, Chen152, Chen22, opatrny14, odelin19, ness18, dengis2024accelerated}. For continuous variable systems such as trapped wavepackets, only dynamical invariants quadratic in momentum are useful \cite{Chen11}. However, such invariants only exist for the special Lewis-Leach class of potentials \cite{LR69, lewis1982direct, Torrontegui11, Chen11}. Consequently, for wavepackets in arbitrary trapping potentials, invariant based STA methods are hardly applicable.} 

\textcolor{black}{In general, STA is mainly geared to closed, stable quantum systems \cite{odelin19} since, being a hamiltonian method, it is apriori unclear, to what extent can STA suppress irreversible bath effects \cite{odelin19, villazon19, yin22}? One should be mindful that the success of any Hamiltonian protocol for a lossy quantum system depends on the typical ratio of the level-width to level-spacing. Yet, there have been numerous extensions of STA techniques to open quantum systems, mostly of either discrete variables or harmonic potentials (invariants quadratic in momentum) \cite{chen2015fast, luo2015dynamical, jing2013inverse, levy2018noise, lu2014fast, ruschhaupt2012optimally, sarandy2007dynamical, wu2015dynamical, maamache2017pseudo, vacanti2014transitionless, delcampo20, wu2017adiabatic, pancotti2020speed, alipour2020shortcuts, santos21, kosloff19, impens2019fast, wu21, mahunta2024shortcuts, boubakour24, zhou24}. Application of the invariant based STA in open quantum systems, as in \cite{chen2015fast, luo2015dynamical, jing2013inverse, levy2018noise, lu2014fast, ruschhaupt2012optimally, sarandy2007dynamical, wu2015dynamical, maamache2017pseudo} has limited applicability for arbitrary (non-harmonic) trapping potentials. Extensions of the CDF method to transitionless driving of open quantum systems, typically composed of discrete variables, may involve non-Hamiltonian or non-Hermitian control \cite{vacanti2014transitionless, wu21}, as opposed to the unitary control for closed-system STA.}

On top of the difficulties in using STA for general open quantum systems, none of the above methods can simultaneously control the irreversible leakage of an unstable wavepacket due to combined non-adiabatic and bath-induced transitions in a realistic model, where the time-dependence of the system Hamiltonian induces changes in the bath-induced leakage. This is the challenging problem we address in this work.

\begin{figure*}[!t]
\centering
\includegraphics[width=1\textwidth]{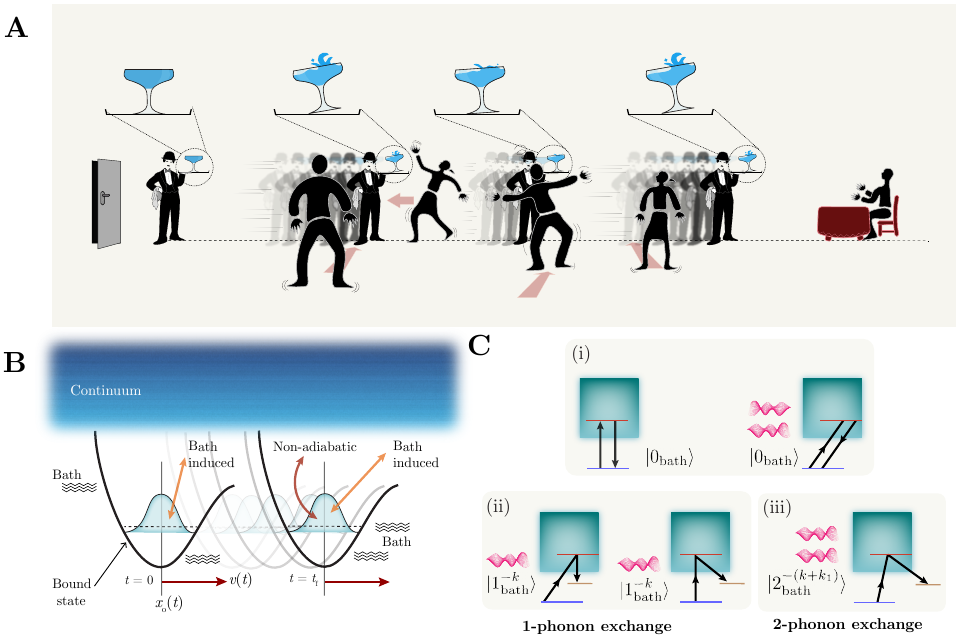}
\caption{\textbf{A}: Classical analog of our problem: A waiter carrying a shallow glass filled up to the brim with water, in a crowded hall. Random kicks from the crowd makes water spill out of the glass even when the waiter is static. Moving with a tilted tray without spilling water from such an unstable system is almost impossible. Neither is evasive maneuvering by the waiter which requires frequent feedback. Instead, the waiter can speed up as much as possible to avoid kicks, with occasional slowdown to keep the energy constraint. \textbf{B}: Schematic view of nonadiabatic transport of a quantum wavepacket through a dissipative medium in a shallow moving trap or motion along a dissipative potential surface. \textbf{C}: Self-energy diagrams contributing to leakage and dissipation: (i) No quanta exchange (purely non-adiabatic -- left panel) and  virtual quanta exchange (purely bath-mediated -- right panel) with the bath. (ii) Real quanta exchange via partly non-adiabatic and partly bath-induced processes  (iii) Real quanta exchange with the bath via excitation non-conserving processes. \textcolor{black}{In panels (i), (ii) and (iii) the horizontally shifted levels indicate advanced (scattered) wavepacket states, the vertical arrows indicate non-adiabatic transitions, while the oblique arrows indicate bath-induced transitions.}}\label{S1-1}
\end{figure*}

To illustrate the difficulty, it is instructive to consider the classical analog of the problem: a waiter carrying a shallow glass of water filled to the brim (Fig. \ref{S1-1} A) in a crowded hall. Random kicks can cause water to be spilled out of the glass even when the waiter stands still, let alone moves. The compensating  counter-diabatic field (CDF) is tantamount to tilting the tray \cite{polkovnikov17}. Yet such tilting  cannot prevent the waiter from spilling  out the water, the glass being a randomly-perturbed unstable system. 

Here we advance an altogether different strategy, dubbed acceleration-controlled quantum dissipative transport (AC-QUDIT), which we rigorously show to be more effective than STA for any randomly-kicked unstable system carried  over a broad range of transport speeds: to minimize the spilling, the waiter should move around at a variable pace on non-Markovian time scales, i.e. change velocity faster than the correlation time of the random kicks. We show how such pace control should be executed, if instead of water the shallow vessel would contain a quantum liquid or wavepacket. 

To this end we introduce motion control of arbitrary dissipating wavepackets by a general non-Markovian description of coupled system-bath quantum dynamics instead of the Wigner-Weisskopf description \cite{wignerweisskopf, scully},  previously used by our group in discrete-variable control \cite{kofman01, zwick14}. Euler-Lagrange (EL) optimization then yields a non-Markovian integro-differential equation of motion of the wavepacket, that accounts for both non-adiabatic and bath effects. For a general Fr{\"o}lich\textcolor{black}{-}type coupling \cite{frolich52, Coalson19, Lampo17}, this results in a nonlinear integro-differential equation which is analytically solvable in a speed regime that, remarkably, allows for significantly non-adiabatic transfer and can even be faster than the bath-excitation propagation, e.g. \textcolor{black}{take place at a} supersonic speed in phonon baths. The more non-adiabatic the motion, the more advantageous is AC-QUDIT compared to STA.  The advocated approach is broadly applicable to atomic and molecular quantum wavepacket transport in the scenarios mentioned above \cite{Coalson19, Lampo17, Mazets05, niedenzu19, Rowe02, ibloch22, ibloch01, opatrny01, deb83, brumer12, tannor07}.
%===============================
%===============================
%-----------------------------------------------------------------------------------

\section*{Results}
\subsection*{General scenario}

\noindent We consider a wavepacket describing a quantum single- or multi-partite system of mass $m$, in a trapping potential $V[x - x_{\circ}(t)]$, whose center (minimum) position $x_{\circ}(t)$ is driven by an external force. It is immeresed in a bath of interacting or free bosons of mass $m_B$ (Fig. 1 B). The system $+$ bath compound is described by the time-dependent Hamiltonian
%===============================
\begin{align}\label{ham}
H(t) &= H_S(t) + H_B + H_{SB}, \nonumber \\
H_S(t) &= \frac{p^2}{2\,m} + V[x - x_{\circ}(t)]~;~ H_B = \sum_{k \neq 0} \Omega_kb_k^{\dagger}b_k.
\end{align}
%===============================
\noindent Here $b^{\dagger}_k$ and $b_k$ are the creation and anihilation operators for a bath excitation, e.g. a phonon with frequency $\Omega_k$ and momentum $k$, while $x$ and $p$ denote the position and conjugate momentum of the system. 

Rather generally, the quantized system-bath interaction Hamiltonian, without resorting to the Lamb-Dicke limit \cite{leibfried03}, nor to the rotating wave approximation \cite{scully}, is taken to be \textcolor{black}{of the Fr{\"o}lich-type \cite{frolich52, Coalson19, Lampo17}:}
\begin{equation}\label{hamib}
       H_{SB} = \sum_{k \neq 0} \big[ g_k\; \textcolor{black}{b_{-k}}\; e^{-i k x} + h.c.],
\end{equation}
\textcolor{black}{$g_k$ being the $k$-mode coupling-constant (see Methods, SI). Here, the system operator $e^{-ikx}$ imparts an overall momentum of $-k$ to the instantaneous wave-packet while the operator $b_{-k}$ simultaneously annihilates a phonon having the same momentum $-k$, thereby ensuring conservation of momentum (see SI IV, Methods A).}

For simplicity, the trap $V[x - x_{\circ}(t)]$ is assumed to be shallow, such that it supports a single bound-state along with a continuum of unbound (scattering) states, although this assumption is non-essential. Then, under non-adiabatic and bath-induced transitions, the system dynamics is similar to the Friedrichs model for resonance phenomena \cite{friedrichs1}. 

%-----------------------------------------------------------------------------------
\subsection*{Transported wavepacket dynamics}

The system-bath dynamics can be described by an expansion of the combined system $+$ bath state at time $t$ as:
%===============================
\begin{equation}\label{genstate}
  \vert \psi (t) \rangle = \sum_{\alpha} \, A_{\alpha}(t)\,e^{-i\omega_{\alpha} t}\,\vert \alpha (t) \rangle \; + \; \sum_{\beta} \, A_{\beta}(t)\,e^{-i\omega_{\beta} t}\,\vert \beta (t) \rangle. 
\end{equation}
%===============================
Here $\lbrace \vert \alpha(t) \rangle \rbrace$ and $\lbrace \vert \beta (t) \rangle \rbrace$ denote the instanatenous Fock product basis-states of $H_S(t) + H_B$ \cite{Coalson19}, corresponding to bound and unbound states of the system, respectively.  For any $l \in \lbrace \alpha\rbrace, \lbrace \beta \rbrace$, $\omega_l$ denotes the eigen-energy ($\hbar = 1$) and $A_l(t)$ the instantaneous probability amplitude of the bound or unbound state $\vert l(t)\rangle$. The instantaneous energy eigenstates are time-dependent, but the corresponding instantaneous eigen-energies are independent of time for a given potential $V(x)$. In this instantaneous eigenbasis, the bound-to-continuum transition amplitudes of the system are dependent on the centre-of-mass position $x_{\circ}(t)$ and velocity (speed) $\dot{x}_{\circ}(t)$ (see Methods A). We can then design an optimal control of the trajectory $x_{\circ}(t)$ so as to maximize the bound-state transfer fidelity. 

We label the instantaneous bound and scattering states of the moving trap by $\vert n(t)\rangle$ and $\vert \epsilon (t) \rangle,$ with $\epsilon > 0$, which are functions of $x-x_{\circ}(t)$. At the start of the quench, the system is assumed to be in the bound state of potential with no excitation in the bath, denoted by $\vert \nu(\textcolor{black}{0})\rangle = \vert n(\textcolor{black}{0}) \rangle \otimes \vert 0_{\rm bath} \rangle$, where $\vert 0_{\rm bath} \rangle$ indicates the vacuum bath state. The Schr{\"o}dinger equation $i\frac{\partial}{\partial t} \vert \psi(t) \rangle = H(t)\vert \psi(t)\rangle$ then yields the set of equations

%===============================
\begin{align}\label{dynamics1}
\dot{A}_{l} &= -\sum_{j} A_{j}\,R_{lj}(t)  - i\sum_{j} A_{j}\,S_{lj}(t) ,\\  
&\forall \; l, j \in \lbrace\alpha\rbrace,\lbrace\beta\rbrace. \nonumber
\end{align}
%===============================
In the first term on the r.h.s. of Eq. (\ref{dynamics1}) $R_{lj}(t) = \langle l(t) \vert \frac{\partial}{\partial t}\vert j(t)\rangle$ represent non-adiabatic transition rates due to the motion while in the second term $S_{lj}(t) = \langle l(t)\vert H_{SB} \vert j(t)\rangle$ denote the bath-mediated transitions induced by the coupling $H_{SB}$.  

\textcolor{black}{At $t> 0$, the system-phonon scattering entangles \textit{advanced}, \textit{retarded} and \textit{instantaneous} bound and excited states of the system wavepacket and the many-body bath states, which are required for momentum conservation (see SI IV). The survival probability (fidelity)  of the instantaneous bound state of the trapped wavepacket is the same as the Loschmidt-echo probability \cite{Coalson19, znidaric06}} of finding the system in the \textcolor{black}{instantaneous} bound state with no bath excitations, \textcolor{black}{$\vert \nu(t)\rangle = \vert n(t)\rangle\otimes\vert 0_{\rm bath}\rangle$}, at time $t$ \textcolor{black}{(see Methods A, SI IV, VI)} . To evaluate this probability, we integrate out all probability amplitudes in the bound and unbound sectors except that of the initial state. 

In our non-perturbative approximation we retain only terms of leading order in $\dot{x}_{\circ}$ and $H_{SB}$, in the rate of change of the survival probability amplitude of the initial wavepacket \cite{Cohen-Tannoudji_QM, Cohen-Tannoudji_Atom-Photon}. This standard approximation in decay theory is valid whenever the coupling strengths (system-bath + non-adiabatic) are weaker than the inverse time-scales of the corresponding reservoir (bath or continuum) dynamics \cite{Cohen-Tannoudji_QM, Cohen-Tannoudji_Atom-Photon, breuer02, keitel1995resonance, kofman01, riera2021quantum, nielsen2019critical, gordon07} (see SI V). Then the self-energy diagrams for the rate of change of the Loschmidt echo amplitude consist of only diagrams shown in Fig. \ref{S1-1} C (i). These involve simultaneous system-and-bath (de)excitations which are negligible at long time-scales but are important at short times, when the system energy-uncertainities are substantial. The resulting expression for the survival probability at $t = t_f$, obtained from the Wigner-Weisskopf method, is a resummation of all second-order processes in Fig. \ref{S1-1} C (i) \cite{Coalson19, boyanovsky2011perturbative}, akin to a Dyson series and hence represents a \textit{non-linear} response (see SI V):

%===============================
\begin{subequations}
\begin{align}
 \mathcal{P}(t_f)  & =  \exp\Big(-J[x_{\circ},\dot{x}_{\circ}]\Big),\label{cost}\\
J[x_{\circ},\dot{x}_{\circ}] & =   \;\text{Re}\int d\epsilon\,\Big[\; \int\limits_0^{t_f}dt_1\;\int\limits_0^{t_f}\;dt_2\;\Big\lbrace \,\underbrace{\gamma_{n\epsilon}(t_1)\,\gamma_{n\epsilon}^*(t_2)}_{\rm non-adiabatic} \nonumber\\ 
                             & + \frac{L}{2\pi}\int dk \underbrace{\Delta_{n\epsilon}^k(t_1)\,\Delta_{n\epsilon}^{k *}(t_2)}_{\rm bath-induced} \Big\rbrace\Big]. \label{cost1}
\end{align}
\end{subequations}
%===============================
\noindent Here the integrals over $\epsilon$ and $k$ indicate the cumulative effects of all $n \rightarrow \epsilon$ (bound-to-continuum) transitions and all $k$-modes of the bath respectively \textcolor{black}{and} $L$ is the confining length of the medium (bath). The first term on the r.h.s. of Eq. (\ref{cost1}) describes non-adiabatic transitions in Fig. \ref{S1-1} C (i-left) in terms of the leakage rates $\gamma_{n\epsilon}(t) = e^{-i\omega_{\epsilon n} t}\,\langle n(t) \vert \dot{H}_S(t)\vert \epsilon(t)\rangle/\omega_{\epsilon n}  = \dot{x}_{\circ}(t) \big(\frac{\mu_{n\epsilon}}{\omega_{\epsilon n}}\big) \, e^{-i\omega_{\epsilon n} t}$ where $\big(\frac{\mu_{n\epsilon}}{\omega_{\epsilon n}}\big)$ denotes the nonadiabatic coupling strength with frequencies $\omega_{\epsilon n} = \omega_{\epsilon} - \omega_{n}$ and $\mu_{n\epsilon} = -\langle n(t)\vert \frac{\partial V(q)}{\partial q}\vert \epsilon(t)\rangle $ ; $q = x - x_{\circ}(t)$.
%===============================
%\begin{equation}
%    \mu_{n\epsilon} = -\langle n(t)\vert \frac{\partial V(q)}{\partial q}\vert \epsilon(t)\rangle\,,\, q = x - x_{\circ}(t).
%\end{equation}
%===============================
The second term describes the bath-mediated processes in Fig. \ref{S1-1} C (i-right) in terms of the transition rates that are proportional to the system-bath coupling strengths $g_k$ for bath mode $k$: $ \Delta^k_{n\epsilon}(t) = e^{-ikx_{\circ}(t)}g_k\,\textcolor{black}{\langle n(t)\vert\; e^{-ikq}\;\vert \epsilon(t)\rangle} \,e^{-i(\Omega_k +\omega_{\epsilon n})t}$ integrated over bath-mode wave-vectors $k$. The survival probability for dissipationless transport corresponds to the vanishing of the bath-induced terms in Eq. (\ref{cost1}). In deriving the above expression, we have used the fact that the volume of $k$-space per allowed value of $k$ is $\Delta k =2\pi/L$.

\textcolor{black}{Upon resorting to the} finite-time Fourier transform $f_{t_f}(\omega) = \int_0^{t_f}d\tau \,e^{-i\omega \tau} f(\tau)$, we can rewrite the first \textcolor{black}{(non-adiabatic)} term  on the r.h.s. of (\ref{cost1}) as : $\int d\epsilon \big\vert\frac{\mu_{n\epsilon}}{\omega_{\epsilon n}}\big\vert^2 \big\vert (\dot{x}_{\circ})_{t_f}(\omega_{\epsilon n})\big\vert^2 \geq 0$ while the second \textcolor{black}{(bath-induced)} term becomes $\frac{L}{2\pi}\int d\epsilon\,dk\,\big\vert (\Delta_{n\epsilon}^k)_{t_f} (\omega_{\epsilon n})\big\vert^2 \geq 0$. The positivity of both terms confirms the decay of the survival probability in time. The non-adiabatic contribution to $J[x_{\circ},\dot{x}_{\circ}]$ is identified as the power-spectral density (PSD) of the trap speed, integrated over the entire range of the continuum index $\epsilon$ of the system.

%---------------------------------------------------------------------
\subsection*{Control of non-adiabatic and dissipation losses by optimized acceleration}

The maximum survival probability at $t = t_f$ corresponds to the minimized value of $J[x_{\circ},\dot{x}_{\circ}]$, subject to a physical constraint, chosen here to be related to the kinetic energy $E_K$ supplied for the transport \cite{clausen10}. For a given $E_K$, the Lagrange multiplier $\lambda$ determines the total cost functional for Euler-Lagrange (EL) optimization as 
%===============================
\begin{eqnarray}\label{constraint}
    J_{\rm tot}[x_{\circ},\dot{x}_{\circ}] & = & J[x_{\circ},\dot{x}_{\circ}] + \lambda\, J_1[\dot{x}_{\circ}] ; \nonumber\\
     J_1[\dot{x}_{\circ}]  & = &  \int\limits_0^{t_f} d\tau \; [\dot{x}_{\circ}(\tau)]^2 - E_K .
\end{eqnarray}
%===============================
The stationarity condition $\delta J_{\rm tot} = 0$ results in a \textit{non-linear} and \textit{non-local} EL equation, obtained from Eqs. (\ref{cost1}--\ref{constraint}) \textcolor{black}{(see SI VII)}. This \textit{non-linear} integro-differential equation (Methods) cannot be solved exactly. Its brute-force numerical solution can be impractical in many cases. Approximate analytical solutions are only possible in the speed range, $\vert \dot{x}_{\circ}(t) \vert < v_s = \big\vert (\omega_{\epsilon n} + \Omega_k)/k\big\vert$ for a given transition frequency $\omega_{\epsilon n}$ and Fourier harmonic $\Omega_k$ of the $k$-mode bath response \textcolor{black}{(see Methods, SI)}. We then obtain to lowest order in this high-speed factor, the following linearized integro-differential EL equation which is our central result (Methods \textcolor{black}{B}):
%===============================
\begin{equation}\label{ELfinal}
\lambda\,\ddot{v}(t) = - \eta(t) - \zeta(t)\,v(t) + \int\limits_0^{t_f}\,d\tau\,\phi(t - \tau)\,v(\tau).
\end{equation}
%===============================
Here $v(t) = \dot{x}_{\circ}(t)$, $\eta(t) = \int_0^{t_f}\;ds \; \frac{L}{2\pi}\int d\epsilon\,dk\,(\omega_{\epsilon n} + \Omega_k)\,k\,\vert \widetilde{g}_{n\epsilon}^{(k)} \vert ^2 \,\cos\Big[(\omega_{\epsilon n} + \Omega_k)$ $(t - s)\Big]$, $\zeta(t) = \int_0^{t_f}\;ds \;\frac{L}{2\pi}\int d\epsilon\,dk\, k^2\,\vert \widetilde{g}_{n\epsilon}^{(k)} \vert ^2$ $\cos\Big[(\omega_{\epsilon n} + \Omega_k)(t - s)\Big]$, $\widetilde{g}_{n\epsilon}^{(k)} = g_k\,\langle n(t)\vert e^{-ikq}\vert \epsilon(t)\rangle$  and $\phi(t) = \int d\epsilon\, \vert\mu_{n\epsilon}\vert^2 \cos\big[\omega_{\epsilon n}\,t\big]$.
We have arrived at this \textit{linear} integro-differential equation (\ref{ELfinal}) from the fully non-linear EL equation obtained from minimization of Eq. (\ref{constraint}) by a method adopted in frequency discriminator circuits for FM demodulation \cite{haykin01}. 

\textcolor{black}{Equation} (\ref{ELfinal}) determines the optimal acceleration control of a dissipated, nonadiabatic moving trapped wavepacket. It is a major generalization of the universal Kofman-Kurizki (KK) formula of non-Markovian dynamical control for discrete variables \cite{kofman01,opatrny01}. \textcolor{black}{In} the K-K formula the control field only acts as an amplitude modulation filter of the bath response spectrum whereas, here $\ddot{v} = \dddot{x}_{\circ}$ converts frequency variations of $k x_{\circ}(t)$ to amplitude variations of $k v(t)$ in the response function \textcolor{black}{which is} proportional to $\zeta(t)$ while $\phi(t)$ undergoes \textcolor{black}{the} usual amplitude filtering via $v(t)$. 

The salient merit of our AC-QUDIT method is the ability to optimally protect the fidelity against both non-adiabatic leakage and dissipation simultaneously, by trajectory control. We find this optimal trajectory by solving Eq. (\ref{ELfinal}), under boundary conditions $x_{\circ}(0) = 0$, $v(0) = 0$, while allowing for finite final trap-speed $v(t_f)$. \textcolor{black}{This choice of boundary conditions uniquely determines the optimal trajectory $x_{\circ}(t)$, thereby fixing the final trap position $x_{\circ}(t_f)$ (see SI X, XII).} Importantly, even the linearized EL equation (\ref{ELfinal}) depends on $k v(t)$ \textcolor{black}{as we venture} beyond the Lamb-Dicke regime $\vert k x_{\circ} \vert \ll 1$. \textcolor{black}{In fact $\vert k x_{\circ} \vert \geq 1$ values are} essential for effective control of bath-induced dissipation (Methods \textcolor{black}{B}).

In the case of dissipationless transport of the wavepacket through vacuum or near-vacuum, as explored in \cite{ibloch22,ibloch01} (i.e. zero system-bath coupling), the survival probability in Eq.~(5) becomes a functional of the trap velocity only, since $\Delta_{n\epsilon}^k(t)$ vanishes in (\ref{cost1}). In this case $v(t)$ serves as our control parameter instead of $x_{\circ}(t)$ and the corresponding optimal control problem is designed by imposing constraints on the control bandwidth \cite{clausen10}: $J_1[v] = \int_0^{t_f}dt_1\,[\dot{v}(t_1)]^2 - E$ and the total distance travelled: $J_2[v] = \int_0^{t_f}dt_1\,v(t_1) - \textcolor{black}{\mathcal{S}}$ \textcolor{black}{[fixing the final trap-position $x_{\circ}(t_f)$]} for some constant\textcolor{black}{s} $E$ and $\textcolor{black}{\mathcal{S}}$. 

Introducing Lagrange multipliers $\lambda, \lambda_1$, the stationarity of the total cost functional $J_{\rm tot}[v] = J[v] + \lambda J_1[v] + \lambda_1 J_2[v]$ then yields the EL equation: 
%===============================
\begin{equation}\label{elid}
\lambda \ddot{v}(t) = \frac{1}{2}\lambda_1 + \int\limits_0^{t_f}\, ds\, \phi_1(t - s)v(s)
\end{equation}
%===============================
where $\phi_1(t) = \int d\epsilon \,\vert\frac{\mu_{n\epsilon}}{\omega_{\epsilon n}}\vert^2 \cos(\omega_{\epsilon n} t)$. For $\lambda_1 \neq 0$, Eq. (\ref{elid}) is a \textit{linear} Fredholm integro-differential equation of \textcolor{black}{the} second kind which can be solved using methods adopted for solving Eq. (\ref{ELfinal}) (SI \textcolor{black}{X}).

%-------------------------------------------------------------------------
\subsection*{Comparison with STA} 

A compensating counter-diabatic field (CDF)\textcolor{black}{,} $\dot{x}_{\circ}(t)\,p$\textcolor{black}{,} is added in the STA approach, to the system Hamiltonian (\ref{ham}) in order to achieve transitionless transport of the trapped particle \cite{polkovnikov17, polkovnikovreview17}. Such a CDF that breaks time-reversal symmetry, is extremely hard to implement experimentally. Instead, a gauge transformation of the form $p \rightarrow p' =  p + \partial_x f \; ,\; H_{CDF} \rightarrow H_{CDF}' =  H_{CDF} + \partial_t f$ with $f(x, t) = -m\,\dot{x}_{\circ}\,x$, is usually invoked as a more practical compensating term in the form of an effective gravitational field: $-m\,\ddot{x}_{\circ}\,x$  \cite{polkovnikov17, polkovnikovreview17}. Here $H_{CDF}$ indicates the system Hamiltonian with a CDF term.

Such local gauge transformations require a corresponding transformation of the wave functions: $\psi(x,t) \rightarrow \psi(x,t)' =  e^{-i\,f(x,t)}\,\psi(x,t)$. If $\psi(x,t)$ denotes an eigenstate of the untransformed system Hamiltonian $H_S$, the corresponding transformed state $\psi(x,t)'$ is not an eigenfunction of $H_S'$ unless $\partial_t f = 0$\textcolor{black}{. Namely,} only in the case of uniform trap speed, \textcolor{black}{does} the gauge transformation \textcolor{black}{preserve} the eigenstates. Thus, for an accelerating trap, a particle initialized in a bound state of the initial trap position continues to experience non-adiabatic transitions even in presence of the compensating gravitational field. 

%===============================
\begin{figure*}[!t]
\centering
\includegraphics[width=0.49\textwidth,keepaspectratio]{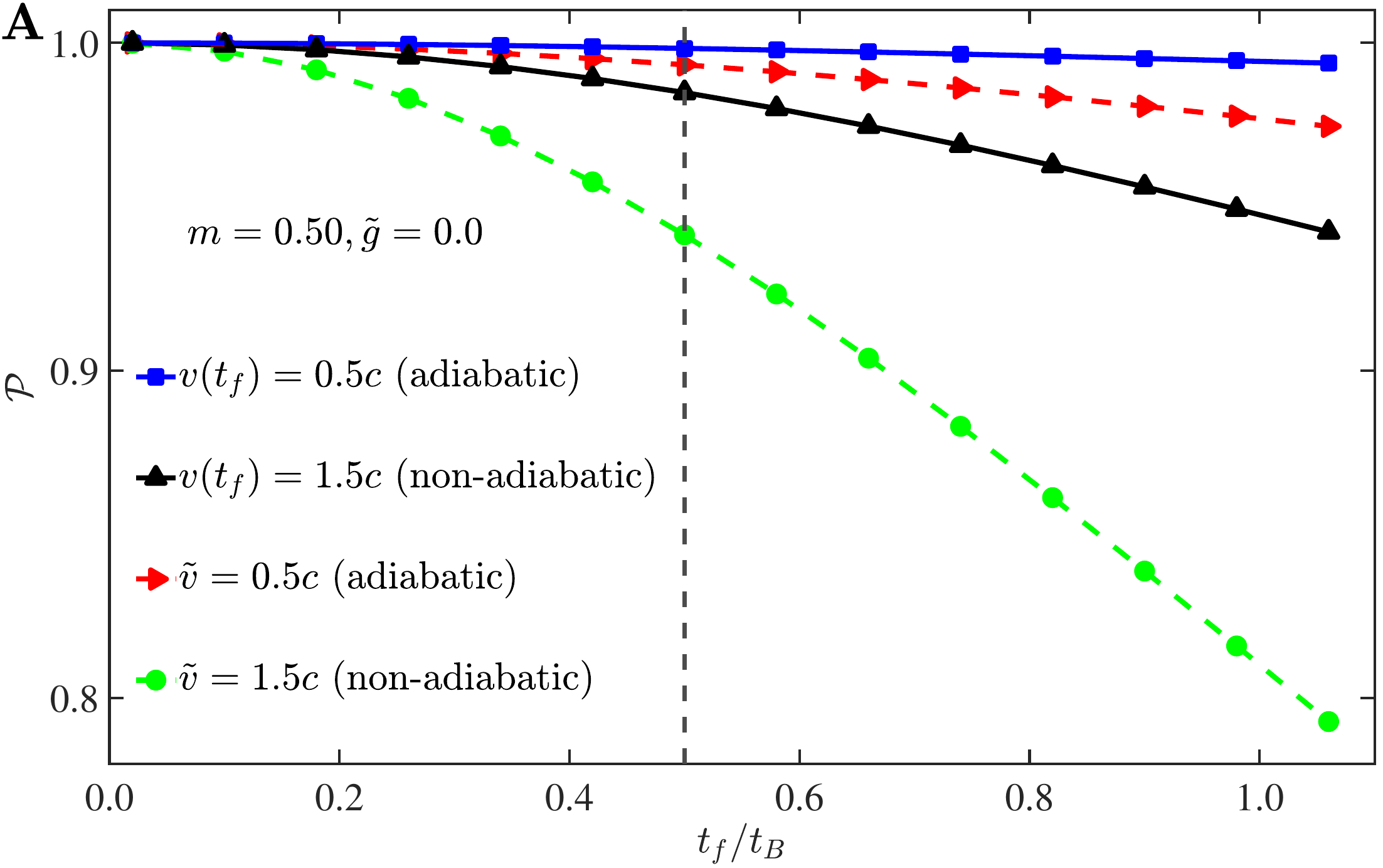}
\includegraphics[width=0.49\textwidth,keepaspectratio]{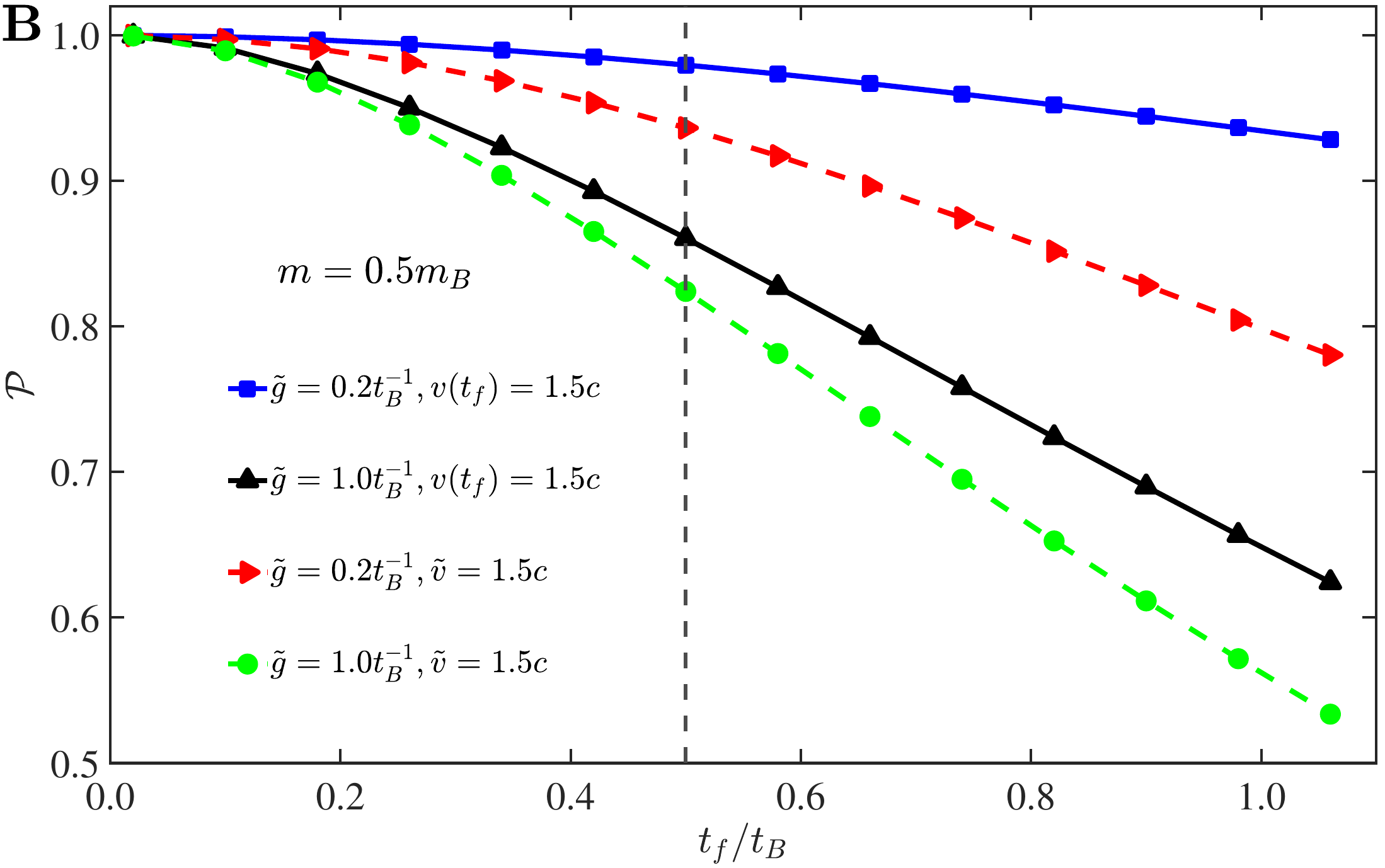}\\
\includegraphics[width=0.49\textwidth,keepaspectratio]{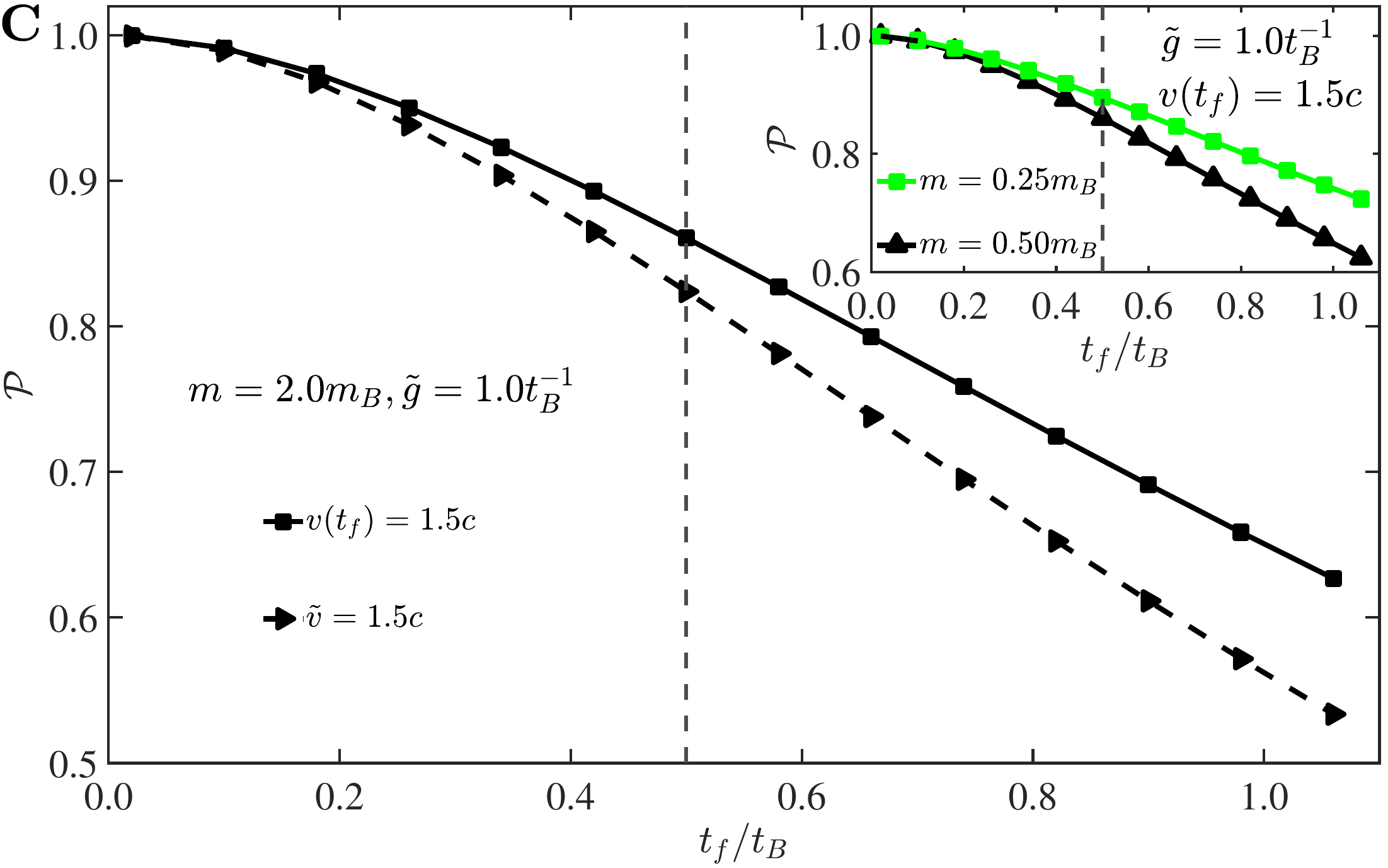}
\includegraphics[width=0.49\textwidth,keepaspectratio]{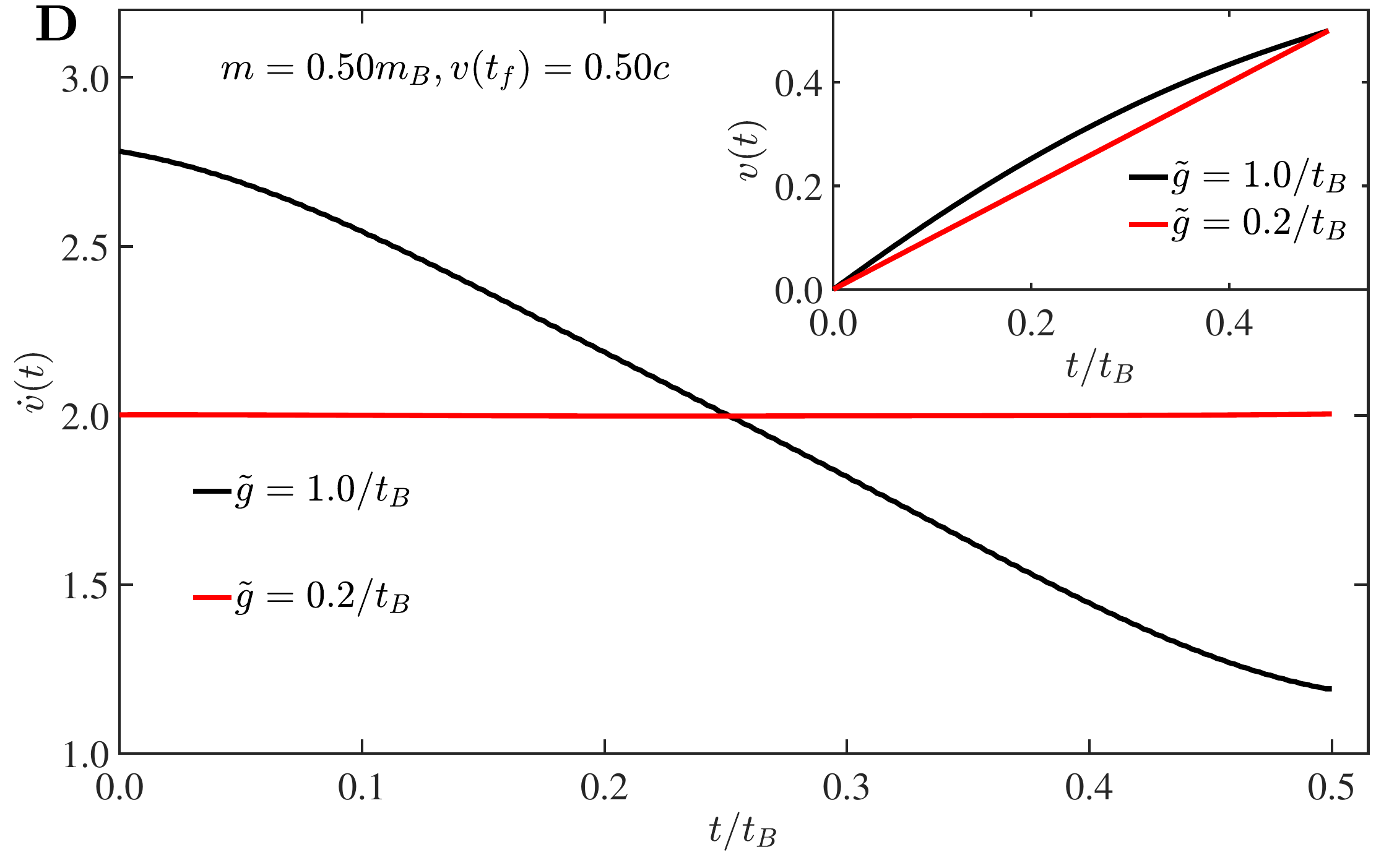}
\caption{\textbf{Comparison of the AC-QUDIT method with constant speed transport:}\textbf{A} Survival probabilities \textcolor{black}{$\mathcal{P}(t_f)$} for dissipation-less transport (through vacuum) at non-adiabatic final speed $\textcolor{black}{v(t_f)} = 1.5c$ (black) as well as adiabatic final speed $\textcolor{black}{v(t_f)} = 0.5c$ (\textcolor{black}{blue}) for $m = 0.5\,m_B$, obtained by our AC-QUDIT method compared to corresponding constant speed motion with speeds $\tilde{v} = 1.5c$ (\textcolor{black}{green}, non-adiabatic) and $\tilde{v} = 0.5c$ (red, adiabatic). In BEC, $c$ is the speed of sound. In vacuum it can be replaced by $1/2a t_c$ where $a$ is the Morse trap-width parameter and $t_c$ the coherence time associated with nonadiabatic leakage from the trap (Methods), $m$ can be measured in any arbitrary unit. Throughout, we have used the Lagrange multipliers $\lambda = 1, \lambda_1 = 1$. \textbf{B} Survival probabilities for transport through BEC (dissipative: $\tilde{g} = 1.0t_B^{-1}$ and $\tilde{g} = 0.2t_B^{-1}$) at non-adiabatic and adiabatic final speeds $v(t_f/t_B) = 1.5c\;{\rm and}\; 0.5c$ respectively for $m = 0.5 m_B$ obtained by our AC-QUDIT method (solid lines) compared to constant speed motion with $\tilde{v} = 1.5c\; {\rm and}\; 0.5c$ (dashed lines). \textbf{C} Survival probabilities obtained by AC-QUDIT method for $m = 2.0 m_B$ (heavier impurity) and $v(t_f/t_B) = 1.5c$ compared with constant speed transfer at $\tilde{v} = 1.5c$ \textcolor{black}{(see Methods C)}. Inset shows the survival probability as a function of the transport duration for $m = 0.25\,m_B$ and $m = 0.5\,m_B$ (lighter impurities), all other parameters remaining same \textcolor{black}{in this case (see Methods C)}. In panels \textbf{A}, \textbf{B} and \textbf{C}, the vertical dashed line indicates the $t_f/t_B = t_c/t_B$ point on the horizontal axis. \textbf{D} Optimal trap acceleration obtained from our method, as a function of time, for $\tilde{g} = 0.2\,t_B^{-1}$ (red line) and $\tilde{g} = 1.0\,t_B^{-1}$ (black line) for $v(t_f) = 1.5\,c$, $m = 0.5\,m_B$. The inset shows the corresponding optimal trap-speeds.
}
\label{F2}
\end{figure*}
%===============================

If the trapping potential is shallow and admits a continuum spectrum, such non-adiabatic transitions lead to irreversible loss of fidelity even in the absence of an external heat bath \textcolor{black}{\cite{kofman01, barone04, wilkinson1997experimental, keshavamurthy2011dynamical}}. This irreversible loss of fidelity cannot be avoided by merely making the velocity and acceleration vanish at the boundary points, as is usually done to achieve transition-less transport via STA \cite{polkovnikov17, polkovnikovreview17} \textcolor{black}{(see Methods F).} 

The introduction of a compensating counter-diabatic field (CDF) changes the non-adiabatic transition rates $\gamma_{n\epsilon}(t)$ in Eq. (\ref{cost1}) as follows :
%===============================
\begin{equation}\label{e1}
      \Gamma_{n\epsilon}(t) = \gamma_{n\epsilon}(t) - i m \dot{v}(t)D_{n\epsilon}e^{-i\omega_{\epsilon n}t}.
\end{equation}
%===============================
where $D_{n\epsilon} = \langle n (t) \vert \,q\, \vert \epsilon (t) \rangle$, is the  time-independent dipole transition matrix element. By contrast, $\Delta_{n\epsilon}^k(t)$ remains invariant since CDF does not involve bath-induced transitions. The first term on the r.h.s. of (\ref{cost1}) is then transformed to ${\rm Re}\int d\epsilon\int_0^{t_f}dt_1\int_0^{t_f}dt_2 \,\Gamma_{n\epsilon}(t_1)\Gamma_{n\epsilon}^{*}(t_2) = \int d\epsilon \big\vert \int_0^{t_f}dt_1\,\Gamma_{n\epsilon}(t_1) \big\vert^2 > 0 $, which can be rewritten as a sum of terms proportional to the trap-speed PSD, corresponding trap-acceleration PSD and the cross-spectral density (CSD) of between the trap-speed and acceleration, albeit the overall  integral over $\epsilon$ (see Methods \textcolor{black}{E} for detailed expressions).

%===============================
In general the CDF method would yield lower values of $\mathcal{P}(t_f)$ than that obtained by our approach whenever
%===============================
\begin{equation}\label{condition1}
\int\limits_0^{t_f}dt_1\int\limits_0^{t_f}dt_2\;\Gamma_{n\epsilon}(t_1)\Gamma_{n\epsilon}^{*}(t_2) > \int\limits_0^{t_f}dt_1\int\limits_0^{t_f}dt_2\;\gamma_{n\epsilon}(t_1)\gamma_{n\epsilon}^{*}(t_2),
\end{equation}
%===============================
which holds provided the sum of the speed-acceleration CSD and the acceleration PSD is positive (see Methods \textcolor{black}{E}).

\subsection*{Numerical results}

While the AC-QUDIT method described by the linearized Eq. (\ref{ELfinal}) is general, with broad applicability in AMO physics as it can be derived for any trapping potential and bosonic bath, we have numerically illustrated the method for an impurity in a moving Morse-potential trap immersed in a BEC (see Methods \textcolor{black}{C} and SI \textcolor{black}{I -- V, VIII, IX}). Expressing all length scales in units of the phonon coherence length $\xi$, mass in units of the phonon mass $m_B$ and time in units of the characteristic time $t_B = \xi/\sqrt{2}c$ (see Methods \textcolor{black}{C}) where $c$ is the speed of sound in BEC \cite{nielsen2019critical}, we study the dependence of the survival probability (fidelity) $\mathcal{P}(t_f)$ on the final transport time $t_f$ for different values of the impurity mass $m$ and final trap speeds $v(t_f)$. Survival probability at transport durations shorter than the characteristic time of non-adiabatic leakage $t_c$ (Methods \textcolor{black}{C}) as well as $t_B$ are relevant for quantum information processing.

%\subsubsection{Comparision with constant-speed transport}
\textit{Comparision with constant-speed transport:} Analytical results of Eq.~(\ref{ELfinal}), confirmed by numerical solution\textcolor{black}{s} of the fully non-linear Eq.~(\ref{EL1}) (see Methods \textcolor{black}{B}, SI \textcolor{black}{XI, XIII}) show that AC-QUDIT invariably ensures higher fidelty than constant-velocity transport, all parameters being equal. 
In the absence of dissipation (e.g. for atoms trapped in a tweezer moving in vacuum) the advantage of AC-QUDIT becomes salient for appreciable non-adiabatic leakage rates (Fig. \ref{F2} A) (Methods \textcolor{black}{D}). Under bath-induced dissipation, AC-QUDIT is more advantageous compared to \textcolor{black}{the} constant velocity transport, for faster (less adiabatic) transfer (Fig. \ref{F2} B). The final speed $v(t_f)$ and the constant speed $\tilde{v}$ are here chosen to be non-adiabatic when they are higher than $c$, the speed of sound in the BEC (\textcolor{black}{see} Methods \textcolor{black}{D}). 

%===============================
\begin{figure*}[!t]
\centering
\includegraphics[width=0.478\textwidth,keepaspectratio]{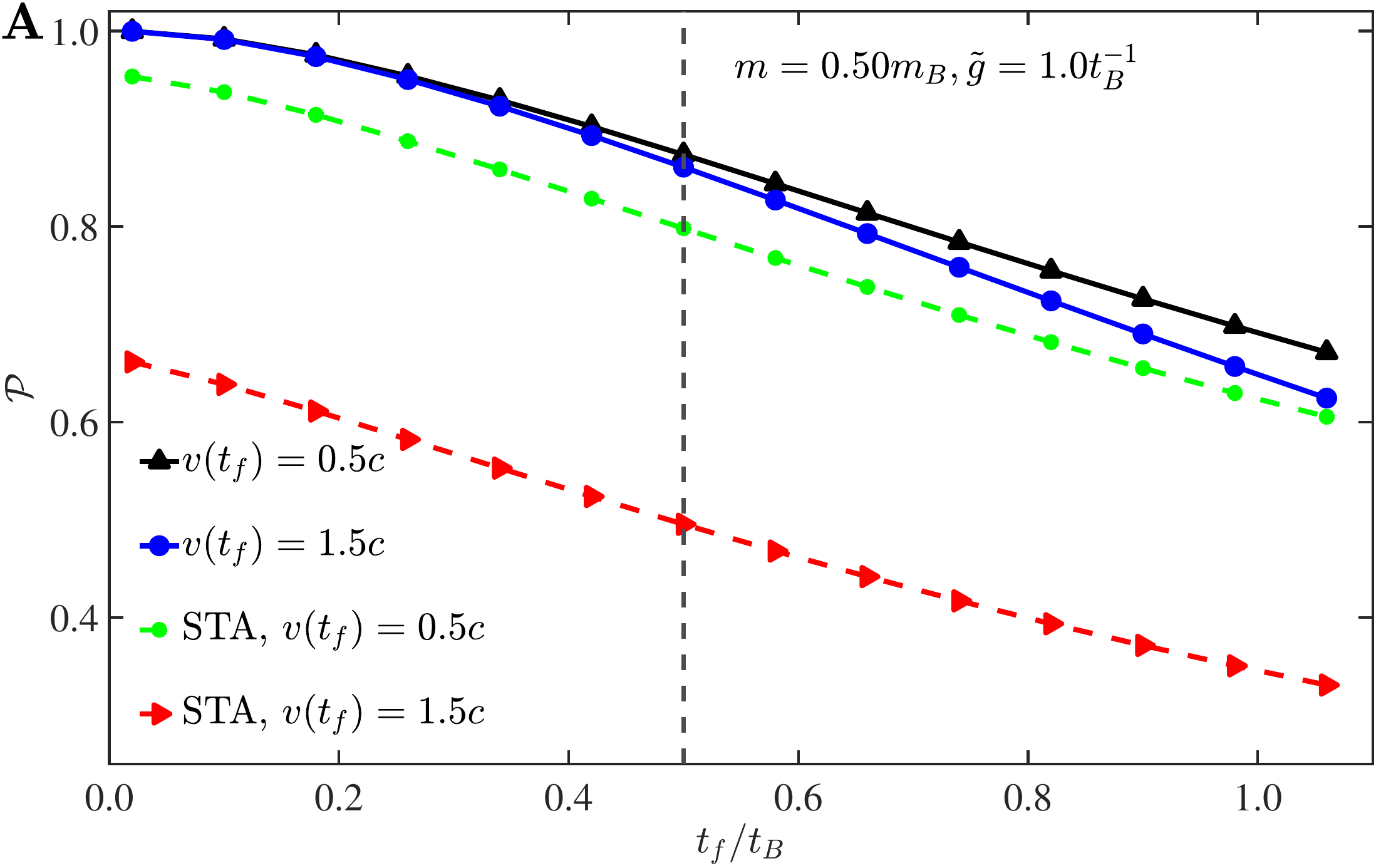}
\includegraphics[width=0.478\textwidth,keepaspectratio]{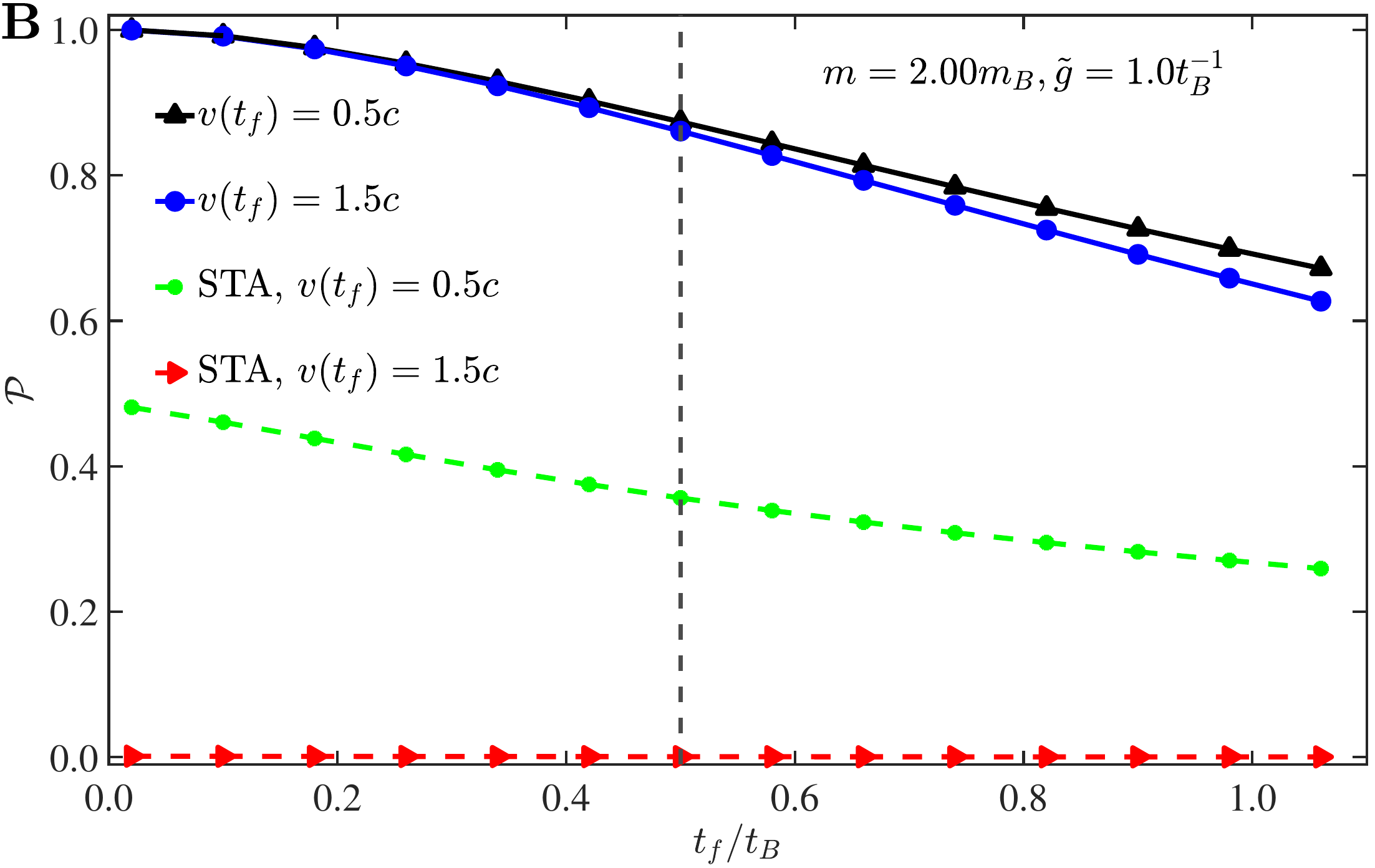}\\
\includegraphics[width=0.478\textwidth,keepaspectratio]{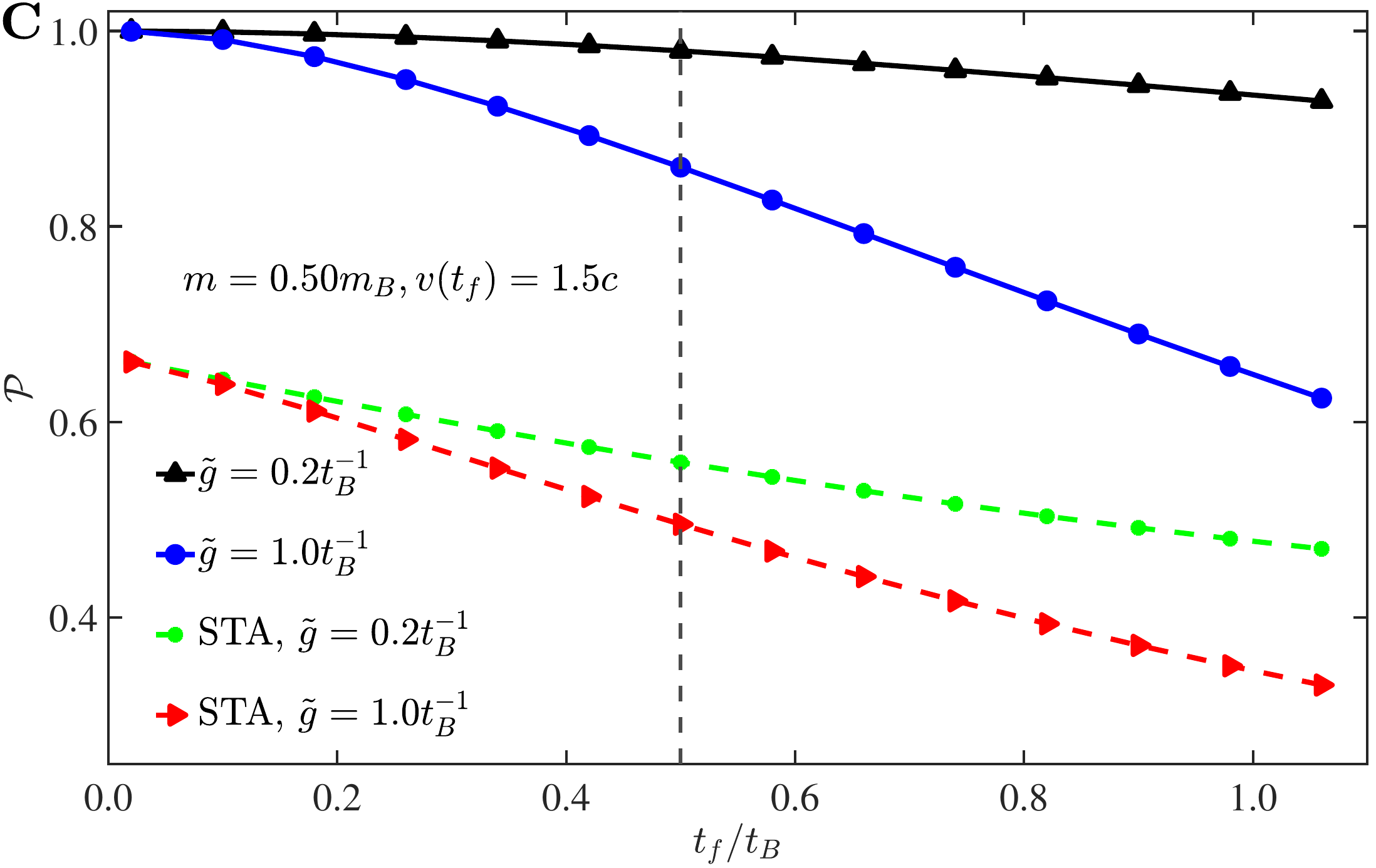}
\includegraphics[width=0.478\textwidth,keepaspectratio]{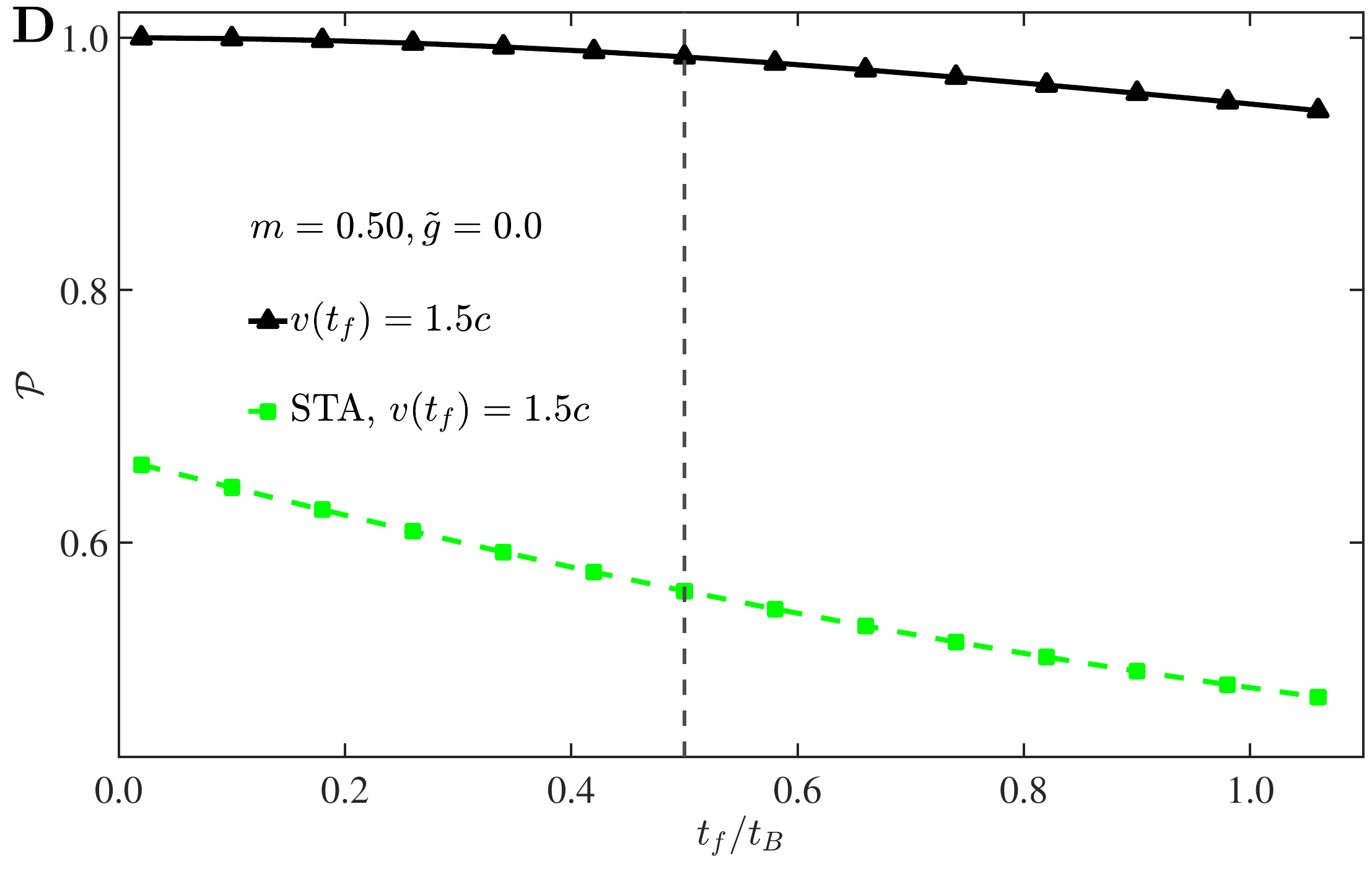}
 \caption{\textbf{Comparison of the AC-QUDIT method with STA:} \textbf{A} Survival probabilities as a function of $t_f/t_B$ obtained by CDF (STA) (dashed lines) and AC-QUDIT (solid lines) for $m = 0.5\,m_B$, $v(t_f) = 0.50\,c$ and $v(t_f) = 1.5\,c$. \textbf{B} Same as panel \textbf{A} with $m = 2\,m_B$ and final speeds $v(t_f) = 0.5 c$ and $v(t_f) = 1.5 c$. \textbf{C} $\mathcal{P}(t_f/t_B)$ for two different system-bath coupling strengths $\tilde{g} = 0.2\,t_B^{-1}$ [black solid (AC-QUDIT) and green dashed lines (STA)] and $\tilde{g} = 1.0\,t_B^{-1}$ [blue solid (AC-QUDIT) and red dashed lines (STA)] with $m = 0.5\,m_B$ and $v(t_f) = 1.5\,c$. \textbf{D} Survival probability for transport through vacuum ($\tilde{g} = 0$) at non-adiabatic final speed $v(t_f) = 1.5c$ i.e. $1.5/2at_c$ ($a$ is the Morse trap-width parameter and $t_c$ the coherence time associated with nonadiabatic leakage from the trap),  $m = 0.5 m_B$ (can be in any arbitrary unit) obtained by AC-QUDIT (black) and with CDF (green). As before we have used the Lagrange multipliers $\lambda = 1$, $\lambda_1 = 1$. In all the panels the vertical dashed lines indicate the $t_f/t_B= t_c/t_B$ point on the time axis.}\label{F3}
\end{figure*}
%===============================

Defining the $k$-independent coupling strength $\tilde{g}$ as $g_k = \tilde{g}\,U(k)$, $U(k)$ being a suitable function of $k$ (Methods), we find that an increase in $\tilde{g}$ i.e. an increase in the bath-induced rates $\Delta_{n\epsilon}^k$ results in a corresponding decrease in the survival probability. Remarkably, AC-QUDIT turns out to be advantageous compared to constant velocity transport for any finite coupling strength (see Fig. 2B).  

The survival probability decreases with the increase in impurity mass $m$ (scaled to the phonon mass $m_B$) since the non-adiabaticity grows with $m^2$ ($\frac{\mu_{n\epsilon}}{\omega_{\epsilon n}} \propto m $) (SI). Still, AC-QUDIT yields higher fidelty than constant speed transfer, for a given $m$ (see Fig. 2C).

Importantly, for appreciable \textcolor{black}{system-bath} coupling $\tilde{g}$, the prescribed AC-QUDIT acceleration \textcolor{black}{becomes} nonlinear in time. \textcolor{black}{In Fig. 2D we observe that the stronger the coupling $\tilde{g}$, the larger is the deviation of the trap-acceleration from the constant value which is optimal for weak couplings. From the inset displaying the corresponding speeds we conclude that the potential travels faster through the medium as the coupling gets stronger, arriving at a longer distance within the same time (as indicated by the area below the speed curves). Since the final speed is adiabatic, the integral term in Eq. (\ref{ELfinal}) can be neglected, and the displayed behavior is then due to the first two terms on the r.h.s. of this equation, where the functions $\eta$ and $\zeta$ become appreciable as the coupling grows. By contrast, for weak coupling, the constancy of acceleration implies that $\ddot{v}(t)=0$, so that all the terms on the r.h.s. of Eq. (\ref{ELfinal}) become negligible.}

%\subsubsection{Comparision with STA}
\textit{Comparision with STA:} The survival probabilities obtained by AC-QUDIT surpass those obtained by CDF, as shown numerically in Fig. \ref{F3}, in accordance with the rigorous theoretical conditions developed before (see Eq. \textcolor{black}{(\ref{condition1})}, Methods \textcolor{black}{E}). \textcolor{black}{Since} the CDF method is designed for any arbitrary \textcolor{black}{trajectory}, for illustration in Fig. \ref{F3} we have used the optimal trajectory obtained from our method which minimzes both non-adiabatic and bath-induced leakages simultaneously. Any other arbitrary trajectory may result in even lower values of \textcolor{black}{the} survival probability under CDF in \textcolor{black}{the} presence of dissipation, due to non-optimal suppression of bath-induced effects. 

AC-QUDIT becomes significantly advantageous compared to CDF primarily for non-adiabatic final trap speed $v(t_f)>c$ in the dissipative case (Fig. \ref{F3} A). This is comparable with the advantage of AC-QUDIT relative to STA for non-adiabatic speeds in \textcolor{black}{dissipationless} transport (Fig. \ref{F3} D).
Upon increasing the impurity mass $m$ the advantage of AC-QUDIT becomes appreciable even for final speeds, $v(t_f)<c$ (Fig. \ref{F3} B). The reason is that the adverse effect of acceleration-PSD and velocity-acceleration CSD incurred by CDF (Eq. (\ref{exponent1}) ) scales with $m^2$. 

The advantage of AC-QUDIT increases with \textcolor{black}{the} decrease in the dissipative system-bath coupling $\tilde{g}$, (Fig. \ref{F3} C), although the CDF does not change the bath-induced transition rates $\Delta^k_{n\epsilon}$ as discussed before. This reflects the dependence of optimal trajectories in AC-QUDIT on $\tilde{g}$, which cannot be matched by their STA counterparts. Thus, we find that AC-QUDIT yields much higher survival probabilities compared to STA and constant speed transport both in the presence and \textcolor{black}{in the} absence of system-bath coupling. 

\textcolor{black}{Importantly, the CDF method is designed to work only for $v(0) = v(t_f) = 0$ \cite{polkovnikov17, polkovnikovreview17}. We show, by contrast, that our optimal trajectory yields significantly higher survival probabilities than CDF, even for $v(t_f) \neq 0$, proving its superiority and wide range of applicability (see SI for $v(t_f) = 0$)}.

\section*{Discussion}

We have put forth the general AC-QUDIT powerful method of maximizing the fidelity of fast, non-adiabatic quantum transport in dissipative media by means of the optimal velocity-or-acceleration control (\ref{ELfinal}) or its dissipationless counterpart Eq. (\ref{elid}). This method is universally applicable to  the study of diverse topical issues, such as the  evolution  of atomic wavepackets~\cite{Gal2021,Manolo2021}, solitonic transport~\cite{Fu2022}, transport  of  trapped atoms and ions in quantum information processing protocols~\cite{jain24, sterk22, walther12, Rowe02, ibloch22, ibloch01}, quantum refrigeration by trapped impurities  in condensates~\cite{niedenzu19}, radiation from fast charged particles in solids \cite{gevorkyan15} among others. 

All such processes should be completed as fast as possible,  not only  to shorten their duty cycle, but also to minimize the inherently quantum hurdle of wavepacket spread and dissipation /decoherence. Moving field-induced traps (tweezers), can reduce the wavepacket leakage, but, unless the trap is very deep (which may require overly intense fields) \cite{ibloch22}, fast transport triggers non-adiabatic leakage to the continuum.

Remarkably, we have shown that our method yields transport efficiencies $\geq 95\%$ at $t_f \lesssim t_c$, even for a very shallow potential trap that supports a single bound-state (the Friedrichs model), in case of transport through vacuum (Fig. 3D). Similar considerations can apply to the control of molecular dissociation and collisions \cite{deb83} by engineering the potential surface on which the molecular wavepacket moves, and thereby the wavepacket speed and trajectory. 

For transport through BEC, the time scales we have explored are shorter than the typical time for the emission of elementary excitations (polaron formation) in the BEC as well as their lifetimes, but longer than the time-scales on which the assumption of one-dimensionality of the BEC breaks down: the formation and lifetimes of elementary excitations exceed the coherence time of impurity dynamics in BEC, $t_B$ \cite{nielsen2019critical, imambekov12}, while the violation of one-dimensionality is manifest only at times shorter than the inverse transversal trapping frequency, which correspond to elementary excitations with energies high enough to break down the one-dimensionality \cite{schmiedmayer19, torrontegui12}.

The nonlinear EL integro-differential equation (see Methods \textcolor{black}{B, Eqs .(\ref{EL1}),\,(\ref{FM-KK})}) for the optimal wavepacket trajectory has broad applicability in the scenarios discussed above. Its linearized, analytically soluble version (\ref{ELfinal}), agrees with fully-numerical solutions of the former (see SI \textcolor{black}{XI, XIII}) even for fast, supersonic trajectories.  

Our method presents an advantageous alternative to  the conceptually powerful  optimal control  of non-adiabatic transport  by  counter-diabatic driving (alias shortcuts to adiabaticity - STA) \cite{polkovnikov17,berry09,Torrontegui11,Chen11,Chen151, Chen152, opatrny14, odelin19} which  faces limitations when  open  quantum systems are concerned (as discussed in the Introduction \cite{vacanti2014transitionless, wu21}). Instead, we have tackled the formidable problem of quantum transport in dissipative media by dynamical control based on a generalized Wigner-Weisskopf approach that can effectively counter both quantum dissipation and nonadiabaticity of moving wavepackets, resulting in maximal transport fidelity. This method is a paradigmatic generalization \textcolor{black}{of the Kofman-Kurizki (KK) non-Markovian universal control formula for discrete quantum variables, to wavepackets} \cite{kofman01,clausen10,kkbook21}.  

Our general approach is rigorously proven to outperform STA protocols based on CDF for a broad class of transport trajectories and trapping potentials, while the invariant based STA methods are restricted only to the specific, Lewis-Leach class of potential traps \cite{odelin19}. Importantly, for non-adiabatic transfer, our method yields significantly higher transfer fidelity than STA.  

This approach is also conceptually advantageous compared to quantum methods based on feedback \cite{gordon13} since it does not rely on measurements and thus can be included in quantum information processing schemes.

A central insight obtained from our approach is that non-adiabatic leakage constitutes an effective bath, that has acceleration-dependent coloured spectrum, and is added to the standard environmental bath. Thus, even in the absence of a dissipative environment i.e. transport through vacuum, there is inevitable loss of fidelity due to this effective ``bath'' which cannot be \textcolor{black}{avoided} by methods based on STA (\textcolor{black}{see Methods F}). For non-zero system-bath coupling, we have found that it is essential to account for coupling between the trapped wavepacket and the environmental bath, beyond the Lamb-Dicke limit \cite{leibfried03, Lampo17}, in order to find an optimal trajectory that can simultaneously minimize the effect of the two ``baths''. The survival probability (Loschmidt echo) is then determined by the spectra of both baths whose Fourier components are modulated by the time-varying trap-speed. 

To sum up, this approach opens a new avenue towards the control of quantum transport in dissipative media and has the potential for the discovery of unexplored, novel features of the transport.

%------------------------------------------------------------
\section*{Methods}

\textcolor{black}{\subsection{Fidelity and Loschmidt echo}}
\textcolor{black}{Fidelity of a dynamical state is defined as its susceptility to perturbations, given by the overlap of the perturbed and un-perturbed states at a given instant (Loschmidt echo \cite{znidaric06, Coalson19}). In our problem, the perturbed state at $t>0$ corresponds to $\vert \psi(t)\rangle$ which accounts for the effects of phonon-induced and non-adiabatic transitions (`perturbations'), while the unperturbed state at the same instant is $\vert \nu(t)\rangle = \vert n(t)\rangle\otimes\vert 0_{\rm bath}\rangle$ up to a global phase (without transitions induced by the `perturbations').}

\textcolor{black}{In our problem $H_{SB}$ entangles advanced, retarded and instantaneous wavepacket states with excited or ground states of the bath, ensuring conservation of momentum. We note that the reduced density matrix of the system at $t>0$ is a mixture of such advanced, retarded and instantaneous states (a mixed state) since the combined system-bath state is entangled. Detection of a wavepacket state advanced by momentum $k$, implies corresponding collapse of the bath to an excited state having momentum $-k$. Similarly, detection of an instantaneous wavepacket state, ensures that the bath collapses to the ground state $\vert 0_{\rm bath}\rangle$ i.e. the survival probability in the instantaneous bound state (fidelity) is given by the Loschmidt echo which is here the probability of finding the combined system in the many-body state $\vert \nu(t)\rangle$ (see SI IV for details). Only the diagrams in Fig. \ref{S1-1} C (i) corresponding to no quanta exchange (purely non-adiabatic) and  virtual quanta exchange (purely bath-mediated) can result in the state $\vert \nu(t)\rangle$ at $t > 0$. The diagrams in Fig. \ref{S1-1} C (ii) and (iii) corresponding to real quanta exchange with the bath, result in advanced or retarded system states and corresponding excited bath states. A classical analog of this scenario is that of a catcher on a moving train that throws a ball vertically upwards. To an observer on the platform (in the laboratory frame), both the catcher and the ball undergo a horizontal displacement before the falling ball lands back in the catcher's palm (in an instantaneous trap state). If instead the ball was launched up in the train along with a finite horizontal momentum, the catcher may have to move backward or forward in order to catch the ball (in an advanced or retarded trap), thereby compensating for the additional momentum imparted. This reflects the fact that advanced/ retarded states do not contribute to fidelity in the instantaneous state.}

It is important to note that Loschmidt echoes are routinely measured experimentally, in studying quantum many-body dynamics, and is often used as a probe for dynamical phase transitions \cite{xu2020probing, roberts2024manybody, cetina16, braumuller2022probing, tonielli2020ramsey, Coalson19, nielsen2019critical}.

\subsection{Optimal wavepacket trajectory}

In the spirit of the Wigner-Weisskopf method \cite{wignerweisskopf, kofman01}, we first integrate out the unbound sector of $A_l$ in Eq. (\ref{dynamics1}) to obtain a system of differential equations for the bound sector (see SI). Within the bound-sector, we again integrate out all states except $\vert \nu (t) \rangle$ to obtain the Loschmidt echo probability amplitude $A_{\nu}$ (see SI V, VI). 

The resulting integro-differential non-local Euler-Lagrange \textcolor{black}{(EL)} equation has the following form:
%=================================
\begin{subequations}
\begin{eqnarray}\label{EL1}
     \lambda \ddot{x}_{\circ}(t) & = & \int\limits_0^{t_f}dt' \Big\lbrace \dot{x}_{\circ}(t') \int d\epsilon~\frac{\vert \mu_{n\epsilon}\vert^2}{\omega_{\epsilon n}}\sin\left[\omega_{\epsilon n}(t - t')\right] \nonumber\\
                                 &   &  \hspace{3cm} - K(t,t')\Big\rbrace ;
\end{eqnarray}
\begin{eqnarray}\label{FM-KK}
                                K(t,t')  & = & \frac{L}{2\pi}\int dk\,d\epsilon \, k\,\vert \widetilde{g}_{n\epsilon}^{(k)} \vert ^2 \, \sin\Big[\int\limits_{t'}^t\,dt'' \lbrace\omega_{\epsilon n} + \Omega_k \nonumber\\
                                &  & \hspace{3.5cm} + \; k \, \dot{x}_{\circ}(t'')\rbrace\Big].
\end{eqnarray} 
\end{subequations}
%=================================
\noindent The kernel $K(t,t')$ is expanded as a weighted sum of Fourier harmonics with frequencies $(\omega_{\epsilon n} + \Omega_k)$ and is frequency modulated (FM) via the control function $k \dot{x}_{\circ}(t)$ \cite{haykin01}. This hitherto unattempted FM control results from the nonlinearity in $k x_{\circ}$ of the system-bath coupling Hamiltonian, when $e^{-ikx}$ is beyond the Lamb-Dicke limit in Eq. (\ref{hamib}). This $kx_{\circ}$-nonlinearity is essential for controlling the bath-spectrum by changing the trap speed $v(t) = \dot{x}_{\circ}(t)$. If we were restricted to the linear-coupling (Lamb-Dicke) regime where $e^{-i\,k\,x} \rightarrow -i\,k\,x = -i\,k\,[x - x_{\circ}(t)] - i\,k\,x_{\circ}(t)$, then $x_{\circ}(t)$ would act as an external field in the instantaneous system basis, which cannot serve as a handle on the bath-induced transitions between the instantaneous levels.

By time-differentiating both sides of Eq. (\ref{EL1}) we convert the kernel Eq. (\ref{FM-KK}) into a sum of amplitude- and frequency-modulated (AFM) terms of the form
%=================================
\begin{align}\label{HM-KK}
\dot{K}(t,t') & =\frac{L}{2\pi}\int dk \int d\epsilon\; k\,\vert \widetilde{g}_{n\epsilon}^{(k)} \vert ^2 \, \Big[(\omega_{\epsilon n} + \Omega_k) + k\,\dot{x}_{\circ}(t) \Big]\nonumber\\
              & \cos\Big[(\omega_{\epsilon n} + \Omega_k)\int\limits_{t'}^t\,dt''\, \Big\lbrace 1 + \frac{k}{(\omega_{\epsilon n} + \Omega_k)}\,\dot{x}_{\circ}(t'')\Big\rbrace\Big].
\end{align}
%=================================
\noindent In the speed range $\vert v(t)\vert = \vert \dot{x}_{\circ}(t) \vert < v_s = \big\vert (\omega_{\epsilon n} + \Omega_k)/k\big\vert$, we obtain to the lowest order in $\big\vert k v(t)/(\omega_{\epsilon n} + \Omega_k)\big\vert$, the linearized Eq. (\ref{ELfinal}). The boundary value problem (BVP) associated with Eq. (\ref{ELfinal}) is solved by first reducing it to a Fredholm integral equation using a Green's function method \cite{tamarkin27, singh16} (see SI) and then applying the Liouville-Neumann expansion to solve the resulting integral equation \cite{zemyan2012classical}. Numerically, we solve the \textcolor{black}{fully} non-linear \textcolor{black}{EL} equation by successive iterations (see SI \textcolor{black}{XIII}) and verify that the solutions match well with the results obtained analytically from Eq. (\ref{ELfinal}). \textcolor{black}{The corresponding results presented in Figs. \ref{F2} and \ref{F3} satisfy the condition $\vert v(t)\vert < v_s$ (see SI \textcolor{black}{XI}). }

\subsection{A wavepacket trapped in Morse potential and dissipated as in the Fr{\"o}lich model}

In the Fr{\"o}lich model of an impurity interacting with a bath of bosons forming a BEC \cite{frolich52, Coalson19, Lampo17} confined in a volume $\mathcal{V}$, the parameters in Eq. (\ref{ELfinal}) are \textcolor{black}{$g_k = g_{-k} = g_k^* =$} $g_{SB}\sqrt{\frac{n_0}{\mathcal{V}}}\;\Big[\frac{(\xi k)^2}{(\xi k)^2 + 2}\Big]^{\frac{1}{4}}$, $\Omega_k = c\vert k \vert \sqrt{1 + \frac{1}{2}(\xi k)^2}$. Here  $\xi = 1 / \sqrt{2g_Bm_Bn_0}$ represents the coherence length of phonons, $c = 1 /(\sqrt{2}m_B\xi)$ is the BEC speed of sound, $m_B$ is the phonon mass and $n_0$ is the ground state density of the BEC, while $g_{SB}$ and $g_B$ denote the impurity-boson and inter-boson interaction strengths, respectively \cite{Lampo17}. In a $1$-d confinement of the BEC, the volume $\mathcal{V}$ is replaced by the confining length $L$. The Morse potential trap is given by $V[x - x_{\circ}(t)] = D\Big[ e^{-2a\lbrace x - x_{\circ}(t)\rbrace} - 2 e^{-a\lbrace x - x_{\circ}(t)\rbrace} \Big]$. The expressions for the transition matrix elements $\Delta_{n\epsilon}^k(t)$ and $D_{n\epsilon}$, for this choice of the trapping potential, are obtained from literature (see SI).

\paragraph*{Relevant time-scales:} The two distinct leakage terms in Eq. (\ref{cost1}) are associated with two different time-scales. The shortest correlation time of the continuum $t_c$ for non-adiabatic transitions is the inverse of trap-depth \cite{barone04}, here $t_c = 1/D$. The coherence time-scale for impurity dynamics in BEC, $t_B$ is determined by the Bogoliubov coherence length $\xi$, $t_B = \xi/\sqrt{2}c$ \cite{nielsen2019critical,Lampo17}.

\paragraph*{Choice of units and parameters:} We express all length-scales in units of $\xi$, time-scales in units of $t_B$ and mass in units of $m_B$. The coupling strength $\tilde{g}$ is expressed in units of $1/t_B$. We choose $\hbar = 1$, $c = 1$ and $m_B = 1$ for our numerical calculations. In our analysis, we set the Lagrange multiplier to be $\lambda = 1$.
%=================================
\begin{table}[h]
\begin{tabular}{lllrr}
\hline
Parameter & & Unscaled    &  & Scaled  \\
\hline
 Trap-depth & &    $D$        &  & $D' = Dt_B$      \\
 Impurity mass & & $m$        &  & $m' = \frac{m}{m_B}$     \\
 Trap-width parameter & & $a$        &  & $a' = a\xi$     \\
 Continuum correlation time & & $t_c$      &  & $t_c' = \frac{t_c}{t_B}$      \\
 Phonon wave vector &  & $k$        &  & $k' = k\xi$ \\
 Phonon coupling strength & & $g_k$      &  & $g_{k'}' = g_{k\xi}t_B$      \\
\hline
\end{tabular}
\caption{Scaled and unscaled parameters}
\end{table}
%=================================
\noindent In order to have a single bound-state (Friedrich's model), we must have $\sqrt{\frac{2mD}{a^2}} - \frac{1}{2} < 1$ (see SI). In the scaled parameters, this imples $D' < \frac{9 a'^2}{8m'}$, using the fact that $t_B/\xi^2 = 1$ in our units. 

\noindent In order to illustrate our method, we explore the regime $t_c < t_B$ i.e. $t_c' = 1/D' < 1$ or $D' > 1$, so that $1 < D' < \frac{9 a'^2}{8 m'}$. In the plots of Fig.~\ref{F2} and Fig.~\ref{F3}, we have used $D' = 2$, $a' = 1$, $m' = i)\, 0.25$,\; $ii)\,0.5$ and $D' = 2$, $a' = 2\; $, $m' = 2$. Hence $t_c' = 1/D' = 0.5$. In all our computations, we choose the cut-off values of $\epsilon$ and $k'$ as  $\epsilon_{\rm max} = 5$ and
$\vert k' \vert_{\rm max} = 5$.
\noindent Without loss of generality we choose $\tilde{g}= t_B\,g_{SB}\sqrt{\frac{n_0}{2L}}$.

\subsection{Non-adiabaticity condition}

For a closed quantum system, the adiabatic approximation is valid in the limit \cite{lidar05}
%===============================
\begin{equation}\label{ad1}
 \underset{0\leq t\leq t_f}{\rm max}\Bigg\vert\frac{\langle n(t)\vert\frac{\partial H_S(t)}{\partial t}\vert \epsilon(t)\rangle}{\omega_{\epsilon n}} \Bigg\vert \ll \underset{0\leq t\leq t_f}{\rm min} \vert \omega_{\epsilon n} \vert.
\end{equation}
%===============================
For a moving potential trap (without a bath), the condition (\ref{ad1}) becomes, in terms of the non-adiabatic transition matrix elemts $\frac{\mu_{n\epsilon}}{\omega{\epsilon n}}$
%===============================
\begin{equation}\label{ad2}
 \underset{0\leq t\leq t_f}{\rm max}\Bigg\vert v(t) \frac{\mu_{n\epsilon}}{\omega_{\epsilon n}}\Bigg\vert \ll \underset{0\leq t\leq t_f}{\rm min} \vert \omega_{\epsilon n} \vert .
\end{equation}
%===============================
For $m = 0.5\,m_B$ the minimum value of $\omega_{\epsilon n}$ is $\sim 0.84\,t_B^{-1}$ . For $v(t_f) = 0.50\,c$ (subsonic target speed) the l.h.s. of the above inequality (\ref{ad2}) is $0.26\,t_B^{-1}$ which is $\ll {\rm min}\vert \omega_{\epsilon n} \vert$ and hence corresponds to an adiabatic transport. On the other hand for $v(t_f) = 1.5\,c$ (supersonic target speed) the l.h.s. of (\ref{ad2}) is $1.65\,t_B^{-1}$ which is $ > {\rm min}\vert \omega_{\epsilon n} \vert$ corresponding to non-adiabatic transport.

%~~~~~~~~~~~~~~~~~~~~~~~~~~~~~~~~~~~~~~~~~~~~~
\subsection{Condition for superiority of AC-QUDIT over the CDF method}
With the introduction of a CDF, the non-adiabatic transition amplitudes get modified to $\Gamma_{n\epsilon}(t)$ (see Eq. \textcolor{black}{(\ref{e1})}). We then have:

%===============================
\begin{align}\label{exponent1}
  & \int\limits_0^{t_f}dt_1\int\limits_0^{t_f}dt_2\, \Gamma_{n\epsilon}(t_1)\Gamma_{n\epsilon}^{*}(t_2) = \underbrace{\Big\vert\frac{\mu_{n\epsilon}}{\omega_{\epsilon n}}\Big\vert^2 \big\vert v_{t_f}(\omega_{\epsilon n})\big\vert^2}_{\text{velocity-PSD}} \nonumber\\
  & \hspace{1cm} + \underbrace{m^2\int d\epsilon \vert D_{n\epsilon} \vert^2 \big\vert \dot{v}_{t_f}(\omega_{\epsilon n})\big\vert^2}_{\text{acceleration-PSD}} \nonumber\\
  & +  \underbrace{2 \int d\epsilon\,{\rm Re}\Big[i\,m \Big(\frac{\mu_{n\epsilon}D_{n\epsilon}^*}{\omega_{\epsilon n}}\Big)c_{t_f}(\omega_{\epsilon n}) \Big]}_{\text{velocity-acceleration-CSD}},
\end{align}
%===============================
where  $\vert \dot{v}_{t_f}(\omega_{\epsilon n})\vert^2$ is the power-spectral density (PSD) of the trap acceleration and $c_{t_f}(\omega_{\epsilon n}) = \int_0^{t_f} dt_1 e^{-i\omega_{\epsilon n}t_1}v(t_1)\int_0^{t_f} dt_2 \Big[e^{-i\omega_{\epsilon n}t_2}\dot{v}(t_2)\Big]^*$ is the complex amplitude of the cross-spectral density (CSD) between the trap-speed and the acceleration of the trap. In order to satisfy Eq. (\ref{condition1}) we must have 
\begin{equation}\label{condition2}
{\rm Re}\Big[2i\,m \Big(\frac{\mu_{n\epsilon}D_{n\epsilon}^*}{\omega_{\epsilon n}}\Big)c_{t_f}(\omega_{\epsilon n})\Big] + m^2 \vert D_{n\epsilon}\vert^2 \big\vert \dot{v}_{t_f}(\omega_{\epsilon n})\big\vert^2 > 0
\end{equation}
%===============================
which holds for the entire parameter regime that we have explored.

\textcolor{black}{\subsection{Inadequacy of CDF to avoid irreversible loss of fidelity}}

\textcolor{black}{Unstable bound-state wave-packets undergo irreversible leakage to the continuum \textcolor{black}{\cite{kofman01, barone04, wilkinson1997experimental, keshavamurthy2011dynamical}} as described by Eqs. (\ref{cost}) and (\ref{cost1}). The exponent on the r.h.s. of Eq. (\ref{cost}) , $J[x_{\circ},\dot{x}_{\circ}]$, depends on the time-integral over the entire trajectory as shown in Eq. (\ref{cost1}). The non-adiabatic contribution to $J[x_{\circ},\dot{x}_{\circ}]$ (the first term on the r.h.s. of Eq. (\ref{cost1})), transforms to Eq. (\ref{exponent1}) in the presence of a CDF (STA), which is essentially $\Big\vert \int_0^{t_f}\Gamma_{n\epsilon}(s) ds \Big\vert^2 \geq 0$. Likewise, the phonon-mediated term is $\Big\vert \int_0^{t_f}\frac{L}{2\pi}\int\, dk\Delta^k_{n\epsilon}(s) ds \Big\vert^2 \geq 0$. Since the coupling $H_{SB}$ depends on the canonical position of the trapped-particle $x$, taking the trap velocity $v(t_f) = 0$ and trap-acceleration $\dot{v}(t_f) = 0$ at $t = t_f$ is not sufficient to cancel the phonon contribution to the irreversible loss of fidelity. This shows that irreversible loss of fidelity cannot be avoided via CDF. At best one might strive for $ \Big\vert \int_0^{t_f}\Gamma_{n\epsilon}(s) ds \Big\vert = 0$ by this method. Yet Eq. (9) suggests that this would require $\gamma_{n\epsilon}(t) = im\dot{v}(t)D_{n\epsilon}e^{-i\omega_{n\epsilon}(t)}$ for all $t > 0$ and $\epsilon >0$. This in turn requires that $v(t)\frac{\mu_{n\epsilon}}{\omega_{\epsilon n}}e^{-i\omega_{\epsilon n}t} =  im\dot{v}(t)D_{n\epsilon}e^{-i\omega_{n\epsilon}(t)}$, i.e. $v(t) = v(0)\exp[-i\int_0^tds \Big(\frac{\frac{\mu_{n\epsilon}}{\omega_{\epsilon n}}e^{-i\omega_{\epsilon n}t}}{m D_{n\epsilon}e^{-i\omega_{n\epsilon}(t)}}\Big)]$ for all $t>0$ and $\epsilon >0$. It is however impossible for a single velocity profile to satisfy an infinite set of equations of the form shown above for each value of $\epsilon > 0$.} 

\textcolor{black}{This proves the fact that the irreversible loss of fidelity cannot be avoided by merely setting $v(t_f) = 0$ and $\dot{v}(t_f) = 0$ at the end point $t = t_f$.}

\vspace{1cm}

\section*{Acknowledgments} GK is supported by the US-Israel NSF-BSF Program \textcolor{black}{and} DFG \textcolor{black}{(FOR 2724)}. IM is supported by the grant SFB F65 ``Taming Complexity in PDE system'' by the Austrian Science Fund (FWF), the FQXI program on ``Fueling quantum field machines with information'' and the Wiener Wissenschafts- und Technologie-Fonds (WWTF) project No MA16-066 (SEQUEX). \textcolor{black}{I.M. acknowledges support from the European Research Council via ERC/AdG ``Emergence in Quantum Physics" (EmQ) under Grant Agreement No. 101097858
and the Austrian Science Fund (FWF) via  SFB F65 ``Taming complexity in PDE systems". XC is supported by the Project Grant No. PID2021-126273NB-I00 funded by MCIN/AEI/10.13039/50110/0011033 and by ``ERDF A way of making Europe'' and ``ERDF Invest in your Future''.} AC thanks Mayank Shreshtha for designing Fig.~1. BA acknowledges the support, encouragement and assistance of Hillel Aharoni. AC and BA thank Pritam Chattopadhyay, Saikat Sur and Ankita Ganguly for \textcolor{black}{discussions}.

\section*{Author contributions} GK with IM and XC  have conceived and formulated the problem. AC developed the theory and AC and BA developed the numerical methods. AC and BA obtained the analytical and numerical solutions and identified the noteworthy  physical results.  All authors have jointly written the paper.

\section*{Data Availability}
Data supporting the numerical findings of this paper are available from the corresponding author upon request.

\section*{Code Availability}

The code that support the ﬁndings of this study are available from the corresponding author upon request.

%%%%%%%%%%%%%%%%%%%%%%%%%%%%%%%%%%%%%%%%%%%%%%%
%%%%%%%%%%%%%%%%%%%%%%%%%%%%%%%%%%%%%%%%%%%%%%%
%%%%%%%%%%%%%%%%%%%%%%%%%%%%%%%%%%%%%%%%%%%%%%%

\newpage
\clearpage
\onecolumngrid

\renewcommand{\theequation}{S\arabic{equation}}
\setcounter{equation}{0}
\renewcommand{\bibnumfmt}[1]{[s#1]}
\renewcommand{\citenumfont}[1]{s#1}
\renewcommand\thefigure{S-\arabic{figure}}
\setcounter{figure}{0}
\setcounter{secnumdepth}{4}

\section*{Supplementary Information: Quantum Transport Protected by Acceleration From Nonadiabaticity and Dissipation}  % Force line breaks with \\

%~~~~~~~~~~~~~~~~~~~~~~~~~~~~~~~~~~~~~~~~~~~~~~~~~~~%
\section{The Morse potential}

 In our numerical analysis, we focus on a moving Morse-trap where,

\begin{equation}\label{Morse}
       V[x - x_{\circ}(t)] = D\Big[ e^{-2a\lbrace x - x_{\circ}(t)\rbrace} - 2 e^{-a\lbrace x - x_{\circ}(t)\rbrace} \Big].
\end{equation}
Here $D$ denotes the depth of the potential, $a$ is a parameter that assumes non-zero positive values determining the width of the Morse-trap, $x_{\circ}(t)$ is the time-dependent center of the well \cite{morse29}.

\vspace{0.2cm}

\textcolor{black}{\section{Instantaneous Eigenstates (eigen basis) of a wavepacket in a moving Morse potential}}
\textcolor{black}{At each time-instant $t$ the} Hamiltonian $H_S(t)$ admits both bound (discrete) and conitnuous energy eigen-states \cite{lima05,lima06, deffner15} \textcolor{black}{which are defined as the instantaneous energy eigenstates}. The discrete sector, enumerated by integer quantum numbers $n$, has discrete eigen-frequencies given by \cite{lima05,lima06, deffner15}

\begin{equation}\label{f1}
       \omega_{n} = -\frac{a^2}{2m}(N - n)^2,
\end{equation}
where $n$ ranges from 0 to the integer part of $N$, which is given by

\begin{equation}
       \Big(N + \frac{1}{2}\Big)^2 = \frac{2\,m\,D}{a^2}.
\end{equation}

\noindent The frequency dispersion relation for the continuous part is given by \cite{lima05, lima06, deffner15}

\begin{equation}\label{f2}
     \omega_{\kappa} = \frac{a^2}{2m}\kappa^2 \; , \; \; \forall \; \kappa \in [0, \infty],
\end{equation}
where $\kappa$ denotes the continuum index of the unbound impurity state. The corresponding discrete and continuous \textcolor{black}{eigenfunctions}, $\Phi[n, x - x_{\circ}(t)], \;\Phi[\kappa, x - x_{\circ}(t)]$ respectively, are given by

\begin{equation}\label{diseig}
    \Phi(n, z(t)) = \mathcal{N}_{n}\;[z(t)]^{N-n}\;e^{-z(t)/2}\;M[-n, 2N - 2n + 1, z(t)]
\end{equation}
and
\begin{equation}\label{coneig}
    \Phi(\kappa, z(t)) = \mathcal{N}(\kappa)\;[z(t)]^{-i\kappa}\;e^{-z(t)/2}\;U[-N -i\kappa, 1 - 2i\kappa, z(t)],
\end{equation}
where 
\begin{equation}\label{zeig}
    z(t) = (2 N + 1)\,e^{-a[x - x_{\circ}(t)]},
\end{equation}
$M(a,b,z)$ and $U(a,b,z)$ are Kummer functions of first and second kind respectively and

\begin{equation}\label{normn}
      \mathcal{N}_{n} = \Big[ \frac{(2N - 2n)\Gamma(2N - n + 1)}{n! \Gamma(2N - n + 1)^2}\Big]^{\frac{1}{2}}
\end{equation}
while $\mathcal{N}(\kappa)$ is determined using $\langle \Phi[\kappa, z(t)] \vert \Phi[\kappa', z(t)]\rangle = \delta(\kappa - \kappa')$, as \cite{lima05,lima06,deffner15}
\begin{equation}\label{normk}
      \mathcal{N}(\kappa) = \frac{\vert \Gamma(-N - i\kappa)\vert}{\pi}\sqrt{\kappa \sinh(2\pi\kappa)}.
\end{equation}
\textcolor{black}{Both $\Phi[n, z(t)]$ and $\Phi[\kappa, z(t)]$} are real-valued functions of its arguments \cite{lima05, lima06, deffner15}. Motion of the trap-center $x_{\circ}(t)$ causes the time-variation of these eigenfunctions due to their dependence on $x - x_{\circ}(t)$. \textcolor{black}{However,} the instantaneous discrete and continuous eigenvalues are independent of the potential-center and hence \textcolor{black}{of} $t$.

We denote the single instantaneous bound-state in our problem by $\vert n(t)\rangle$ and the set of instantaneous continuum states by \textcolor{black}{$\lbrace \vert \kappa (t); \forall \kappa > 0\rangle \rbrace$}. \textcolor{black}{Following \cite{kurizki01}, we represent the integrals over the continuum modes as summations, for simplicity, but account for their continuous character whenever evaluation of the integrals is necessary.} 

In the instantaneous eigenbasis of the moving Morse-trap, the \textcolor{black}{system (wavepacket)} Hamiltonian can be written as

\begin{equation}\label{hamiint}
    H_S(t) = \omega_{\circ}\vert n(t)\rangle\langle n(t) \vert + \sum_{\kappa} \omega_{\kappa}\vert\kappa(t)\rangle\langle\kappa(t)\vert.
\end{equation}

%===========================
\section{Fr\"olich coupling in the instantaneous eigenbasis}

\textcolor{black}{The system-bath interaction Hamiltonian $H_{SB}$ has time-dependent matrix-elements in the instantaneous eigen basis of $H_S(t)$ and is of the general form:}

\begin{eqnarray}\label{ham2}
     H_{SB} & = & \sum_{k} \Big[ \textcolor{black}{b_{-k}}\; \Big\lbrace g_k\, e^{-ikx_{\circ}(t)}\Big\rbrace e^{-i k \lbrace x - x_{\circ}(t)\rbrace} + \textcolor{black}{b_{-k}^{\dagger}}\; \Big\lbrace g_k\, e^{-i k x_{\circ}(t)}\Big\rbrace ^* e^{i k \lbrace x - x_{\circ}(t)\rbrace}\Big] \nonumber\\
               & = & \sum_{k ,\kappa} \Big[ \textcolor{black}{b_{-k}}\; \Big\lbrace d_{n\kappa}^k(t)\,\vert n(t)\rangle\langle \kappa(t) \vert + d_{\kappa n}^k(t)\,\vert \kappa\rangle\langle n(t) \vert \Big\rbrace  + h.c.\Big]\nonumber\\
               &   & + \;\sum_{k ,\kappa,\epsilon} \Big[ \textcolor{black}{b_{-k}}\; \Big\lbrace d_{\epsilon\kappa}^k(t)\,\vert \epsilon(t)\rangle\langle \kappa(t) \vert + d_{\kappa \epsilon}^k(t)\,\vert \kappa(t)\rangle\langle \epsilon(t) \vert \Big\rbrace  + h.c.\Big]. \nonumber\\
\end{eqnarray}

\noindent \textcolor{black}{Here, the phonon-induced transition matrix elements are defined as
\begin{equation}
     d_{rs}^{k}(t) = \Big\lbrace g_k\, e^{-i k x_{\circ}(t)}\Big\rbrace \; d_{rs}^{k}.
\end{equation}
with
\begin{equation}
     d_{rs}^{k} = \langle r(t)\vert\; e^{-ik \lbrace x - x_{\circ}(t)\rbrace}\;\vert s(t)\rangle
\end{equation}
being time-independent.}

\noindent In \textcolor{black}{Eq.} (\ref{ham2}), the indices $\kappa,\epsilon$ label instantaneous eigen-states in the continuum spectrum. \textcolor{black}{The Hamiltonian} (\ref{ham2}) includes both standard Rotating Wave Approximation-(RWA) terms as well as terms which do not comply with RWA (non-RWA). The non-RWA terms play significant role in short-time dynamics \textcolor{black}{but} become negligible in the long-time limit.

%~~~~~~~~~~~~~~~~~~~~~~~~~~~~~~~~~~~~~~~~~~~~~~~~~~~%

\textcolor{black}{\section{Trapped Wavepacket Dynamics}\label{Wavepacket}}

\textcolor{black}{The state $\vert n(t)\rangle$  denotes the \textit{instantaneous} bound eigenstate (eigenfunction) of the system Hamiltonian at time $t$, given by: $\Phi[n, z(t)]$ where $z(t) \propto e^{-a \lbrace x - x_{\circ}(t)\rbrace}$ and $x - x_{\circ}(t)$ is the time-dependent relative position of the trapped-particle with respect to the \textit{instantaneous} trap center $x_{\circ}(t)$, as described in details in Eqs. (\ref{diseig}, \ref{coneig}, \ref{zeig}) ($x$ is the absolute position of the trapped particle). }

\textcolor{black}{Now, the action of a single instance of the system-bath coupling (Fr\"olich) Hamiltonian (see Eq. 2 of the manuscript) on the product state $\vert n(t)\rangle\otimes\vert 0_{\rm bath}\rangle$ results in: 
\begin{equation}\label{entangle}
      H_{SB}\vert n(t)\rangle\otimes\vert 0_{\rm bath}\rangle = \sum\limits_{k\neq0}g_k^*\, e^{ikx}\vert n(t)\rangle\otimes\vert 1^{-k}_{\rm bath}\rangle,
\end{equation}
where $\vert 1^{-k}_{\rm bath}\rangle$ indicates the many-body bath state with a single-phonon excitation having momentum $-k$. The system state $e^{ikx}\vert n(t)\rangle = e^{ikx}\Phi[n, z(t)]$ can be expressed in the form 
\begin{equation}\label{advanced}
e^{ikx_{\circ}(t)}\phi[x - x_{\circ}(t)]. 
\end{equation}
where $\phi[x - x_{\circ}(t)] = e^{ik\lbrace x -x_{\circ}(t)\rbrace}\Phi[n, z(t)]$ is a function of the relative coordinate $x - x_{\circ}(t)$. This implies that the wavepacket $e^{ikx}\vert n(t)\rangle$ has an additional momentum $k$ associated with a fixed value of the trap-centre coordinate $x_{\circ}$, expressing the overall conservation of momentum (advanced/retarded wavepacket) \cite{frohlich1954electrons, polkovnikovreview17}. On the other hand $\vert n(t)\rangle$ being the instantaneous eigenstate has no such momentum \cite{polkovnikovreview17}. This may also be understood by noting that the scattered wavepacket $e^{ikx}\vert n(t)\rangle = e^{ikx}\Phi[n, z(t)]$ is a gauge transformed eignefunction of the trap, for which the transformed canonical momentum is shifted by $k$ \cite{Sakurai, polkovnikovreview17}. }

\textcolor{black}{Alternatively, $e^{ikx}\vert n(t)\rangle$ may be thought of as an instantaneous eigenstate for a potential trap whose centre is moving with a speed $\dot{x}_{\circ}(t) + k/m$. To see this explicitly, we consider the coordinate transformation $q = x - x_{\circ}$, under which quantum mechanical wavefunctions transform as:  }

\textcolor{black}{
\begin{equation}\label{co1}
      \psi(t, q = x-x_{\circ}) = e^{i\,m\,\big\lbrace x\,\dot{x}_{\circ} - x_{\circ}\,\dot{x}_{\circ} + \frac{1}{2}\int_0^t\,\dot{x}_{\circ}^2\big\rbrace}\psi(t,x),
\end{equation}
in order to satisfy the Schr{\"o}dinger equation \cite{becor05, rouchon03}. Omitting the time-dependent phase factors which are kixed at a particular instant $t$, the above equation can be simplified as:
\begin{equation}\label{co2}
      \psi(t,q)  =  e^{i\,m\,x\,\dot{x}_{\circ}}\;\psi(t,x). 
\end{equation}
Now for the scattered wavepacket $e^{ikx}\vert n(t)\rangle = e^{ikx}\Phi[n, z(t)]$, we invoke a coordinate transformation $q_1 = x + \frac{k}{m}t$ so that the wavepacket transforms to
\begin{equation}\label{retard}
             e^{im(-\frac{k}{m})x}\;e^{ikx}\;\Phi[n, z(t)] : = \Phi[n, z_1(t)]
\end{equation}
where $z_1(t) \propto e^{-a\lbrace q_1 - (\dot{x}_{\circ} + \frac{k}{m}t)\rbrace}$. $\Phi[n, z_1(t)]$ is evidently the instantaneous eigenstate (upto an instantaneous global phase) for the potential $V[q_1 - (x_{\circ} + \frac{k}{m}t)]$ i.e. a potential moving with speed $\dot{x}_{\circ} + k/m$ (advaned/retarded trap). By similar arguments one can show that the states $\lbrace e^{ikx}\vert \epsilon(t)\rangle\rbrace$ describe eigenstates of advanced or retarded potential traps. Eq. (\ref{entangle}) then suggests that the action of the coupling $H_{SB}$ on the initial state $\vert\psi(0)\rangle = \vert \nu(0)\rangle = \vert n(0)\rangle\otimes\vert 0_{\rm bath}\rangle$, results in a many-body entangled state, made up of Fock-states with eigenstates of advanced or retarded traps moving with speed $\dot{x}_{\circ} + k/m$ . Equivalently, $H_{SB}$ couples instantaneous eigenfunctions with advanced or retarded wavepackets moving with speed $k/m$ along with the creation of a phonon with momentum $-k$, that maintains the conservation of momentum in the many-body scattering process. }
 
\vspace{0.2cm}
\noindent \textcolor{black}{On the other hand, the non-adiabatic transitions connect instantaneous eigenstates, in traps moving with the same trap-speed, with no phonon-excitation and as such, conserve the overall momentum in the many-body dynamics.}

\vspace{0.2cm}
%~~~~~~~~~~~~~~~~~~~~~~~~~~~~~~~~~~~~~~~~~~~~~~~~~~~~~~~~~~~~~~~~~~~~~~~~~~~~~~~~~~~~~~
 \textcolor{black}{Since the set of instantaneous eigenstates $\lbrace \vert n(t)\rangle, \vert \epsilon(t)\rangle\;;\; \forall \epsilon > 0\rbrace$ form a complete basis for the system at any instant of time $t$ and the bath states can be expanded in the Fock basis $\forall k \neq 0$, without loss of generality we expand the many-body (entangled) state at time $t > 0$ in a product basis (using Einstein's summation convention) as:
\begin{eqnarray}\label{genstate}
     \vert \psi(t) \rangle & = & \Big[\alpha_{\circ}(t)\,e^{-i\omega_{\circ}t}\,\vert n(t)\rangle \otimes \vert 0_{\rm bath}\rangle +\,  \sum\limits_k\alpha_{1k}(t)\,e^{-i(\omega_{\circ} + \Omega_k)t}\,\vert n(t)\rangle \otimes \vert 1^{-k}_{\rm bath}\rangle  \nonumber\\
                           &   & +\, \sum\limits_{k,k_1}\alpha_{2kk_1}(t)\,e^{-i(\omega_{\circ} + \Omega_{k} + \Omega_{k_1})t}\,\vert n(t)\rangle \otimes \vert 2^{-(k+k_1)}_{\rm bath}\rangle + ....\Big] \nonumber\\
                           &   & + \sum\limits_{\epsilon}\Big[\,\beta_{\epsilon}(t)\,e^{-i\omega_{\epsilon}t}\,\vert \epsilon(t)\rangle \otimes \vert 0_{\rm bath}\rangle +\,  \sum\limits_k\beta_{\epsilon 1k}(t)\,e^{-i(\omega_{\epsilon} + \Omega_k)t}\,\vert \epsilon(t) \rangle \otimes \vert 1^{-k}_{\rm bath}\rangle \nonumber\\
                    &   & +\, \sum\limits_{k,k_1}\beta_{\epsilon2kk_1}(t)\,e^{-i(\omega_{\epsilon} + \Omega_{k} + \Omega_{k_1})t}\,\vert \epsilon(t)\rangle \otimes \vert 2^{-(k+k_1)}_{\rm bath}\rangle + ....\Big],         
\end{eqnarray}}

\noindent \textcolor{black}{where we have used the fact that $\Omega_k$ is symmetric in $k$ (see Methods). The} quantity within the first \textcolor{black}{pair of square brackets} on the r.h.s. of (\ref{genstate}) corresponds to the ``bound'' sector of the instantaneous \textcolor{black}{ system states in many-body product basis, while that inside the second pair of square brackets} indicate the ``unbound'' sector. We note that the non-adiabatic transitions due to the motion of the trap cannot directly induce bath excitations.

\vspace{0.2cm}

\textcolor{black}{\section{Wigner-Weisskopf Treatment}}

\noindent In order to make the calculation simple and transparent, we define infinite dimensional tensors: $\bar{\beta}_{\epsilon}$, $\bar{\beta}_{\epsilon 1k}$, $\bar{\beta}_{\epsilon 2 kk_1}$,.... and $\bar{\alpha}_{1k}$, $\bar{\alpha}_{1kk_1}$ , ... etc ,  where $0 \leq \epsilon < \infty$, $0 < k,k_1, \dots < \infty $. Using these\textcolor{black}{, we} define higher-order (super)-tensors of the form \begin{equation}\label{ketvector}
      \bm{B}(t) := 
\begin{bmatrix}
\bar{\beta}_{\epsilon}\\
\bar{\beta}_{\epsilon 1k}\\
\bar{\beta}_{\epsilon 2kk_1}\\
. \\
. \\
. \\
\end{bmatrix}
\hspace{1cm};\hspace{1cm}
\bm{A}(t) := 
\begin{bmatrix}
\alpha_{\circ}  \\
\bar{\alpha}_{1k}\\
\bar{\alpha}_{2kk_1}\\
. \\
. \\
. \\
\end{bmatrix}
\end{equation}
in terms of which the Schr{\"o}dinger equation can be expressed as:

\begin{eqnarray}\label{dyn4}
       \frac{d}{dt}\bm{A} & = & -i\bm{F}^{\dagger}\bm{B}\nonumber\\
       \frac{d}{dt}\bm{B} & = & \bm{M}\bm{B} - i\bm{F}\bm{A},
\end{eqnarray}
where $\bm{M}$ and $-i\bm{F}^{\dagger}$ are super-tensor operators of the form :
 
\begin{equation}\label{tridiag}
 \begin{bmatrix}
-\bm{\tilde{\gamma}_{j\epsilon}}(t) & -i\bm{\tilde{\Delta}^{k}_{j\epsilon}}(t) &  &  & \\
-i\bm{\tilde{\Delta}^{k\dagger}_{\epsilon j}}(t) & - \bm{\tilde{\gamma}_{j\epsilon}}(t) & -i\sqrt{2}\,\bm{\tilde{\Delta}^{k_1}_{j\epsilon}}(t) &  &  \\
0 & -i\sqrt{2}\,\bm{\tilde{\Delta}^{k_1 \dagger}_{\epsilon j}}(t)  & - \bm{\tilde{\gamma}_{j\epsilon}}(t) &  &\\
  &  & \ddots & \ddots & \ddots \\
\end{bmatrix} \hspace{0.5cm} ; \hspace{0.5cm} j\in \lbrace n, \lbrace \epsilon > 0 \rbrace \rbrace .
\end{equation}
The tensor products in equations (\ref{dyn4}) use the standard convention where repeated indices are contracted (summed). \textcolor{black}{In (\ref{tridiag}), the operators $\bm{\tilde{\gamma}_{j\epsilon}}(t)$ denote} phonon-number preserving, non-adiabatic transitions \big[either bound-to-continuum ($j = n$) in case of $-i\bm{F}^{\dagger}$  or continuum-to-continuum (\textcolor{black}{$j \in \lbrace \epsilon > 0\rbrace$}) in case of $\bm{M}$\big] with \textcolor{black}{matrix} elements of the form

\begin{equation}\label{ad3}
      \gamma_{j\epsilon}(t) = \langle j(t) \vert \frac{\partial}{\partial t}\vert \epsilon(t)\rangle \, e^{-i\omega_{\epsilon j} t} = \frac{\langle j(t)\vert\frac{\partial H_S(t)}{\partial t}\vert \epsilon(t)\rangle}{\omega_{\epsilon j}} \, e^{-i\omega_{\epsilon j} t} = \dot{x}_{\circ}(t)\frac{\mu_{j\epsilon}}{\omega_{\epsilon j}}\, e^{-i\omega_{\epsilon j} t}, 
\end{equation}
where, $\omega_{\epsilon j} = \omega_{\epsilon} - \omega_j$ and $\mu_{j\epsilon} = \int\limits_{-\infty}^{\infty} dq \; \Phi[j, q]\Big[ 2\,a\,D\lbrace e^{-2aq} - e^{-aq}\rbrace\Big]\Phi[\epsilon, q]$ \textcolor{black}{and} we have used the variable substitution $q = x - x_{\circ}(t)$]. 

We note that \textcolor{black}{(\ref{ad3}) cannot strictly} describe continuum-continuum non-adiabatic transitions, \textcolor{black}{since it may tend to diverge, having a vanishingly small denominator} \cite{polkovnikov17}. This problem \textcolor{black}{may be} avoided by introducing a ``virtual gap'' in the continuous spectrum through re-defining the continuum \textcolor{black}{eigenfunctions} $\Phi(\kappa,z(t))$ as We\'yl eigen-differential wave packets, which behave \textcolor{black}{like discrete eigenfunctions} \cite{maamacheprl08, maamachepra08}. One can then arrive at the limiting (finite) value of the non-adiabatic transition rates for a gapless spectrum \cite{maamacheprl08, maamachepra08}. However, these transitions are insignificant \textcolor{black}{on the time-scales} we wish to explore.

The operators $\bm{\tilde{\Delta}^{k}_{j\epsilon}}(t)$ indicate transtions induced by the (de)excitation of a single phonon due to the Fr\"olich coupling, with matrix elements 

\begin{equation}\label{frolich}
\Delta^k_{j\epsilon}(t) = d^k_{j\epsilon}(t)\,e^{-i\omega_{\epsilon j}t}\,e^{-i\Omega_k t}.
\end{equation}
 
\textcolor{black}{We next perform the Wigner-Weisskopf calculation of the dynamics in two steps. First, we formally solve the second equation in (\ref{dyn4}), to find}

\begin{eqnarray}\label{sol1}
     \bm{B}(t) & = & T\,e^{-\int\limits_0^t\,ds \bm{M}(s)}\bm{B}(0) - i\int\limits_0^t \; ds\; T\;e^{-\int\limits_s^t\,d\tau \bm{M}(\tau)}\,\bm{F}(s)\;\bm{A}(s), 
\end{eqnarray}
where $T$ denotes the chronological time-ordering operator. Since we assume that initially the \textcolor{black}{wavepacket} was trapped in the Morse potential, we must have $\bm{B}(0) = \bm{0}$. Then \textcolor{black}{(\ref{sol1})} reduces to

\begin{eqnarray}\label{sol2}
     \bm{B}(t)  =  -\;i\int\limits_0^t \, ds\; \bm{U_M}(t,s)\;\bm{F}(s)\;\bm{A}(s),
\end{eqnarray}
where we have defined $\bm{U_M}(t,s) = T\,\exp[-\int\limits_s^t\,d\tau \bm{M}(\tau)]$. Substituting (\ref{sol2}) \textcolor{black}{in} the r.h.s. of the first equation in (\ref{dyn4}), we then have

\begin{eqnarray}\label{dyn5}
       \frac{d}{dt}\bm{A}(t) & = & -\int\limits_0^t \, ds\, \Big[\bm{F}^{\dagger}(t)\bm{U_M}(t,s)\bm{F}(s)\Big]\,\bm{A}(s).
\end{eqnarray}
Thus, the effect of the unbound sector is formally integrated out as in \textcolor{black}{the} standard Wigner-Weisskopf treatment and (\ref{dyn5}) describes the exact dynamics of the bound sector. \textcolor{black}{In the lowest order of approximation}, we replace $\bm{U_M}(t,s)$ by an identity operator and $\bm{A}(s)$ by $\bm{A}(t)$ in the kernel of (\ref{dyn5}) to have 

\begin{equation}\label{dyn6}
     \frac{d}{dt}\bm{A}(t)   =  -\int\limits_0^t \, ds\,\bm{F}^{\dagger}(t)\bm{F}(s)\,\bm{A}(t).
\end{equation} 

\textcolor{black}{We next} find an equation for the probability amplitude $\alpha_{\circ}(t)$ from (\ref{dyn6}) \textcolor{black}{in the lowest order,} by integrating out all other coefficients in the vector $\bm{A}(t)$ in the same manner as before, while using the initial condition that  the bound-sector coefficients associated with excited phonon-states were identically zero at the beginning of the quench. 

The rationale behind retaining only the lowest order contribution in the expressions for $\frac{d}{dt}\bm{A}(t)$ and hence $\frac{d}{dt}\alpha_{\circ}(t)$ is that the higher-order effects of systematic evolution become negligible upon averaging over a continuous and effectively infinite range of the bath or continuum frequencies \cite{Cohen-Tannoudji_QM, Cohen-Tannoudji_Atom-Photon, breuer02, Khalfin, keitel1995resonance, riera2021quantum, nielsen2019critical, gordon07}. This is true for couplings that are not strong enough to resolve the dynamics within the characteristic time-scales of this averaging. Since the transitions governed by $\bm{U_{M}}$ in (S27) spread the wave packet throughout the continuum, we can estimate that effects beyond the leading order would be observable within a time-scale of the order of the inverse of the continuum width (energy uncertainty \cite{Babu00}), which can be much smaller than the measurement resolution. For phonon-induced transitions, this charachteristic time can be estimated as $\tau_B = (r_B/\xi)^2\, t_B$ where $r_B$ denotes the charachteristic length-scale of inter-atomic interactions, $t_B$ being the coherence time of impurity dynamics in BEC as before \cite{nielsen2019critical}.  Since $r_B$ can be orders of magnitude smaller than the bath coherence length $\xi$, we have $\tau_B \ll t_B$ \cite{nielsen2019critical}. Both non-adiabatic and phonon-mediated coupling strengths explored here are weaker than the inverse of the charachteristic times discussed above so that our theory is accurate for times $t > {\rm max} \lbrace \tau_B\,,\, \Delta\omega^{-1} \rbrace$, $\Delta\omega$ being the continuum width.

\textcolor{black}{The resulting dynamical equation for the amplitude $\alpha_{\circ}(t)$ [defined in (\ref{genstate})], after this two-step Wigner-Weisskopf protocol, while retaining only the lowest order terms in the trap-speed and system-bath coupling (see Eq. (\ref{dyn6})) has the form: } 

\textcolor{black}{
\begin{equation}
     \frac{d\alpha_{\circ}}{dt}  =  -\int\limits_0^t\,ds \; \Sigma(t,s)\alpha_{\circ}(t),
\end{equation}
where the kernel (self-energy) is given by, $\Sigma(t,s) = \Big\lbrace \,\gamma_{n\epsilon}(t)\,\gamma_{n\epsilon}^*(s) + \Delta_{n\epsilon}^k(t)\,\Delta_{n\epsilon}^{k *}(s) \Big\rbrace $ (repeated indicies are summed). With the initial condition $\alpha_{\circ}(0) = 1$ we then solve the above equation exactly, to have}

\textcolor{black}{
\begin{eqnarray}\label{alpha}
     \alpha_{\circ}(t) & = & \exp{\Big[-\int\limits_0^t\,dt_1\int\limits_0^{t_1}\,dt_2 \;\Sigma(t_1,t_2)\Big]}\nonumber\\
                       & = & 1 - \int\limits_0^t\,dt_1\int\limits_0^{t_1}\,dt_2 \;\Big\lbrace \,\gamma_{n\epsilon}(t_1)\,\gamma_{n\epsilon}^*(t_2) + \Delta_{n\epsilon}^k(t_1)\,\Delta_{n\epsilon}^{k *}(t_2) \Big\rbrace\nonumber\\
                       &   & + \int\limits_0^t\,dt_1\int\limits_0^{t_1}\,dt_2\int\limits_0^{t_2}\,dt_3\int\limits_0^{t_3}\,dt_4 \;\Big\lbrace \,\gamma_{n\epsilon}(t_1)\,\gamma_{n\epsilon}^*(t_3) + \Delta_{n\epsilon}^k(t_1)\,\Delta_{n\epsilon}^{k *}(t_3)\Big\rbrace\nonumber\\
                       &   & \hspace{5.5cm}\Big\lbrace \,\gamma_{n\epsilon}(t_2)\,\gamma_{n\epsilon}^*(t_4) + \Delta_{n\epsilon}^k(t_2)\,\Delta_{n\epsilon}^{k *}(t_4) \Big\rbrace\nonumber\\
                       &   & + \hspace{0.5cm} \dots \dots \dots ,
\end{eqnarray}
where in the last-step we have used the fact that $\Sigma(t_1,t_2)$ is symmetric in $t_1$ and $t_2$.}

\textcolor{black}{Since $\Sigma(t_1,t_2)$ is quadratic in $\dot{x}_{\circ}$ and $H_{SB}$, the above expression is a resummation of all second-order diagrams akin to a Dyson series \cite{boyanovsky2011perturbative, Coalson19},.  This Wigner-Weisskopf non-perturbative approach is widely used in quantum optics \cite{Coalson19, boyanovsky2011perturbative, scully, sargent, louisell}.} Hence, the survival probability $\vert \alpha_{\circ}(t_f)\vert^2$ in (\ref{prob1}) goes beyond linear response, since it takes into account the cumulative effect of an infinite sequence of second-order self-energies.

\textcolor{black}{\section{Dynamical Fidelity and Loschmidt echo}}

\textcolor{black}{From the discussion in Sec. (\ref{Wavepacket}) and Methods A of the manuscript, it follows that a non-zero phonon excitation in the general state corresponds to advanced or retarded wavepackets. Thus fidelity (survival probability), defined as the transition probability between two pure quantum states \cite{nielsenchuang, liang19}  \big[in our case : perturbed state $\vert \psi (t) \rangle$ and unperturbed state $\vert n(t)\rangle \otimes \vert 0_{\rm bath} \rangle$\big] is given by $\vert \alpha_{\circ}(t)\vert^2$, i.e. the probability of finding the many-body state in $\vert n(t)\rangle\otimes\vert 0_{\rm bath}\rangle$ (Loschmidt-echo). All other many-body states having non-zero phonon excitations, corresponding to advanced or retarded wavepacket states, are orthogonal to the state $\vert n(t)\rangle\otimes\vert 0_{\rm bath}\rangle$ so that projecting on to $\vert n(t)\rangle\otimes\vert 0_{\rm bath}\rangle$ we eliminate all other phonon states that do not contribute to the fidelity.  }
 
%the probability of detecting the instantaneous bound state $\vert n(t)\rangle$ is evidently given by

\vspace{0.2cm}
\textcolor{black}{The reduced density matrix $\rho_S(t) = {\rm Tr}_{\rm bath}\big(\vert \psi(t)\rangle\langle \psi(t)\vert\big)$ is a mixed state composed of advanced or retarded (gauge-transformed) and instantaneous system eigenstates, since $\vert \psi(t)\rangle$ is an entangled system-bath state [Sec. (\ref{Wavepacket})]. For mixed states (density matrices) \cite{mendonca08, liang19}, the most common fidelity measure is the Uhlmann-Josza fidelity $F(\rho_S(t),\sigma) = \Big({\rm Tr}\sqrt{\sqrt{\rho_S(t)}\sigma\sqrt{\rho_S(t)}}\Big)^2$, including its version when one of the density matrices is pure (Schumacher's fidelity: $\langle\psi\vert\rho\vert\psi\rangle$), defined as the maximal transition probability between purifications of the two density matrices \cite{mendonca08, liang19, Schumacher95}. Yet such a fidelity measure relies on the assumption that the two density matrices being compared, are derived from identically enlarged Hilbert spaces, which is not always the case \cite{liang19}. Moreover, being the maximum over purifications, such a fidelity measure over a reduced Hilbert space may be greater than the relevant fidelities for two pure states \cite{liang19}. In our problem, the instantaneous and scattered wavepackets $\vert n(t) \rangle$ and $e^{ikx}\vert n(t) \rangle$ may have a non-zero overlap in general, due to their finite width. However, the bath states $\vert 0_{\rm bath}\rangle$, $\vert 1^{-k}_{\rm bath}\rangle$ are all orthogonal, hence the many body states $e^{ikx}\vert n(t) \rangle \otimes \vert 1^{-k}_{\rm bath}\rangle $ and $\vert n(t)\rangle \otimes \vert 0_{\rm bath}\rangle$ are orthogonal. Therfore, tracing out the bath degrees of freedom reduces the distinguishability between scattered and unscattered contributions to the instantaneous states $\vert n(t)\rangle$. As a result, the corresponding Schumacher's fidelity $\langle n(t)\vert \rho_{S}(t)\vert n(t)\rangle$ may include contributions from the scattered (advanced/retarded states) and exceed the true survival probability in the instantaneous state $\vert n(t)\rangle$, required for quantum state preservation.''}

\textcolor{black}{Note that the set of states $\mathcal{B}:=\lbrace  \vert n(t)\rangle \otimes \vert 1^{-k}_{\rm bath}\rangle , \vert n(t)\rangle \otimes \vert 2^{-(k+k_1)}_{\rm bath}\rangle .... ,  \vert \epsilon(t)\rangle \otimes \vert 0_{\rm bath}\rangle,\vert \epsilon(t)\rangle \otimes \vert 1^{-k}_{\rm bath}\rangle , \vert \epsilon(t)\rangle \otimes \vert 2^{-(k+k_1)}_{\rm bath}\rangle ...$ \; $\forall \epsilon > 0 , k\neq 0\rbrace$ defines a hyper-surface in the system-bath Fock space, orthogonal to the instantaneous state $\vert n(t)\rangle \otimes \vert 0_{\rm bath}\rangle$. On the other hand, the instantaneous scattered states $\lbrace e^{ikx}\vert n(t) \rangle \otimes \vert 1^{-k}_{\rm bath}\rangle , e^{ikx}\vert n(t)\rangle \otimes \vert 2^{-(k+k_1)}_{\rm bath}\rangle .... e^{ikx}\vert n(t) \rangle \otimes \vert 1^{-k}_{\rm bath}\rangle , e^{ikx}\vert n(t)\rangle \otimes \vert 2^{-(k+k_1)}_{\rm bath}\rangle ....\rbrace$ are also orthogonal to $\vert n(t)\rangle \otimes \vert 0_{\rm bath}\rangle$ and hence lie entierly on the hyper-surface $\mathcal{B}$. }

\vspace{0.2cm}

\textcolor{black}{So from Eq. (\ref{alpha}) of the SI, the survival probability (Loschmidt echo) at $t = t_f$, is:}

\begin{equation}\label{prob1}
    \mathcal{P}(t_f) = \big\vert \alpha_{\circ}(t_f) \big\vert ^2 = \exp\Big[- 2 \;\text{Re}\; \int\limits_0^{t_f}dt_1\;\int\limits_0^{t_1}\;dt_2\;\textcolor{black}{\sum\limits_{k, \epsilon}}\Big\lbrace \,\gamma_{n\epsilon}(t_1)\,\gamma_{n\epsilon}^*(t_2) + \Delta_{n\epsilon}^k(t_1)\,\Delta_{n\epsilon}^{k *}(t_2) \Big\rbrace\,\Big],
\end{equation}
The first term in the integrand of equation (\ref{prob1}) accounts for the non-adiabatic contribution to the loss of fidelity while the second term indicates phonon-mediated contributions to the same, arising from the non-RWA terms in \textcolor{black}{ (\ref{ham2})}.
\textcolor{black}{Using}, 

\begin{equation}\label{Gamma}
\text{Re}\big[\gamma_{n\epsilon}(t_1)\,\gamma_{n\epsilon}^*(t_2) \big] = \dot{x}_{\circ}(t_1)\dot{x}_{\circ}(t_2) \sum\limits_{\epsilon} \frac{\vert \mu_{n\epsilon} \vert^2}{\omega_{\epsilon n}^2} \cos\big[\omega_{\epsilon n}(t_1 - t_2)\big],
\end{equation}
and

\begin{equation}\label{Delta}
\color{black}{\text{Re}\big[\Delta_{n\epsilon}^k(t_1)\,\Delta_{n\epsilon}^{k *}(t_2)\big] = \sum\limits_{k \epsilon} \vert g_k d^k_{n\epsilon} \vert^2 \cos\big[(\omega_{\epsilon n} + \Omega_k)(t_1 - t_2) + k\lbrace x_{\circ}(t_1) - x_{\circ}(t_2)\rbrace \big]}
\end{equation}
are symmetric in the dummy variables $t_1$ \textcolor{black}{,} $t_2$ \textcolor{black}{and thus} we can rewrite the nested integral in (\ref{prob1}) as a double integral to get

\begin{equation}\label{prob2}
       \mathcal{P}(t) = \exp\Big(-J[x_{\circ},\dot{x}_{\circ}]\Big),
\end{equation}
where
\begin{eqnarray}\label{cost1}
    J[x_{\circ},\dot{x}_{\circ}] & = & \int\limits_0^{t_f}dt_1 \int\limits_0^{t_f}dt_2\; \dot{x}_{\circ}(t_1)\,\dot{x}_{\circ}(t_2)\,\sum_{\epsilon}\, \frac{\vert \mu_{n\epsilon} \vert^2}{\omega_{\epsilon n}^2} \cos\Big[\omega_{\epsilon n}(t_1 - t_2)\Big] \nonumber\\
                                 &   & + \, \int\limits_0^{t_f}dt_1 \int\limits_0^{t_f}dt_2\; \sum_{k,\epsilon}\, \vert g_k \, d_{n\epsilon}^k \vert ^2 \cos\Big[(\omega_{\epsilon n} + \Omega_k)(t_1 - t_2) \nonumber\\
                                 &   & \hspace{5cm} + k \big\lbrace x_{\circ}(t_1) - x_{\circ}(t_2)\big\rbrace\Big].
\end{eqnarray}

%==========================
\section{Non-local Euler-Lagrange (EL) equation}

Since $J[x_{\circ},\dot{x}_{\circ}]$ has the form of a double integral, we can calculate the first variation $\delta J$ as outlined in \cite{edelen69}. To this end, using the identity $\cos(A - B) = \cos(A)\cos(B) + \sin(A)\sin(B)$, we rewrite equation (\ref{cost1}) as 

\begin{eqnarray}
    J[x_{\circ},\dot{x}_{\circ}] & = & \int\limits_0^{t_f}dt_1 L_0\big[t_1,\dot{x}_{\circ}(t_1), K_0[t_1,\dot{x}_{\circ}] \big] + \sum_{k,\epsilon}\,\int\limits_0^{t_f}dt_1 L_1\big[t_1,x_{\circ}(t_1), K_1[x_{\circ}] \big] \nonumber\\
    &  &  \; +  \sum_{k,\epsilon}\,\int\limits_0^{t_f}dt_1 L_2\big[t_1,x_{\circ}(t_1), K_2[x_{\circ}] \big].
\end{eqnarray}

\textcolor{black}{Here}
\begin{equation}
L_0\big[t_1,\dot{x}_{\circ}(t_1), K_0[t_1,\dot{x}_{\circ}] \big] = \dot{x}_{\circ}(t_1)K_0[t_1,\dot{x}_{\circ}],
\end{equation}
\begin{equation}
K_0[t_1,\dot{x}_{\circ}]  = \int\limits_0^{t_f}dt_2\,\dot{x}_{\circ}(t_2)\,\sum_{\epsilon}\, \frac{\vert \mu_{n\epsilon} \vert^2}{\omega_{\epsilon n}^2} \cos\Big[\omega_{\epsilon n}(t_1 - t_2)\Big],
\end{equation}
\begin{equation}
L_1\big[t_1,x_{\circ}(t_1), K_1[x_{\circ}] \big] =  \vert g_k \, d_{n\epsilon}^k \vert ^2\, \cos\Big[(\omega_{\epsilon n} + \Omega_k)t_1 + k x_{\circ}(t_1)\Big] K_1[x_{\circ}],
\end{equation}
\begin{equation}
K_1[x_{\circ} \big] = \int\limits_0^{t_f} dt_2 \,\cos\Big[(\omega_{\epsilon n} + \Omega_k)t_2 + k x_{\circ}(t_2)\Big]
\end{equation}
\begin{equation}
L_2\big[t_1,x_{\circ}(t_1), K_2[x_{\circ}] \big] = \vert g_k \, d_{n\epsilon}^k \vert ^2\, \sin\Big[(\omega_{\epsilon n} + \Omega_k)t_1 + k x_{\circ}(t_1)\Big] K_2[x_{\circ}]
\end{equation}
and
\begin{equation}
K_2[x_{\circ} \big] = \int\limits_0^{t_f} dt_2 \,\sin\Big[(\omega_{\epsilon n} + \Omega_k)t_2 + k x_{\circ}(t_2)\Big].
\end{equation}
Thus,
\begin{eqnarray}\label{el1}
\delta J[x_{\circ}, \dot{x}_{\circ}] & = & \int\limits_0^{t_f} dt_1\,\Big[\frac{\partial L_0}{\partial \dot{x}_{\circ}(t_1)}\delta\dot{x}_{\circ}(t_1) + \frac{\partial L_0}{\partial K_0}\delta K_0 \Big] \nonumber\\
&  & + \sum_{k,\epsilon}\,\int\limits_0^{t_f} dt_1\,\Big[\frac{\partial L_1}{\partial x_{\circ}(t_1)}\delta x_{\circ}(t_1) + \frac{\partial L_1}{\partial K_1}\delta K_1 \Big] \nonumber\\
&  & + \sum_{k,\epsilon}\,\int\limits_0^{t_f} dt_1\,\Big[\frac{\partial L_2}{\partial x_{\circ}(t_1)}\delta x_{\circ}(t_1) + \frac{\partial L_2}{\partial K_2}\delta K_2 \Big].
\end{eqnarray}
\textcolor{black}{We evaluate}
\begin{eqnarray}\label{na1}
\int\limits_0^{t_f} dt_1\,\frac{\partial L_0}{\partial \dot{x}_{\circ}(t_1)}\delta\dot{x}_{\circ}(t_1) & = & - \int\limits_0^{t_f} dt_1\, \frac{d}{d t_1}\Big[\frac{\partial L_0}{\partial \dot{x}_{\circ}(t_1)}\Big]\delta x_{\circ}(t_1)\nonumber\\
                 & = & \int\limits_0^{t_f}dt_1\int\limits_0^{t_f}dt_2 \;\dot{x}_{\circ}(t_2)\sum_{\epsilon}\, \frac{\vert \mu_{n\epsilon} \vert^2}{\omega_{\epsilon n}} \sin\Big[\omega_{\epsilon n}(t_1 - t_2)\Big]\delta x_{\circ}(t_1)
\end{eqnarray}
and
\begin{equation}\label{refeq}
\int\limits_0^{t_f} dt_1\,\frac{\partial L_0}{\partial K_0}\delta K_0 = -\int\limits_0^{t_f}dt_1\int\limits_0^{t_f}dt_2\,\sum_{\epsilon}\, \frac{\vert \mu_{n\epsilon} \vert^2}{\omega_{\epsilon n}}\dot{x}_{\circ}(t_1)\sin\Big[\omega_{\epsilon n}(t_1 - t_2)\Big]\delta x_{\circ}(t_2).
\end{equation}
Interchanging the dummy variables $t_1$ and $t_2$ in the r.h.s. of \textcolor{black}{(\ref{refeq})} equation we have
\begin{eqnarray}\label{na2}
\int\limits_0^{t_f} dt_1\,\frac{\partial L_0}{\partial K_0}\delta K_0 & = & -\int\limits_0^{t_f}dt_2\int\limits_0^{t_f}dt_1\,\sum_{\epsilon}\, \frac{\vert \mu_{n\epsilon} \vert^2}{\omega_{\epsilon n}}\dot{x}_{\circ}(t_2)\sin\Big[\omega_{\epsilon n}(t_2 - t_1)\Big]\delta x_{\circ}(t_1) \nonumber\\
    & = & \int\limits_0^{t_f}dt_1\int\limits_0^{t_f}dt_2 \;\dot{x}_{\circ}(t_2)\sum_{\epsilon}\, \frac{\vert \mu_{n\epsilon} \vert^2}{\omega_{\epsilon n}} \sin\Big[\omega_{\epsilon n}(t_1 - t_2)\Big]\delta x_{\circ}(t_1).
\end{eqnarray}
So,
\begin{align}\label{na3}
\int\limits_0^{t_f} dt_1\,\Big[\frac{\partial L_0}{\partial \dot{x}_{\circ}(t_1)}\delta\dot{x}_{\circ}(t_1) & + \frac{\partial L_0}{\partial K_0}\delta K_0 \Big]     \nonumber\\
 & =  2 \int\limits_0^{t_f}\Bigg[dt_1\int\limits_0^{t_f}dt_2 \;\dot{x}_{\circ}(t_2)\sum_{\epsilon}\, \frac{\vert \mu_{n\epsilon} \vert^2}{\omega_{\epsilon n}}\sin\Big\lbrace\omega_{\epsilon n}(t_1 - t_2)\Big\rbrace\Bigg]\delta x_{\circ}(t_1).
\end{align}
Similarly we have,
\begin{align}\label{ph1}
\sum_{k,\epsilon}\,\int\limits_0^{t_f}dt_1 \Big[\frac{\partial L_1}{\partial x_{\circ}(t_1)}\delta x_{\circ}(t_1) & + \frac{\partial L_1}{\partial K_1}\delta K_1 \Big] \nonumber\\
 & = -2\int\limits_0^{t_f} dt_1\,\Bigg[\sum\limits_{k \epsilon} k\vert g_k d^k_{n \epsilon} \vert^2\int\limits_0^{t_f} dt_2\,\sin\Big\lbrace (\omega_{\epsilon n} + \Omega_k) t_1 + k x_{\circ}(t_1) \Big\rbrace\nonumber\\
 & \hspace{2cm}\cos\Big\lbrace (\omega_{\epsilon n} + \Omega_k) t_2 + k x_{\circ}(t_2) \Big\rbrace\Bigg]\delta x_{\circ}(t_1)
\end{align}
and
\begin{align}\label{ph2}
\sum_{k,\epsilon}\,\int\limits_0^{t_f}dt_1 \Big[\frac{\partial L_2}{\partial x_{\circ}(t_1)}\delta x_{\circ}(t_1) & + \frac{\partial L_2}{\partial K_2}\delta K_2 \Big] \nonumber\\
 & = 2\int\limits_0^{t_f} dt_1\,\Bigg[\sum\limits_{k \epsilon} k\vert g_k d^k_{n \epsilon} \vert^2\int\limits_0^{t_f} dt_2\,\cos\Big\lbrace (\omega_{\epsilon n} + \Omega_k) t_1 + k x_{\circ}(t_1) \Big\rbrace\nonumber\\
 & \hspace{2cm}\sin\Big\lbrace (\omega_{\epsilon n} + \Omega_k) t_2  + k x_{\circ}(t_2) \Big\rbrace\Bigg]\delta x_{\circ}(t_1).
\end{align}
\textcolor{black}{The first} variation of the constraint functional $J_1[\dot{x}_{\circ}]$ gives
\begin{equation}\label{varconst}
\lambda \delta J_1[\dot{x}_{\circ}] = -2\lambda\int\limits_0^{t_f} dt_1 \big[\ddot{x}_{\circ}(t_1)\big]\delta x_{\circ}(t_1).
\end{equation}
Using \textcolor{black}{(\ref{el1}),} (\ref{na3}), (\ref{ph1}), (\ref{ph2}), (\ref{varconst}) and the identity: $\sin(A - B) = \sin(A)\cos(B)$ \\ $- \cos(A)\sin(B)$ we arrive at the following form of $\delta J_{\rm tot}[x_{\circ}, \dot{x}_{\circ}]$:
\begin{eqnarray}\label{fullvar}
\delta J_{\rm tot}[x_{\circ}, \dot{x}_{\circ}] & = & 2\int\limits_0^{t_f}dt_1\Bigg[\int\limits_0^{t_f}dt_2 \;\dot{x}_{\circ}(t_2)\sum_{\epsilon}\, \frac{\vert \mu_{n\epsilon} \vert^2}{\omega_{\epsilon n}}\sin\Big\lbrace\omega_{\epsilon n}(t_1 - t_2)\Big\rbrace\nonumber\\
& & -\int\limits_0^{t_f} dt_2\,\sum\limits_{k \epsilon} k\vert g_k d^k_{n \epsilon} \vert^2\sin\Big\lbrace (\omega_{\epsilon n} + \Omega_k) (t_1 - t_2) + k \lbrace x_{\circ}(t_1) - x_{\circ}(t_2)\rbrace \Big\rbrace\nonumber\\
& & - \lambda \ddot{x}_{\circ}(t_1) \Bigg]\delta x_{\circ}(t_1).
\end{eqnarray} 
The \textit{non-local} Euler-Lagrange \textcolor{black}{(EL)} equation is then obtained from the condition:
\begin{equation}\label{el2}
\delta J_{\rm tot}[x_{\circ}, \dot{x}_{\circ}] = 0.
\end{equation}

\textcolor{black}{\section{Bound-to-continuum non-adiabatic transitions}}
Since we have a single bound state in our problem, the quantum number $n = 0$ in our case. For explicit evaluation of the optimal trajectory we need to calculate \textcolor{black}{the squared transition matrix element} $\vert \mu_{0\epsilon} \vert^2$. \textcolor{black}{From} the definition of $\mu_{n\epsilon}$ [see (\ref{ad3})] we have
\begin{eqnarray}\label{nonadmu1}
     \mu_{0\epsilon} & = & 2aD\int\limits_{0}^{\infty} \, \frac{dz}{a\,z}\;\Phi(0, z)\Big[\Bigg\lbrace \frac{z}{(2 N + 1)}\Bigg\rbrace^2 - \frac{z}{(2 N + 1)}\Big]\Phi(\epsilon, z)
\end{eqnarray}
where, \textcolor{black}{equation} (\ref{zeig}), we have used the variable substitution $z = (2 N + 1)\,e^{-a q}$. Simplifying, we have
\begin{eqnarray}\label{nonadmu2}
      \mu_{0\kappa} & = & \frac{2D}{(2 N + 1)^2}\int\limits_0^{\infty} dz \; \phi _0(z)\; z \;\phi_{\kappa}(z) \nonumber\\
             &  &  - \frac{2D}{(2 N + 1)}\int\limits_0^{\infty} dz \; \phi _0(z)\; \phi_{\kappa}(z)\nonumber\\
      & := & \frac{2D}{(2 N + 1)^2}\,I_1 - \frac{2D}{(2 N + 1)}\,I_2,
\end{eqnarray}
where in the last step we have \textcolor{black}{denoted the first and second integrals on the r.h.s. as $I_1$ and $I_2$. Using} (\ref{diseig} -- \ref{zeig}) and expressing the Kummer functions of \textcolor{black}{the} second-kind, $U(a,b,z)$\textcolor{black}{,} in terms of the Whittaker's function $W_{\lambda\mu}(z)$ \cite{abramowitz72} we can \textcolor{black}{evaluate} $I_1$ and $I_2$ \textcolor{black}{\cite{dixit15}}, as 
\begin{eqnarray}\label{I1-4}
       I_1 & = & \mathcal{N}_0\mathcal{N}(\kappa)\;\Gamma(N + 2 + i\kappa)\Gamma(N + 2 - i\kappa).
\end{eqnarray}
and
\begin{eqnarray}\label{I2-2}
     I_2 & = & \mathcal{N}_0\mathcal{N}(\kappa)\;\Gamma(N + 1 + i\kappa)\Gamma(N + 1 - i\kappa).
\end{eqnarray}

\noindent Substituting (\ref{I1-4}) and (\ref{I2-2}) on the r.h.s. of (\ref{nonadmu2}) we \textcolor{black}{obtain},

\begin{eqnarray}\label{nonadmufinal}
   \mu_{0\kappa} & = & \frac{2D\mathcal{N}_0\mathcal{N}(\kappa)}{(2 N + 1)^2}\;\Big[\;\Gamma(N + 2 + i\kappa)\;\Gamma(N + 2 - i\kappa) \nonumber\\
                             &   & - \;(2 N + 1)\;\Gamma(N + 1 + i\kappa)\;\Gamma(N + 1 - i\kappa)\;\Big].
\end{eqnarray}
\textcolor{black}{\section{Bound-to-continuum bath-mediated transition:}}

The bound-to-continuum phonon-mediated transition matrix element $d^k_{n\epsilon}$ \textcolor{black}{can be calculated as }\cite{lima05}:

\begin{eqnarray}
    d^k_{n\epsilon} & = & \langle n(t) \vert e^{-ik q} \vert \epsilon(t) \rangle \nonumber\\
                    & = &   \langle \epsilon(t) \vert e^{ik q} \vert k (t) \rangle ^*\nonumber\\
                    & = & \Big[A_n(\epsilon,k)\, _3F_2(-n, N - n + i\epsilon - ik, N - n - i\epsilon -ik;-n - ik, 2N - 2n + 1; 1)\Big]^*,\nonumber\\
\end{eqnarray}
where
\begin{eqnarray}
  A_n(\epsilon,k) & = & \sqrt{\frac{(2N - 2n)\Gamma(2N - n + 1)\epsilon \sinh(2\pi\epsilon)}{n!}}\nonumber\\
                  &   & \times \frac{(2N + 1)^{ik}(-1)^n\sinh(\pi k)\vert \Gamma(-N + i\epsilon)\vert}{i\pi^2 \Gamma(2N - 2n + 1)}\nonumber\\
                  &   & \times \Gamma(1 + n + ik)\Gamma(N - n - i\epsilon - ik)\Gamma(N - n + i\epsilon - ik)
\end{eqnarray}
and $_3F_2$ is the generalized hypergeometric function \cite{lima05}. In our problem $n = 0$, which gives $ _3F_2(0, N + i\epsilon - ik, N - i\epsilon -ik;- ik, 2N + 1; 1) = 1$ and thus $d^k_{n = 0\epsilon} = A_{n = 0}^*(\epsilon,k)$. \textcolor{black}{The} dipole transition matrix element $D_{n\epsilon}$ \textcolor{black}{is} \cite{lima05}:

\begin{eqnarray}
   D_{n\epsilon} & = & D_{\epsilon n}^* \\
   D_{\epsilon n} & = & \frac{(-1)^{n+1}\vert \Gamma(-N + i\epsilon)\vert}{\pi[(N - n)^2 + \epsilon^2]}\sqrt{\frac{\epsilon \sinh(2\pi\epsilon) (2N - 2n)}{n!\Gamma(2N-n+1)}}\vert\Gamma(1 + N + i\epsilon)\vert^2.
\end{eqnarray}

\section{Solution of the Boundary Value Problem using Green's Function}\label{Sec.BVP_GF_Theory}

\textcolor{black}{The Euler-Lagrange (EL) equation, Eq. (11), is a second-order integro-differential equation, which admits two arbitrary constants, specified by the two BCs: $x_{\circ}(0) = {\rm constant}$ and $x_{\circ}(t_f) = {\rm constant}$. In order to arrive at the simplified version Eq. (7)  we have applied an additional time-derivative on both sides of Eq. (11). This comes with a penalty of requiring a third BC: Eq. (7) is second order in $v(t)$ and thus admits two independent BCs of $v(t)$. Together with the equation 
\begin{equation}\label{1st}
   \dot{x}_{\circ}(t) = v(t),
\end{equation}
we then have three independent BCs in the solution for the optimal trajectory $x_{\circ}(t)$. Upon fixing the initial speed, $v(0)$, in addition to $x_{\circ}(0)$ and $x_{\circ}(t_f)$, we then \textbf{\textit{uniquely}} specify the optimal trajectory including the final trap-speed $v(t_f) = \dot{x}_{\circ}(t_f)$.}

\textcolor{black}{On the other hand, for comparison with CDF, the useful BCs are to set the initial and final trap speeds to be $0$  i.e. we choose $v(0)= v(t_f) = 0$ [s17]. Fixing the initial trap position $x_{\circ}(0)$ ($=0$ in our case), we then  \textbf{\textit{uniquely}} determine the optimal trap-trajectory from Eq. (7) and (\ref{1st}), which inevitably fixes the final trap-position $x_{\circ}(t_f)$. One may instead choose to fix the initial and final trap positions along with the choice of $v(0) = 0$, which would then fix $v(t_f)$.}

\vspace{0.2cm}
\textcolor{black}{In order to show the superiority of our method over CDF, we have chosen the fixed final trap speed to be non-zero, a case where CDF fails but our method works well.}

\vspace{0.2cm}
\textcolor{black}{For transport without a bath, $v(t)$ serves as the generalized coordinate for the EL optimization (see Eq. (8)). In this case, the BCs fix $v(0) = {\rm constant}$ and $v(t_f) = {\rm constant}$ instead of $x_{\circ}(0)$ and $x_{\circ}(t_f)$ . Integrating over the optimal velocity profile we then get the optimal trajectory, which is \textbf{\textit{uniquely}} determined by the initial position $x_{\circ}(0)$, which in turn fixes $x_{\circ}(t_f)$ as before. The corresponding survival probabilities are plotted in Fig. 3D. In this case, the control problem becomes solvable only after introducing the additional constraint on the total distance covered by the trap-center, $J_2[v]$, which fixes $x_{\circ}(t_f)$ while $x_{\circ}(0)$ is set by the BC.}

\vspace{0.2cm}
\textcolor{black}{To obtain the optimal trajectory $x_{\circ}(t)$, we choose to solve the boundary value problem (BVP):}

\begin{equation}\label{BVP-1}
\lambda\,\ddot{v}(t) = - \eta(t) - \zeta(t)\,v(t) + \int\limits_0^{t_f}\,ds\,\phi(t - s)\,v(s) := \mathcal{M}(t),
\end{equation}
\begin{equation}\label{BVP-2}
   v(0) = 0 \hspace{1cm};\hspace{1cm} v(t_f) = 0.
\end{equation}
\textcolor{black}{along with (\ref{1st}) with the initial condition $x_{\circ}(0) = 0$.}
First, we reduce the BVP (\ref{BVP-1}) \& (\ref{BVP-2}), \textcolor{black}{to a Fredholm integral equation \cite{tamarkin27, singh16} using the Green's function for the problem}

%Note that, we still have to provide another boundary condition while integrating the solution $v(t)$, .
\begin{equation}\label{hom-1}
\lambda\,\ddot{v}(t) = \mathcal{M}(t),
\end{equation}
\begin{equation}\label{hom-2}
   v(0) = 0 \hspace{1cm};\hspace{1cm} v(t_f) = 0.
\end{equation}
\textcolor{black}{given by} \cite{arfken13}

\begin{equation}\label{Sol0}
G(t,s) = \begin{cases}
 -\frac{1}{\lambda\,t_f}\, t\,(t_f-s)\,& 0 \leq t< s\\
 \\
-\frac{1}{\lambda\,t_f}\, s\,(t_f-t)& s  < t \leq t_f.
\end{cases}
\end{equation}
Then the solution to the BVP (\ref{hom-1}) \& (\ref{hom-2}) is given by 
\begin{equation}\label{Sol1}
v(t) = \int\limits_0^{t_f}\,ds\,G(t,s)\,\mathcal{M}(s).
\end{equation}
Using the definition of $\mathcal{M}(t)$ given in (\ref{BVP-1}) we then have to solve the Fredholm integral equation of \textcolor{black}{the} second kind:

\begin{eqnarray}\label{Solfinal-1}
v(t) = \mathcal{F}(t) + \int\limits_0^{t_f}\, ds_1\, K(t,s_1)v(s_1), 
\end{eqnarray}
where
\begin{equation}\label{Solfinal-2}
\mathcal{F}(t) = -\int\limits_0^{t_f}\,ds\,G(t,s)\,\eta(s),
\end{equation}
\begin{equation}\label{Solfinal-3}
K(t,s_1) = \mathcal{H}(t,s_1) - G_1(t,s_1),
\end{equation}
\begin{equation}\label{Solfinal-4}
G_1(t,s_1) = G(t,s_1)\,\zeta(s_1)
\end{equation}
and
\begin{equation}\label{Solfinal-5}
\mathcal{H}(t,s_1) = \int\limits_0^{t_f}\,ds\,G(t,s)\,\phi(s - s_1).
\end{equation}
Thus, using the Green's function we have reduced the BVP (\ref{BVP-1}) \& (\ref{BVP-2}) to a Fredholm integral equation (\ref{Solfinal-1}). This integral equation can be solved using the Liouville-Neumann series as

\begin{equation}\label{Solution-1} 
  v(t) = \lim_{n \rightarrow \infty} \sum_{j = 0}^n v^{(j)}(t),
\end{equation}
where 
\begin{eqnarray}\label{Solution2}
v^{(0)}(t) & = &  \mathcal{F}(t), \nonumber\\
v^{(j)}(t) & = &  \int\limits_0^{t_f}\, ds_1\, K(t,s_1)v^{(j -1)}(s_1).
\end{eqnarray}
\textcolor{black}{Integrating (\ref{Solution-1}) over $t$, with the initial condition $x_{\circ}(0) = 0$, we obtain the optimal trajectory $x_{\circ}(t)$.}
%===============================================
%\subsection{Solution of the BVP using Green's Function for finite boundary condition}

We use a modified Green's Function \textcolor{black}{when} the \textcolor{black}{BCs} are $v(0) = 0$, $ v(t_f) = c_1 \neq 0$. In this case, we rewrite the (\ref{BVP-1}) in terms of a new variable 
\begin{equation}\label{Q}
Q(t) = v(t)-c_1\frac{t}{t_f}.
\end{equation}
(\ref{BVP-1}) then becomes
\begin{equation}\label{dQ}
\lambda\,\ddot{Q}(t) = - \eta(t) - \zeta(t) \left(Q(t) + c_1\frac{t}{t_f}\right) + \int\limits_0^{t_f}\,ds\,\phi(t - s) \left(Q(s) + c_1\frac{s}{t_f}\right).
\end{equation}
Here, $Q(t)$ satisfies the homogeneous boundary conditions $Q(0)=0$ and $Q(t_f)=0$.\textcolor{black}{We} solve this equation following the \textcolor{black}{same} method \textcolor{black}{as for (\ref{BVP-1}) and (\ref{BVP-2})}.

%\vspace{1cm}
%===============================================
\section{Numerical method to compute survival probability} \label{numerical1}

Our main objective is to compute \textcolor{black}{the} survival probability $\mathcal{P}(t)$ at different times $t$ for different parameters using (\ref{prob2}) and (\ref{cost1}) \textcolor{black}{and hence to compute} $v(t)$ and $x_{\circ}(t)$ which are required to \textcolor{black}{calculate}  $\mathcal{P}(t)$. To comptute $v(t)$ we use the series solution  method \textcolor{black}{outlined} in (\ref{Solution-1}). \textcolor{black}{Namely, we numerically evaluate the r.h.s. of (\ref{Solfinal-1}) using the Green's function given in (\ref{Sol0}) and hence find the Liouville-Neumann series solution of (\ref{Solfinal-1}) as described in (\ref{Solution-1}).}
%___________________________________
\begin{figure}[t]
\centering
\includegraphics[width=0.48\textwidth,keepaspectratio]{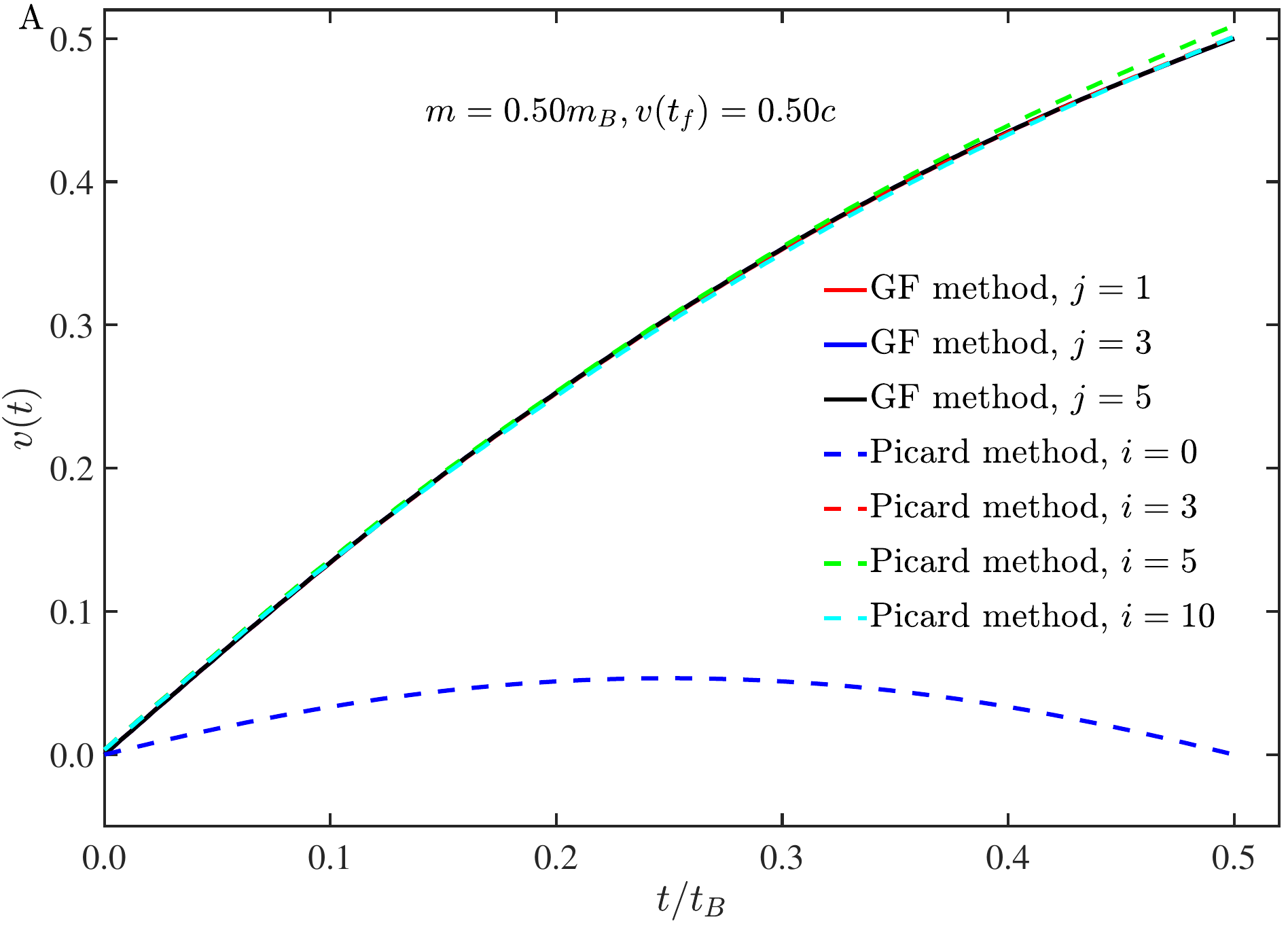}
\includegraphics[width=0.48\textwidth,keepaspectratio]{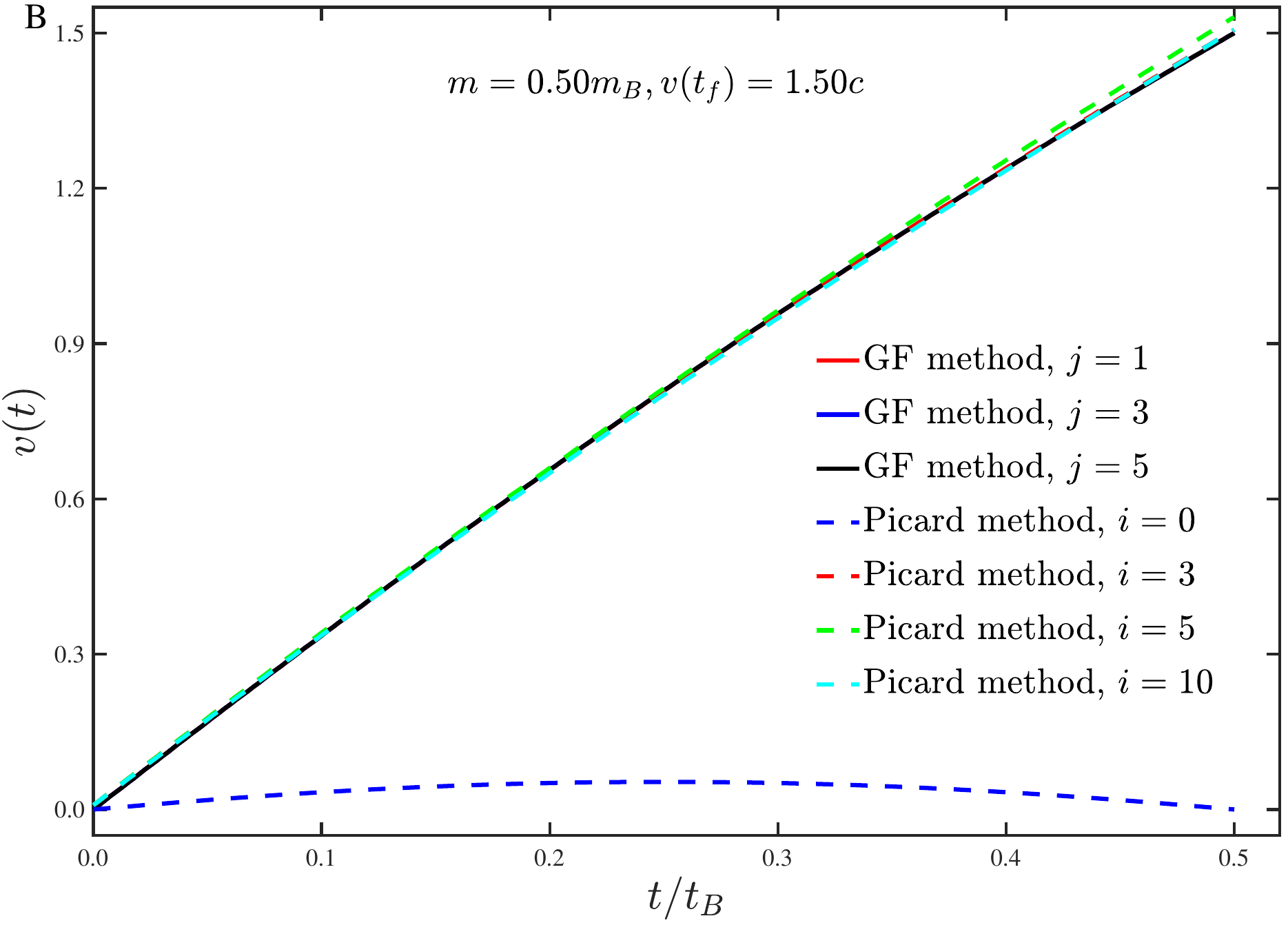}
\caption{Comparison of the convergence of the trap-velocity profile obtained from the Green's function (GF) method and the Picard method with iteration number $j$ and $i$, respectively. Note that the convergence to a stable velocity profile is achived much faster in case of the GF method. For $j>3$ and $i>5$ both the methods yield same time dependence of the optimal trap-velocity $v(t)$. The convergence of the velocity profile $v(t)$ in GF and Picard methods are illustrated for two different values of the final velocities: \textbf{A:} $v(t_f) = 0.5\,c$ and \textbf{B:} $v(t_f) = 1.5\,c$ with $m = 0.5\,m_B$.
}
\label{fig:Picard}
\end{figure}
%___________________________________

\begin{figure}[h!]
\includegraphics[width=\textwidth,keepaspectratio]{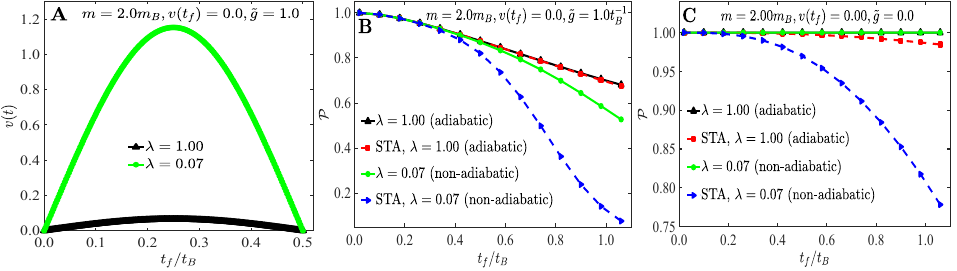}%
\caption{\textcolor{black}{\textbf{Optimal transport with} $\bm{v(t_f) = 0}$ \textbf{(A)} Optimal trap speed for $m = 2.0\,m_B$, $\tilde{g} = 1.0t_B^{-1}$ and all other parameters as given in Methods C of the main text, with $\lambda = 1$ (black) and $\lambda = 0.07$ (green) \textbf{(B)} Corresponding survival probabilities. \textbf{(C)} Survival probabilities for non-adiabatic transport through vacuum ($\tilde{g} = 0$). }}
\label{fig:vtf0}
\end{figure}

\textcolor{black}{From} (\ref{Solution-1}) and (\ref{Solution2}), the solution for $v(t)$ depends on the number of terms $j$, \textcolor{black}{being} a series solution. Thus, \textcolor{black}{we first determine the optimal $j$ required to obtian $v(t) = \dot{x}_{\circ}(t)$}. As shown in Fig.~\ref{fig:Picard}, \textcolor{black}{for $\lambda=1, m=0.5\,m_B,$ and $a=1$ we find that} the solution converges rapidly for $j \ge 3$. \textcolor{black}{Such rapid convergence occurs} also for other sets of parameters used \textcolor{black}{here}. \textcolor{black}{We thus} set $j=50$ for all the parameters used, in order to ensure a stable solution for $v(t)$. \textcolor{black}{Having obtained} $v(t)$, we numerically integrate it to obtain $x_{\circ}(t)$. With \textcolor{black}{this} optimal solution for $v(t)$ and $x_{\circ}(t)$ we numerically \textcolor{black}{evaluate the r.h.s. of} \ref{cost1} to obtain  $J[x_{\circ},\dot{x}_{\circ}]$, which \textcolor{black}{yields the} survival probability $\mathcal{P}(t)$. \textcolor{black}{Note that, using the above procedure we have calculated the optimal trajectories both in case of $v(t_f ) = 0$ and $v(t_f) = c_1 \neq 0$. In the latter case we perform the transformations (\ref{Q}) and (\ref{dQ}) before applying the numerical protocol described above. Once the optimal trajectory is obtained under this transformed scheme, we easily obtain the actual optimal trajectory by applying a back transformation of the form $v(t) = Q(t) + c_1\frac{t}{t_f}$ and use it to calculate $\mathcal{P}(t_f)$ as before. The corresponding numerical results are shown in the main text. For $v(t_f) = 0$ we directly solve (\ref{Solfinal-1}) using our numerical protocol to obtain the optimal trajectory and hence $\mathcal{P}(t_f)$ as shown in Fig. \ref{fig:vtf0}. }

%___________________________________
%___________________________________
%===============================================

\textcolor{black}{The BCs $v(0) = 0$ and $v(t_f) = 0$ yields an adiabatic optimal trajectory for $\lambda = 1$ with our chosen set of parameters, as shown in Fig. \ref{fig:vtf0} A (black curve). Even for such an adiabatic trajectory, we get survival probability nearly equal to that obtained with CDF, although our method yields slightly higher values (see Fig. \ref{fig:vtf0} B, black and red curves). To find a non-adiabatic optimal trajectory for this parameter regime, for fast transport of wavepackets, we can tune $\lambda$, which controls the total energy input for the transport. We find that for $\lambda = 0.07$ the optimal trajectory is non-adiabatic (Fig. \ref{fig:vtf0} A, green curve) with $v(t_f) = 0$ and the corresponding survival probabilities are much higher in our method than in CDF, both in the presence  and in the absence of a bath (Fig. \ref{fig:vtf0} B, C). This result confirms the optimality of our method over a wide range of parameter values.}

{\color{black}
  The derived optimal velocity profiles all satisfy the condition $\vert \dot{x}_{\circ}(t) \vert = \vert v(t) \vert < v_s = \big\vert (\omega_{\epsilon n} + \Omega_k)/k\big\vert$ necessary for the validity of the FDA and hence of the linearized Eq. (7) of the manuscript. To see this we compare the maximum value of the derived optimal velocity $\vert v(t)\vert_{\rm max}$ with the minimum value of $v_s$ i.e. $v_s\vert_{\rm min}$. We note that $v_s$ is a function of $k$ and $\epsilon$ while $\vert v(t)\vert_{\rm max}$ is independent of $k$, $\epsilon$.
%___________________________________
\begin{figure}[h!]
\centering
\includegraphics[width=0.48\textwidth,keepaspectratio]{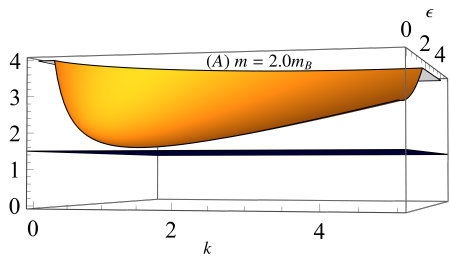}
\includegraphics[width=0.48\textwidth,keepaspectratio]{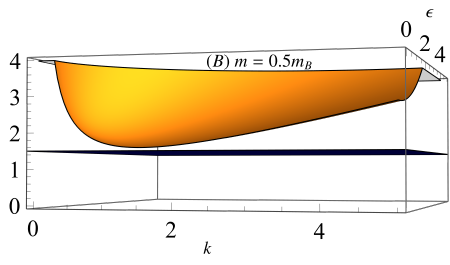}
\caption{\textcolor{black}{\textbf{Validity of the linearized Eq. 7:} \textbf{(A)} Plot of $v_s$ and $\vert v(t)\vert_{\rm max}$ for $m = 2.0\, m_B$ and $\tilde{g} = 1.0\,t_B^{-1}$. The yellow surface shows $v_s$ while the plane (black) indicates $\vert v(t)\vert_{\rm max}$. \textbf{(B)} Same for $m = 0.5\, m_B$. All other parameters are as described in Methods C.}}
\label{fig:CondtnChk}
\end{figure}
%___________________________________
From Fig. (\ref{fig:CondtnChk}) we conclude that 

\begin{equation}
       \vert v(t)\vert_{\rm max} < v_s\vert_{\rm min}   
\end{equation}

\noindent and hence

\begin{equation}
\vert \dot{x}_{\circ}(t)\vert =  \vert v(t)\vert <  v_s \; \forall\; k,\,\epsilon,\,t.
\end{equation}

Importantly, we have shown in Fig. \ref{fig:Picard} that a numerical solution of the full non-linear integro-differential equation Eq. (11) with appropriate BCs using successive Picard iterations, converge to the solutions obtained from the linearized Eq. (7) derived using FDA. This result itself confirms the validity of the FDA and hence the condition $\vert v(t)\vert < v_s$ in our results. }

\begin{figure}[t]
\centering
\includegraphics[width=\textwidth,keepaspectratio]{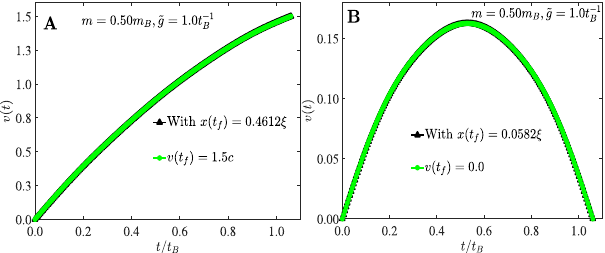}
\caption{\textcolor{black}{(A) Comparison of the velocity profile $v(t)$ obtained from two boundary conditions. The profile $v(t)$ obtained for $v(0)=0$ and $v(t_f)=1.5c$ matches exactly with $v(t)$ obtained for $x_{\circ}(0)=0, v(0)=0,$ and  $x_{\circ}(t_f)=x_f=\int_{0}^{t_f} v(t) dt = 0.4612 \xi$. (B) Same for $v(0)=0$ and $v(t_f)=0$ with $x_{\circ}(t_f)=x_f=0.0582 \xi$.}}
\label{fig:NewBC_Pos}
\end{figure}

\textcolor{black}{\section{Equivalence of different boundary conditions: Uniqueness of the optimal trajectory}}
{\color{black}
In Sec.~\ref{Sec.BVP_GF_Theory} we have presented the Green's function for the BCs $v(0)=0$ and $v(t_f)=0$. Alternatively, we here impose BCs on $x(0), v(0),$ and $x(t_f)$ and explicitly show how appropriate choices of these quantities yield identical velocity profiles, $v(t)$ (and hence identical time dependence of survival probabilities) for the two types of BCs. 

As $\dot{x}_{\circ}(t)=v(t)$, Eq.~(\ref{hom-1}) can be written as 
%---------------
\begin{equation}
\dddot{x}_{\circ}(t) = \frac{1}{\lambda}\mathcal{M}(t)\equiv g(t).
\label{Eq.3Diff}
\end{equation}
%---------------
We solve the above equation under the BCs: $x_{\circ}(0)=0, v(0)=0,$ and $x_{\circ}(t_f)=x_f$. The Green's function for a third order differential equation with the given boundary conditions is obtained as follows~\cite{Thesis_GF}. Integrating Eq.~(\ref{Eq.3Diff}) with respect to $t$ we get
%---------------
\begin{align}
\ddot{x}_{\circ}(t) &= \dot{v}(t)=d_2  + \int_{0}^{t} g(s)ds.
\end{align}
%---------------
where $d_2$ is a constant, to be determined by the BCs. Subsequent integrations over $t$ yields
%---------------
\begin{align}
\dot{x}_{\circ}(t)  &= v(t)=d_1+d_2 t  + \int_{0}^{t} (t-s)g(s)ds.\\
x_{\circ}(t)        &      =d_0+ d_1 t + \frac{1}{2} d_2 t^2+ \frac{1}{2} \int_{0}^{t} (t-s)^2 g(s)ds.
\label{Eq.SolDiff}
\end{align}
%---------------
where $d_0$ and $d_1$ are similar constants determined by the BCs. Imposing the BCs $x_{\circ}(0)=0, v(0)=0,$ and $x_{\circ}(t_f)=x_f$ we then get 
\[
d_0=0, d_1=0, ~\text{and}~ d_2= \frac{2x_f}{t_f^2}-\frac{1}{t_f^2}\int_{0}^{t_f} (t_f-s)^2 g(s)ds
\]
which yields
%---------------
\begin{align}
v(t) &=\frac{2x_f t}{t_f^2} -\frac{t}{t_f^2}\int_{0}^{t_f} (t_f-s)^2 g(s)ds  + \int_{0}^{t} (t-s)g(s)ds. \nonumber\\
\implies v(t) &=\frac{2x_f t}{t_f^2}+\int_{0}^{t} \left[ (t-s) - \frac{t}{t_f^2}(t_f-s)^2 \right] g(s)ds +\int_{t}^{t_f} \left[ -\left(\frac{t}{t_f^2}\right) (t_f-s)^2 \right] g(s)ds.
\label{Eq.GFDiff}
\end{align}
%---------------
Thus, the Green's function for the problem is
%---------------
\begin{equation}\label{Eq.NGF_Pos}
G(t,s) = \begin{cases}
\left[ (t-s) - \frac{t}{t_f^2}(t_f-s)^2 \right]\,& 0 \leq s \leq t \leq t_f\\
 \\
\left[ -\left(\frac{t}{t_f^2}\right) (t_f-s)^2 \right] \,& 0 \leq t \leq s \leq t_f.
\end{cases}
\end{equation}
%---------------
and the corresponding optimal velocity profile is given by

\begin{equation}\label{GFvopt}
v(t) =\frac{2x_f t}{t_f^2} + \int\limits_0^{t_f}\,ds\,G(t,s)\,g(s).
\end{equation}
Eq. (\ref{GFvopt}) can be written in the form of  Eq. (\ref{Solfinal-1}) i.e. a Fredholm integral equation of second kind in $v$, by absorbing the term $\frac{2x_f t}{t_f^2}$ in the inhomogeneous part $\mathcal{F}(t)$. Then Eq. (\ref{GFvopt}) can be solved directly following the method described in Sec.~\ref{Sec.BVP_GF_Theory}.

To compare the survival probabilities obtained for a given $t_f$ under the BCs: (i) $v(0)=0$ ,$v(t_f)=c_1\neq 0$, $x_{\circ}(0) = 0$ and (ii) $x_{\circ}(0)=0, v(0)=0,$ $x_{\circ}(t_f)=x_f \neq 0$, we first find the optimal velocity $v^{(i)}(t)$ corresponding to (i) and evaluate the corresponding $x_{\circ}^{(i)}(t_f)$ as the area under the velocity profile. We then find the optimal velocity $v^{(ii)}(t)$ corresponding to BCs (ii) for the same final trap centre position $x_{\circ}^{(i)}(t_f)$ i.e. by setting $x_f = x_{\circ}^{(i)}(t_f)$. In Fig.~\ref{fig:NewBC_Pos} we explicitly show that both these BCs yield identical velocity profiles  $v^{(i)}(t)$ and $v^{(ii)}(t)$ and hence the corresponding optimal transport tajectories $x_{\circ}(t)$ given by the area under the velocity curves and survival probabilities $\mathcal{P}(t_f)$ given by (\ref{prob2}--\ref{cost1}) are identical. We thus conclude that the BCs (i) and (ii) yield equivalent results confirming the uniqueness of the derived optimal trajectory as explained in Sec.~\ref{Sec.BVP_GF_Theory}.

%set $x_{f}=\int_{0}^{t_f} v(t) dt$. Here $v(t)$ is the velocity profile obtained for the given $t_f$ under the boundary condition $v(0)=0$ and $v(t_f)=c_1\neq 0$. We find that both these boundary conditions yield identical velocity profile as shown in Fig.~\ref{fig:NewBC_Pos}. Thus, the choice of the nature of the boundary conditions does not affect the dependence of the survival probability on $t_f$.

}

%___________________________________
%___________________________________

\vspace{0.2cm}

%===============================================
\section{Solution of the nonlinear problem using Picard method of successive approximations}

We have implemented a Picard iteration scheme to solve the full non-linear integro-differential equation, without invoking the Frequency Discrimantor Approximation (FDA) \textcolor{black}{(see text)}. Using \textcolor{black}{the} Picard method\textcolor{black}{, we wish} to solve a two-point \textcolor{black}{BVP} of the form
%___________________________________
\begin{align}
\ddot{v}(t) = f(t,v,\dot{v})
\end{align}
%___________________________________
with \textcolor{black}{BCs} $v(t_0)=\alpha$ and $v(t_f)=\beta$ \textcolor{black}{along with (\ref{1st}) with $x_{\circ}(0) = 0$}. Here, $t_0$ and $t_f$ represent the initial and the final boundary points respectively. \textcolor{black}{From} the Picard-Lindelof theorem, we know that if $f$ is \textcolor{black}{a locally Lipschitz function $v$ and $ \dot{v}$,} then for any $\gamma \in R$  the initial value problem
%___________________________________
\begin{align}\label{Eq:picard1}
\ddot{v}(t) = f(t,v,\dot{v}); v(t_0)=\alpha, \dot{v}(t_0)=\gamma
\end{align}
%___________________________________
\textcolor{black}{has} a unique solution \textcolor{black}{in an} interval about $t = t_0$. Introducing the variable $u = \dot{v}$ we obtain the \textcolor{black}{equivalent set of first order differential equations:}
%___________________________________
\begin{align}
\dot{v} =& u \\
\dot{u} =& f(t,v,u) \\
v(t_0)  = &\alpha, u(t_0)=\gamma.
\end{align}
%___________________________________
\textcolor{black}{We} can express \textcolor{black}{(\ref{Eq:picard1}) in the equivalent integral form}
%___________________________________
\begin{align}
v(t)=\alpha + \gamma (t-t_0) + \int_{t_0}^{t} (t-s)f(s,v(s),\dot{v}(s)) ds 
\end{align}
%___________________________________
which we then solve for $\gamma$ using the \textcolor{black}{BC,} $v(t_f) = \beta$ 
%___________________________________
\begin{align}
\gamma=\frac{1}{t_f-t_0}\left(\beta-\alpha - \int_{t_0}^{t_f} (t_f-s)f(s,v(s),\dot{v}(s)) ds\right).
\end{align}
%___________________________________
\noindent We \textcolor{black}{next use Picard iterations} to obtain successive approximations to the value of \textcolor{black}{$\gamma$ as:}
%___________________________________
\begin{align}
v^{(0)}(t) = & \alpha \\
u^{(0)}(t) = & \frac{\beta-\alpha}{t_f-t_0} \\
\gamma^{(0)}(t) = & \frac{\beta-\alpha}{t_f-t_0}\\
u^{(n+1)}(t) = &  \alpha + \int_{t_0}^{t_f} u^{(n)}(s) ds\\
v^{(n+1)}(t) = &  \gamma^{(n)} + \int_{t_0}^{t_f}f(s,v^{(n)},u^{(n)}(s)) ds\\
\gamma^{(n+1)}= &\frac{1}{t_f-t_0}\left(\beta-\alpha - \int_{t_0}^{t_f} (t_f-s)f(s,p^{(n)},u^{(n)}(s)) ds\right) 
\end{align}
%___________________________________
with $n=1,2,3 \cdots$. In our case, 
\[
f(t,v,\dot{v}) = \frac{1}{\lambda}\left[ - \eta_1(t) - \zeta_1(t)v(t) + \int\limits_0^{t_f}\,ds\,\phi(t - s)\,v(s)   \right]
\] 
with the following defining equations for $\eta_1(t), \zeta_1(t)$
%-----------------------------
\begin{eqnarray}
\eta_1(t)&=& \int\limits_0^{t_f}\;ds \; \frac{L}{2\pi}\int\,d\epsilon\,dk\, k\,(\omega_{\epsilon n} + \Omega_k)\,\vert g_k \, d_{n\epsilon}^k \vert ^2 \,\cos\Big[(\omega_{\epsilon n} + \Omega_k)(t - s) + k \lbrace x_{\circ}(t)-x_{\circ}(s)\rbrace \Big],\nonumber\\
\zeta_1(t)&=& \int\limits_0^{t_f}\;ds \; \frac{L}{2\pi}\int\,d\epsilon\,dk\, k^2\,\vert g_k \, d_{n\epsilon}^k \vert ^2\cos\Big[(\omega_{\epsilon n} + \Omega_k)(t - s) + k \lbrace x_{\circ}(t)-x_{\circ}(s)\rbrace \Big].
\end{eqnarray}
%_______________________________________________________________
%In deriving the above expression, we have used the fact that the volume of $k$ space per allowed $k$ value is $\Delta k =2\pi/L$ so that any function $f(k)$ can be conviently expressed as $\sum_k f(k) = \frac{L}{2\pi} \sum_k f(k) \Delta k$ which, in the limit as $\Delta k \to 0$ approaches $\frac{L}{2\pi}\int f(k) dk$.

%\newpage

%%%%%%%%%%%%%%%%%%%%%%%%%%%%%%%%%%%%%%%%%


\begin{thebibliography}{99}
\bibliographystyle{unsrt}

%1
\bibitem{nielsenchuang}{Nielsen, M. A. \& Chuang, I. L. {\textit{Quantum computation and quantum information} (Cambridge University Press, Cambridge, 2010)}}.

%2
\bibitem{degen17}{Degen, C. L., Reinhard, F. \& Cappellaro, P. Quantum sensing. {\textit{Rev. Mod. Phys.}} {\bf 89}, 035002 {(2017)}}.

%4
\bibitem{kkbook21}{Kurizki, G. \& Kofman, A. G. {\textit{Thermodynamics and Control of Open Quantum Systems} (Cambridge university press, Cambridge, 2022)}}.

%5
\bibitem{breuer02}{Breuer, H-P. \& Petruccione, F. {\textit{The theory of open quantum systems} (Oxford university press, Oxford, 2002)}}.

%6
\bibitem{clausen10}{Clausen, J., Bensky, G. \& Kurizki, G. Bath-Optimized Minimal-Energy Protection of Quantum Operations from Decoherence. {\textit{Phys. Rev. Lett.}} {\bf 104}, 040401 {(2010)}}.

%7
\bibitem{viola98}{Viola, L. \& Loyd, S. Dynamical suppression of decoherence in two-state quantum systems. {\textit{Phys. Rev. A}} {\bf 58}, 2733 {(1998)}}.

%8
\bibitem{gamow1928}{Gamow, G. Zur Quantentheorie des Atomkernes. {\textit{Z. Phys.}} {\bf 51}, 204 {(1928)}}.

%9
\bibitem{kofman01}{Kofman, A. G. \& Kurizki, G. Universal Dynamical Control of Quantum Mechanical Decay: Modulation of the Coupling to the Continuum. {\textit{Phys. Rev. Lett.}} {\bf 87}, 270405 {(2001)}}.

%10
\bibitem{wilkinson1997experimental}{Wilkinson, S. R., Bharucha, C. F., Fischer, M. C., Madison, K. W., Morrow, P. R., Niu, Q., Sundaram, B. \& Raizen, M. G. Experimental evidence for non-exponential decay in quantum tunnelling. {\textit{Nature}} {\bf 387}, 575--577 {(1997)}}.

%11
\bibitem{barone04}{Barone, A., Kurizki, G. \& Kofman, A. G. Dynamical Control of Macroscopic Quantum Tunneling. {\textit{Phys. Rev. Lett.}} {\bf 92}, 200403 {(2004)}}.


\bibitem{buchleitner2002non}{Buchleitner, A., Delande, D. \& Zakrzewski, J. Non-dispersive wave packets in periodically driven quantum systems.  {\textit{Phys. Rep.}}, {\bf 368}, 409--547 {(2002)}}.

%12
\bibitem{Coalson19}{Boyanovsky, D., Jasnow, D., Wu, X-L. \& Coalson, R. C. Dynamics of relaxation and dressing of a quenched Bose polaron. {\textit{Phys. Rev. A}} {\bf 100}, 043617 {(2019)}}.

%13
\bibitem{Lampo17}{Lampo, A., Lim, S. H., García-March,  M. {\'A}. \&  Lewenstein, M. Bose polaron as an instance of quantum Brownian motion. {\textit{Quantum}} {\bf 1}, 30 {(2017)}}.

%14
\bibitem{Mazets05}{Mazets, I. E., Kurizki,  G., Katz, N. \& Davidson,  N. Optically Induced Polarons in Bose-Einstein Condensates: Monitoring Composite Quasiparticle Decay. {\textit{Phys. Rev. Lett.}} {\bf 94}, 190403 {(2005)}}.

%33
\bibitem{niedenzu19}{Niedenzu, W., Mazets, I., Kurizki, G. \& Jendrzejewski, F. Quantum refrigerator for an atomic clock.  {\textit{Quantum}} {\bf 3}, 155 {(2019)}}.

%15
\bibitem{jain24}{Jain, S. et. al. Penning micro-trap for quantum computing. {\textit{Nature}} {\bf 627}, 510 -- 514 {(2024)}}.


\bibitem{sterk22}{Sterk, J. D. et al. Closed-loop optimization of fast trapped-ion shuttling with sub-quanta excitation. {\textit{Npj Quantum Inf.}} {\bf 8}, 68 {(2022)}}.

\bibitem{walther12}{Walther, A. et al. Controlling Fast Transport of Cold Trapped Ions. {\textit{Phys. Rev. Lett.}} {\bf 109}, 080501 {(2012)}}.

\bibitem{Rowe02}{Rowe, M. A.  et al. Transport of quantum states and separation of ions in a dual RF ion trap. Preperint at https://doi.org/10.48550/arXiv.quant-ph/0205094 (2002)}.

\bibitem{ibloch22}{Klostermann, T., Cabrera, C. R., von Raven, H., Wienand, J. F., Schweizer, C., Bloch, I. \& Aidelsburger, M. Fast long-distance transport of cold cesium atoms. {\textit{Phys. Rev. A}} {\bf 105}, 043319 {(2022)}}.

\bibitem{ibloch01}{Greiner, M., Bloch, I.,  H\"ansch, T. W. \& Esslinger, T. Magnetic transport of trapped cold atoms over a large distance. {\textit{Phys. Rev. A}} {\bf 63}, 031401 {(2001)}}.

%16
\bibitem{opatrny01}{Opatrn\'y, T. \& Kurizki, G. Matter-Wave Entanglement and Teleportation by Molecular Dissociation and Collisions. {\textit{Phys. Rev. Lett.}} {\bf 86}, 3180 {(2001)}}.

%17
\bibitem{deb83}{Deb, B. \& Kurizki, G. Formation of Giant Quasibound Cold Diatoms by Strong Atom-Cavity Coupling. {\textit{Phys. Rev. Lett.}} {\bf 83}, 714 {(1999)}}.


\bibitem{brumer12}{Shapiro, M. \& Brumer, P. {\textit{Quantum control of molecular processes} (John Wiley \& Sons, Weinheim, 2012)}}.


\bibitem{tannor07}{Tannor, D. J. {\textit{Introduction to Quantum Mechanics: A time-dependent perspective} (University Science Books, Sausalito, 2007)}}.

%17
\bibitem{berry09}{Berry, M. V. Transitionless quantum driving. {\textit{J. Phys. A: Math. Theor.}} {\bf 42}, 365303 {(2009)}}.

%18
\bibitem{polkovnikov17}{Sels, D. \& Polkovnikov,  A. Minimizing irreversible losses in quantum systems by local counterdiabatic driving. {\textit{Proc. Natl. Acad. Sci. U.S.A.}} {\bf 114}, E3909 {(2017)}}.

%19
\bibitem{Torrontegui11}{Torrontegui, E., Ib\'a\~nez, S., Chen, X., Ruschhaupt, A., Gu\'ery-Odelin, D. \& Muga, J. G.  Fast atomic transport without vibrational heating. {\textit{Phys. Rev. A}} {\bf 83}, 013415 {(2011)}}.

\bibitem{chen2011optimal}{Chen, X., Torrontegui, E., Stefanatos, D., Li, J-S. \& Muga, J. G.  Optimal trajectories for efficient atomic transport without final excitation. {\textit{Phys. Rev. A}} {\bf 84}, 043415 {(2011)}}.

%21
\bibitem{Chen11}{Chen, X., Torrontegui, E. \& Muga, J. G. Lewis-Riesenfeld invariants and transitionless quantum driving. {\textit{Phys. Rev. A}} {\bf 83}, 062116 {(2011)}}.

%22
\bibitem{Chen151}{Zhang, Q., Chen, X. \& Gu{\'e}ry-Odelin, D. Fast and optimal transport of atoms with nonharmonic traps. {\textit{Phys. Rev. A}} {\bf 92}, 043410 {(2015)}}.
 
%23
\bibitem{Chen152}{Ban, Y., Chen, X., Muga, J. G. \& Sherman,  E. Y. Quantum state engineering of spin-orbit-coupled ultracold atoms in a Morse potential. {\textit{Phys. Rev. A}} {\bf 91}, 023604 {(2015)}}.

%25
\bibitem{Chen22}{Zhang, Q., Chen, X. \& Gu{\'e}ry-Odelin, D. Robust Control of Linear Systems and Shortcut to Adiabaticity Based on Superoscillations. {\textit{Phys. Rev. Applied}} {\bf 18}, 054055 (2022)}.

%26
\bibitem{opatrny14}{Opatrn{\'y}, T. \&  M{\o}lmer, K. Partial suppression of nonadiabatic transitions. {\textit{New J. Phys.}} {\bf 16}, 015025 {(2014)}}.

%27
\bibitem{odelin19}{Gu{\'e}ry-Odelin, D., Ruschhaupt, A., Kiely, A., Torrontegui, E., Mart{\'i}nez-Garaot, S. \& Muga, J. G. Shortcuts to adiabaticity: Concepts, methods, and applications. {\textit{Rev. Mod. Phys.}} {\bf 91}, 045001 {(2019)}}.


\bibitem{ness18}{Ness, G., Shkedrov, C., Floshaim, Y. \& Sagi, Y. Realistic shortcuts to adiabaticity in optical transfer. {\textit{New J. Phys.}} {\bf 20}, 095002 {(2018)}}.


\bibitem{dengis2024accelerated}{Dengis, S., Wimberger, S. \& Schlagheck, P. Accelerated creation of NOON states with ultracold atoms via counterdiabatic driving. {\textit{Phys. Rev. A}} {\bf 111}, L031301 {(2025)}}.


\bibitem{LR69}{Lewis, H. R. \& Riesenfeld, W. B. An Exact Quantum Theory of the Time‐Dependent Harmonic Oscillator and of a Charged Particle in a Time‐Dependent Electromagnetic Field. {\textit{J. Math. Phys.}} {\bf 10}, 1458 {(1969)}}.

\bibitem{lewis1982direct}{Lewis, H. R. \& Leach, P. G. L. A direct approach to finding exact invariants for one‐dimensional time‐dependent classical Hamiltonians. {\textit{J. Math. Phys.}} {\bf 23}, 2371--2374 {(1982)}}.

\bibitem{villazon19}{Villazon, T., Polkovnikov, A. \& Chandran, A. Swift heat transfer by fast-forward driving in open quantum systems. {\textit{Phys. Rev. A}} {\bf 100}, 012126 {(2019)}}.

\bibitem{yin22}{Yin, Z., Li, C., Allcock, J., Zheng, Y., Gu, X., Dai, M., Zhang, S. \& An, S. Shortcuts to adiabaticity for open systems in circuit
quantum electrodynamics. {\textit{Nat. Commun.}} {\bf 13}, 188 {(2022)}}.

\bibitem{chen2015fast}{Chen, Y-H., Xia, Y., Chen, Q-Q. \& Song, J. Fast and noise-resistant implementation of quantum phase gates and creation of quantum entangled states. {\textit{Phys. Rev. A}} {\bf 91}, 012325 {(2015)}}.

\bibitem{luo2015dynamical}{Luo, D-W., Pyshkin, P. V., Lam, C-H., Yu, T., Lin, H-Q., You, J. Q. \& Wu, L-A. Dynamical invariants in a non-Markovian quantum-state-diffusion equation. {\textit{Phys. Rev. A}} {\bf 92}, 062127 {(2015)}}.

\bibitem{jing2013inverse}{Jing, J., Wu, L-A., Sarandy, M. S. \& Muga, J. G. Inverse engineering control in open quantum systems. {\textit{Phys. Rev. A}} {\bf 88}, 053422 {(2013)}}.

\bibitem{levy2018noise}{Levy, A., Kiely, A.,  Muga, J. G., Kosloff, R. \& Torrontegui, E. Noise resistant quantum control using dynamical invariants. {\textit{New J. Phys.}} {\bf 20}, 025006 {(2018)}}.

\bibitem{lu2014fast}{Lu, X-J., Muga, J. G., Chen, X., Poschinger, U. G., Schmidt-Kaler, F. \& Ruschhaupt, A. Fast shuttling of a trapped ion in the presence of noise. {\textit{Phys. Rev. A}} {\bf 89}, 063414 {(2014)}}.

\bibitem{ruschhaupt2012optimally}{Ruschhaupt, A., Chen, X., Alonso, D. \& Muga, J. G. Optimally robust shortcuts to population inversion in two-level quantum systems. {\textit{New J. Phys.}} {\bf 14}, 093040 {(2012)}}.

\bibitem{sarandy2007dynamical}{Sarandy, M. S. and Duzzioni, E. I. and Moussa, M. H. Y. Dynamical invariants and nonadiabatic geometric phases in open quantum systems. {\textit{Phys. Rev. A}} {\bf 76}, 052112 {(2007)}}.

\bibitem{wu2015dynamical}{Wu, S. L., Zhang, X. Y. \& Yi, X. X. Dynamical invariants of open quantum systems. {\textit{Phys. Rev. A}} {\bf 92}, 062122 {(2015)}}.

\bibitem{maamache2017pseudo}{Maamache, M., Djeghiour, O. K., Mana, N. \& Koussa, W. Pseudo-invariants theory and real phases for systems with non-Hermitian time-dependent Hamiltonians. {\textit{Eur. Phys. J. Plus}} {\bf 132}, 1--8 {(2017)}}.

\bibitem{vacanti2014transitionless}{Vacanti, G., Fazio, R., Montangero, S., Palma, G. M., Paternostro, M. \& Vedral, V. Transitionless quantum driving in open quantum systems. {\textit{New J. Phys.}} {\bf 16}, 053017 {(2014)}}.

\bibitem{delcampo20}{Dupays, L., Egusquiza, I. L., del Campo, A. \& Chenu, A. Superadiabatic thermalization of a quantum oscillator by engineered dephasing. {\textit{Phys. Rev. Res.}} {\bf 2}, 033178 {(2020)}}.


\bibitem{wu2017adiabatic}{Wu, S. L., Huang, X. L., Li, H. \& Yi, X. X. Adiabatic evolution of decoherence-free subspaces and its shortcuts. {\textit{Phys. Rev. A}} {\bf 96}, 042104 {(2017)}}.

\bibitem{pancotti2020speed}{Pancotti, N., Scandi, M., Mitchison, M. T. \& Perarnau-Llobet, M. Speed-Ups to Isothermality: Enhanced Quantum Thermal Machines through Control of the System-Bath Coupling. {\textit{Phys. Rev. X}} {\bf 10}, 031015 {(2020)}}.



\bibitem{alipour2020shortcuts}{Alipour, S., Chenu, A., Rezakhani, A. T. \& del Campo, A. Shortcuts to Adiabaticity in Driven Open Quantum Systems: Balanced Gain and Loss and Non-Markovian Evolution. {\textit{Quantum}} {\bf 4}, 336 {(2020)}}.


\bibitem{kosloff19}{Dann, R., Tobalina, A. \& Kosloff, R. Shortcut to Equilibration of an Open Quantum System. {\textit{Phys. Rev. Lett.}} {\bf 122}, 250402 {(2019)}}.


\bibitem{impens2019fast}{Impens, F. \& Gu{\'e}ry-Odelin, D. Fast quantum control in dissipative systems using dissipationless solutions. {\textit{Sci. Rep.}} {\bf 9}, 4048 {(2019)}}.

\bibitem{santos21}{Santos, A. C. \& Sarandy, M. S. Generalized transitionless quantum driving for open quantum systems. {\textit{Phys. Rev. A}} {\bf 104}, 062421 {(2021)}}.


\bibitem{wu21}{Wu, S. L., Ma, W., Huang, X. L. \& Yi, X. Shortcuts to Adiabaticity for Open Quantum Systems and a Mixed-State Inverse Engineering Scheme. {\textit{Phys. Rev. Appl.}} {\bf 16}, 044028 {(2021)}}.


\bibitem{mahunta2024shortcuts}{Mahunta, S. \& Mukherjee, V. Shortcuts to adiabaticity in open quantum critical systems. {\textit{Phys. Rev. B}} {\bf 111}, 064301 {(2025)}}.

\bibitem{boubakour24}{Boubakour, M., Endo, S., Fogarty, T. \& Busch, T. Dynamical invariant based shortcut to equilibration in open quantum systems. {\textit{Quantum Sci. Technol.}} {\bf 10}, 025036 {(2025)}}.

\bibitem{zhou24}{Zhou, M., C{\'a}rdenas-L{\'o}pez, F. A., Dominique, S. \& Chen, X., Optimal Control for Open Quantum System in Circuit Quantum Electrodynamics. {Preprint at https://doi.org/10.48550/arXiv.2412.20149 (2024)}}.


\bibitem{wignerweisskopf}{Weisskopf, V. \& Wigner, E. Berechnung der natürlichen Linienbreite auf Grund der Diracschen Lichttheorie. {\textit{Z. Phys.}} {\bf 63}, 54 {(1930)}}.


\bibitem{scully}{Scully, M. O. \& Zubairy, M. S. {\textit{Quantum Optics}} {Cambridge University Press, Cambridge, (1997)}}.

%32
\bibitem{zwick14}{Zwick, A., \'Alvarez, G. A.,  Bensky, G. \& Kurizki, G. Optimized dynamical control of state transfer through noisy spin chains. {\textit{New J. Phys.}}, {\bf 16}, 065021 {(2014)}}.


\bibitem{frolich52}{Fr\"olich, H. Interaction of electrons with lattice vibrations. {\textit{Proc. R. Soc. A}} {\bf 215}, 219 {(1952)}}.

%34
\bibitem{leibfried03}{Leibfried, D., Blatt, R., Monroe, C. \& Wineland, D. Quantum dynamics of single trapped ions. {\textit{Rev. Mod. Phys.}} {\bf 75}, 281 {(2003)}}.

%35
\bibitem{friedrichs1}{Friedrichs, K. O. On the perturbation of continuous spectra. {\textit{Commun. Pure Appl. Math.}} {\bf 1}, 361 {(1948)}}.


\bibitem{znidaric06}{Gorin, T., Prosen, T., Seligman, T. H. \& {\v{Z}}nidari{\v{c}}, M. Dynamics of Loschmidt echoes and fidelity decay. {\textit{Phys. Rep.}} {\bf 435 }, 33 -- 156 {(2006)}}.

\bibitem{Cohen-Tannoudji_QM}{Cohen-Tannoudji, C., Diu, B. \& Laloe, F. {\textit{Quantum Mechanics, Volume II}} {(Wiley-VCH Verlag GmbH \& Co., Weinheim, 2020)}}.

\bibitem{Cohen-Tannoudji_Atom-Photon}{Cohen-Tannoudji, C., Dupont-Roc, J. \& Grynberg, G. {\textit{Atom-photon interactions: basic processes and applications}} {(Wiley-VCH Verlag GmbH \& Co., Weinheim, 2004)}}.

\bibitem{keitel1995resonance}{Keitel, C. H., Knight, P. L., Narducci, L. M. \& Scully, M. O. Resonance fluorescence in a tailored vacuum. {\textit{Opt. Commun.}} {\bf 118}, 143--153 {(1995)}}.

\bibitem{riera2021quantum}{Riera-Campeny, A., Sanpera, A. \& Strasberg, P. Quantum systems correlated with a finite bath: Nonequilibrium dynamics and thermodynamics. {\textit{PRX Quantum}} {\bf 2}, 010340 {(2021)}}.

\bibitem{nielsen2019critical}{Nielsen, K. K., Ardila, L. A. P., Bruun, G. M. \& Pohl, T. Critical slowdown of non-equilibrium polaron dynamics. {\textit{New J. Phys.}} {\bf 21}, 043014 {(2019)}}.

\bibitem{gordon07}{Gordon, G., Erez, N. \& Kurizki, G. Universal dynamical decoherence control of noisy single-and multi-qubit systems. {\textit{J. Phys. B: At. Mol. Opt. Phys.}} {\bf 40}, S75 {(2007)}}.


\bibitem{boyanovsky2011perturbative}{Boyanovsky, D. \& Holman, R. On the perturbative stability of quantum field theories in de Sitter space. {\textit{JHEP}} {\bf 2011}, 1--37 {(2011)}}.

\bibitem{haykin01}{Haykin, S. {\textit{Communications Systems}} {(John Wiley \& Sons, Inc., New York, 2001)}}.

\bibitem{polkovnikovreview17}{Kolodrubetz, M., Sels, D., Mehta, P. \& Polkovnikov, A. Geometry and non-adiabatic response in quantum and classical systems. {\textit{Phys. Rep.}} {\bf 697}, 1 -- 87 {(2017)}}.

\bibitem{keshavamurthy2011dynamical}{Keshavamurthy, S. \& Schlagheck, P. {\textit{Dynamical tunneling: theory and experiment}} {(CRC Press, Boca Raton, 2011)}}.

\bibitem{Gal2021}{Ness, G., Alberti, A. \& Sagi, Y. Quantum Speed Limit for States with a Bounded Energy Spectrum. {\textit{Phys. Rev. Lett.}} {\bf 129}, 140403 {(2022)}}.


\bibitem{Manolo2021}{Lam, M. R. et. al. Demonstration of Quantum Brachistochrones between Distant States of an Atom. {\textit{Phys. Rev. X} } {\bf 11}, 011035 {(2021)}}.


\bibitem{Fu2022}{Fu, Q., Wang, P., Kartashov, Y. V., Konotop, V. V. \&  Ye, F. Nonlinear Thouless Pumping: Solitons and Transport Breakdown. {\textit{Phys. Rev. Lett.}} {\bf  128}, 154101 {(2022)}}.

\bibitem{gevorkyan15}{Gevorkyan, A. S., Oganesyan, K. B., Rostovtsev, Y. B. \& Kurizki, G. Gamma radiation production using channeled positron annihilation in crystals. {\textit{Laser Phys. Lett.}} {\bf 12}, 076002 {(2015)}}.

\bibitem{imambekov12}{Imambekov, A., Schmidt, T. L. \& Glazman, Leonid I. One-dimensional quantum liquids: Beyond the Luttinger liquid paradigm. {\textit{Rev. Mod. Phys.}} {\bf 84}, 1253--1306 {(2012)}}.


\bibitem{schmiedmayer19}{Schmiedmayer, J. {\textit{Thermodynamics in the Quantum Regime: Fundamental Aspects and New Directions} One-dimensional atomic superfluids as a model system for quantum thermodynamics 823--851} {(Springer, Cham, 2019)}}.

\bibitem{torrontegui12}{Torrontegui, E., Chen, X., Modugno, M., Ruschhaupt, A., Gu\'ery-Odelin, D. \& Muga, J. G. Fast transitionless expansion of cold atoms in optical Gaussian-beam traps. {\textit{Phys. Rev. A}} {\bf 85}, 033605 {(2012)}}.

\bibitem{gordon13}{Gordon, G., Mazets, I. E. \& Kurizki, G. Quantum particle localization by frequent coherent monitoring.  {\textit{Phys. Rev. A}} {\bf 87}, 052141 {(2013)}}.

\bibitem{xu2020probing}{Xu, K. et. al. Probing dynamical phase transitions with a superconducting quantum simulator. {\textit{Sci. Adv.}} {\bf 25 }, eaba4935 {(2020)}}.

\bibitem{roberts2024manybody}{Roberts, G., Vrajitoarea, A., Saxberg, B., Panetta, M. G., Simon, J. \& Schuster, D. I. Manybody interferometry of quantum fluids. {\textit{Sci. Adv.}} {\bf 10 }, eado1069 {(2024)}.}

\bibitem{cetina16}{Cetina, M. et. al. Ultrafast many-body interferometry of impurities coupled to a Fermi sea. {\textit{Science}} {\bf 354 }, 96--99 {(2016)}}.

\bibitem{braumuller2022probing}{Braum{\"u}ller, J. et. al. Probing quantum information propagation with out-of-time-ordered correlators. {\textit{Nat. Phys.}} {\bf 18 }, 172--178 {(2022)}}.

\bibitem{tonielli2020ramsey}{Tonielli, F., Chakraborty, N., Grusdt, F. \& Marino, J. Ramsey interferometry of non-Hermitian quantum impurities. {\textit{Phys. Rev. Res.}} {\bf 2}, 032003 {(2020)}}.

\bibitem{tamarkin27}{Tamarkin, J. D. The Notion of the Green's Function in the Theory of Integro-Differential Equations. {\textit{Trans. Am. Math. Soc.}} {\bf 29}, 755--800 {(1927)}}.

\bibitem{singh16}{Singh, R. \& Wazwaz, A-M. Numerical solutions of fourth-order Volterra integro-differential equations by the Green’s function and decomposition method. {\textit{Math. Sci.}} {\bf 10}, 159--166 {(2016)}}.

\bibitem{zemyan2012classical}{Zemyan, S. M. {\textit{The classical theory of integral equations: A Concise Treatment}} {(Springer Science+Business Media, Heidelberg, 2012)}}.

%51
\bibitem{lidar05}{Sarandy, M. S. \& Lidar, D. A. Adiabatic approximation in open quantum systems. {\textit{Phys. Rev. A}} {\bf 71}, 012331 {(2005)}}.


\end{thebibliography}

\begin{thebibliography}{99}

%1
\bibitem{morse29}{P. M. Morse, {Phys. Rev.} {\bf 34}, 57, {(1929)}}.

\bibitem{lima05}{E. F. de Lima and J. E. M. Hornos, {J. Phys. B: At. Mol. Opt. Phys.}, {\bf 38}, 815 -- 825, {(2005)}}.

%3
\bibitem{lima06}{E. F. de Lima and J. E. M. Hornos, {J. Chem. Phys.}, {\bf 125}, 164110, {(2006)}}.

%4
\bibitem{deffner15}{A. Leonard and S. Deffner, {Chem. Phys.}, {\bf 446}, 18, {(2015)}}.




\bibitem{boyanovsky2011perturbative}{Boyanovsky, Daniel and Holman, Richard, {JHEP}, {\bf 2011}, 1--37, {(2011)}}.



\bibitem{Coalson19}{Boyanovsky, D., Jasnow, D., Wu, X-L. and  Coalson, R. C., Dynamics of relaxation and dressing of a quenched Bose polaron. {Phys. Rev. A}, {\bf 100}, 043617, {(2019)}}.

\bibitem{scully}{Scully, M. O. and Zubairy, M. S.,{\textit{Quantum Optics}, Cambridge University Press, United Kingdom}, {(1997)}}.



\bibitem{sargent}{Sargent, M. III, Scully, M. and Lamb, W. E. {\textit{Laser Physics}}, Addison-Wesley, U.S.A., {(1974)}}.

\bibitem{louisell}{Louisell, W. {\textit{Quantum statistical properties of radiation}}, Wiley, U.S.A., {(1974)}}.




\bibitem{Sakurai}{Sakurai, J. H. and Fu Tuan, S. {\textit{Modern quantum mechanics}}, Addison-Wesley Publishing Company, U.S.A., {(1994)}}.




\bibitem{becor05}{K. Beauchard and J-M. Coron, {J. Funct.}, {\bf 232}, 328, {(2006)}}.

\bibitem{rouchon03}{P. Rouchon, {Control of a quantum particle in a moving potential well, Second IFAC Workshop on Lagrangian and Hamiltonian Methods for Nonlinear Control}, Seville, 2003.}


\bibitem{znidaric06}{T. Gorin, T. Prosen, T. H. Seligman and M. Znidaric, {Phys. Rep.}, {\bf 435 }, 33 -- 156, {(2006)}.}


\bibitem{cetina16}{M. Cetina et. al., {Science}, {\bf 354 }, 96 -- 9, {(2016)}.}

%2


%5
\bibitem{kurizki01}{A. G. Kofman and G. Kurizki, {Phys. Rev. Lett.}, {\bf 87}, 270405, {(2001)}}.


\bibitem{frohlich1954electrons}{H. Fr{\"o}hlich,  {Adv. Phys.}, {\bf 3}, 325--361, {(1954)}}.



\bibitem{polkovnikovreview17}{Kolodrubetz, M., Sels, D. , Mehta, P.  and Polkovnikov, A., Geometry and non-adiabatic response in quantum and classical systems. {Phys. Rep.}, {\bf 697}, 1 -- 87, {(2017)}}.








%6
\bibitem{polkovnikov17}{D. Sels and A. Polkovnikov, {Proc. Natl. Acad. Sci. U.S.A.}, {\bf 114}, E3909, {(2017)}}.


\textcolor{black}{\bibitem{funo2021general}{Funo, Ken and Lambert, Neill and Nori, Franco, {Phys. Rev. Lett.}, {\bf 15}, 150401, {(2021)}}.}

\textcolor{black}{\bibitem{wubs2006gauging}{Wubs, Martijn and Saito, Keiji and Kohler, Sigmund and H{\"a}nggi, Peter and Kayanuma, Yosuke, {Phys. Rev. Lett.}, {\bf 97}, 200404, {(2006)}}.}

\textcolor{black}{\bibitem{saito2007dissipative}{Saito, Keiji and Wubs, Martijn and Kohler, Sigmund and Kayanuma, Yosuke and H{\"a}nggi, Peter, {Phys. Rev. B}, {\bf 75}, 214308, {(2007)}}.}

\textcolor{black}{\bibitem{nielsenchuang}{Nielsen, M. A. and Chuang, I. L., {\textit{Quantum computation and quantum information}, Cambridge University Press, (Cambridge)}, {(2010)}}.}


\textcolor{black}{\bibitem{liang19}{Liang, Yeong-Cherng and Yeh, Yu-Hao and Mendon{\c{c}}a, Paulo EMF and Teh, Run Yan and Reid, Margaret D. and Drummond, Peter D., Quantum fidelity measures for mixed states. {Rep. Prog. Phys.}, {\bf 82}, 076001, {(2019)}}.}

\textcolor{black}{\bibitem{mendonca08}{Mendon\ifmmode \mbox{\c{c}}\else \c{c}\fi{}a, Paulo E. M. F. and Napolitano, Reginaldo d. J. and Marchiolli, Marcelo A. and Foster, Christopher J. and Liang, Yeong-Cherng, Alternative fidelity measure between quantum states. {Phys. Rev. A}, {\bf 78}, 052330, {(2008)}}.}

\textcolor{black}{\bibitem{Schumacher95}{Schumacher, B., Quantum coding, {Phys. Rev. A}, {\bf 15}, 51, {(1995)}}.}

%7
\bibitem{maamacheprl08}{M. Maamache and Y. Saadi, {Phys.Rev.Lett.}, {\bf 101}, 150407, {(2008)}}.

%8
\bibitem{maamachepra08}{M. Maamache and Y. Saadi, {Phys.Rev.A}, {\bf 78}, 052109, {(2008)}}.

\bibitem{Cohen-Tannoudji_QM}{Cohen-Tannoudji, C. and Diu, B. and Laloe, F. , Quantum Mechanics, Volume II, Second Edition, {Wiley-VCH, Germany}, {(2020}).}

\bibitem{Cohen-Tannoudji_Atom-Photon}{Cohen-Tannoudji, C. and Dupont-Roc, J. and Grynberg, G., Atom-photon interactions: basic processes and applications, {Wiley-VCH, Germany}, {(2004)}.}




\bibitem{breuer02}{Breuer, H-P. \& Petruccione, F. {\textit{The theory of open quantum systems} (Oxford university press, Oxford, 2002)}}.



\bibitem{keitel1995resonance}{Keitel, C. H., Knight, P. L., Narducci, L. M. \& Scully, M. O. Resonance fluorescence in a tailored vacuum. {\textit{Opt. Commun.}} {\bf 118}, 143--153 {(1995)}}.



\bibitem{riera2021quantum}{Riera-Campeny, A., Sanpera, A. \& Strasberg, P. Quantum systems correlated with a finite bath: Nonequilibrium dynamics and thermodynamics. {\textit{PRX Quantum}} {\bf 2}, 010340 {(2021)}}.





\bibitem{nielsen2019critical}{Nielsen, K. K., Ardila, L. A. P., Bruun, G. M. \& Pohl, T. Critical slowdown of non-equilibrium polaron dynamics. {\textit{New J. Phys.}} {\bf 21}, 043014 {(2019)}}.



\bibitem{gordon07}{Gordon, G., Erez, N. \& Kurizki, G. Universal dynamical decoherence control of noisy single-and multi-qubit systems. {\textit{J. Phys. B: At. Mol. Opt. Phys.}} {\bf 40}, S75 {(2007)}}.




\bibitem{Khalfin}{Khalfin, L. A., Contribution to the decay theory of a quasi-stationary state, {Sov. Phys. JETP}, {\bf 6}, 1053--1063 {(1958)}.}

\bibitem{Babu00}{Kofman, A. G. and Kurizki, G., Acceleration of quantum decay processes by frequent observations. {Nature}, {\bf 405}, 546--550 {(2000)}.}

%9
\bibitem{edelen69}{D. G. B. Edelen, {Int.J.Engng.Sci}, {\bf 7}, 269, {(1969)}}.

%10
\bibitem{abramowitz72}{M. Abramowitz and I. A. Stegun, {\textit{Handbook of Mathematical Functions With Formulas, Graphs, and Mathematical Tables},Tenth Printing,National Bureau of Standards (USA), Applied Mathematics Series - 55}, 504 -- 505, {(1972)}}.

%11
\bibitem{dixit15}{A. Dixit and V. H. Moll, {\textit{The integrals in Gradshteyn and Ryzhik Part 28: The confluent hypergeometric function and Whittaker functions}, Scientia Series A: Mathematical Sciences}, {\bf 26}, 49 -- 61, {(2015)}}.

%12
\bibitem{tamarkin27}{J. D. Tamarkin, {Trans. Am. Math. Soc.}, {\bf 29}, 755--800, {(1927)}}.

%13
\bibitem{singh16}{R. Singh and A-M. Wazwaz, {Math. Sci.}, {\bf 10}, 159--166, {(2016)}}.

%14
\bibitem{arfken13}{G. B. Arfken, H. J. Weber and F. E. Harris, {\textit{Mathematical Methods for Physicists, A Comprehensive Guide},$7^{\rm th}$ Edition, Elsevier, Academic Press, (USA)}, {(2013)}}.

\bibitem{clausen10}{J. Clausen, G. Bensky and G. Kurizki, {Phys.Rev.Lett.}, {\bf 104}, 040401, {(2010)}}.

\textcolor{black}{\bibitem{Thesis_GF}{S. M. Morrison,{ Application of the Green's Function for Solutions of Third Order Nonlinear Boundary Value Problems}, {University of Tennessee}, {(2007)}}.}


\end{thebibliography}
\end{document}